\documentclass[preprint,aps,prd,groupedaddress,floatfix,
nofootinbib]{revtex4}
\usepackage{graphicx}
\usepackage{epsfig}
\usepackage{graphicx}
\usepackage{longtable}

\usepackage{hyperref}

\usepackage{amsmath,amssymb,bm}

\newcommand{\U}{\text{U}}
\newcommand{\mathd}{\mathrm{d}}
\newcommand{\mathi}{\mathrm{i}}

\newcommand\Tr{{\rm Tr }}

\newcommand\tr{{\rm tr }}

\def\Comment#1{}
\newcommand{\bean}{\begin{eqnarray*}}
\newcommand{\eean}{\end{eqnarray*}}

\newcommand{\gapproxeq}{\lower
.7ex\hbox{$\;\stackrel{\textstyle >}{\sim}\;$}}
\newcommand{\lapproxeq}{\lower
.7ex\hbox{$\;\stackrel{\textstyle <}{\sim}\;$}}

\newcommand\lsim{\mathrel{\rlap{\lower4pt\hbox{\hskip1pt$\sim$}}
    \raise1pt\hbox{$<$}}}
\newcommand\gsim{\mathrel{\rlap{\lower4pt\hbox{\hskip1pt$\sim$}}
    \raise1pt\hbox{$>$}}}
\newcommand{\ba}{\begin{array}}
\newcommand{\ea}{\end{array}}
\newcommand{\nn}{\nonumber}

\newcommand{\be}{\begin{equation}}
\newcommand{\ee}{\end{equation}}
\newcommand{\bear}{\begin{eqnarray}}
\newcommand{\eear}{\end{eqnarray}}

\newcommand{\ket}{\,\rangle}
\newcommand{\bra}{\langle \,}
\newcommand{\cO}{{\cal O}}

\newcommand{\mA}{\mathcal{A}}

\newcommand{\mF}{\mathcal{F}}

\newcommand{\mO}{\mathcal{O}}

\def\bat{\begin{array}{cc}}

\newcommand{\Frac}[2]{\frac{\displaystyle #1}{\displaystyle #2}}
\newcommand{\Int}{\displaystyle{\int}}
\newcommand{\chpt}{$\chi$PT}

\begin{document}
\preprint{\vbox{\hbox{BARI-TH-2012-657 \hfill}}}
\title{Holography, chiral Lagrangian and form factor relations}
\author{P. Colangelo\footnote{Email: \textsf{Pietro.Colangelo@ba.infn.it}}}
\author{J. J.  Sanz-Cillero\footnote{Email: \textsf{Juan.SanzCillero@ba.infn.it}}}
\author{F. Zuo\footnote{Email: \textsf{fen.zuo@ba.infn.it}}}
\affiliation{Istituto Nazionale di Fisica Nucleare, Sezione di Bari, Italy}

\begin{abstract}
We derive all the $\cO(p^6)$ Chiral Perturbation Theory
low-energy constants from a class of gravity dual models of QCD
described by the Yang-Mills and Chern-Simons Lagrangian terms,
with the chiral  symmetry broken through boundary conditions in the infrared.
All the constants of  the odd intrinsic parity  sector are universally
determined by those at $\cO(p^4)$ in the even sector, together with an extra resonance term.
A few relations for the even sector couplings are also extracted.
Our estimates reasonably agree with the few available $\cO(p^6)$ determinations from
alternative phenomenological analyses.
Some of the relations between low-energy constants
are the manifestation, at large distances,
of universal relations that we find between form factors
in the even and odd sectors, e.g., between the $\gamma^*\to\pi\pi$ and
$\pi\to\gamma\gamma^*$ matrix elements.\\
\\
\\


\\

\end{abstract}


 \maketitle

\section{Introduction}

Chiral symmetry is a crucial ingredient for
the understanding of the light quark interactions.
The low-energy dynamics of the pseudo-Goldstone bosons from the spontaneous symmetry
breaking is provided by the corresponding effective field theory (EFT), Chiral Perturbation
Theory~(\chpt), with a perturbative expansion
in powers of light quark masses and external
momenta~\cite{Weinberg:1978kz,Gasser:1983yg,Gasser:1984gg,Gasser:1984ux}.
This allows a systematic description of the long-distance
regime of QCD, at energies below the lightest resonance mass.
The precision required in present
phenomenological applications makes necessary to include
corrections of $\cO(p^6)$. While many two-loop $\chi$PT   calculations
have been already carried out~\cite{Bijnens:1999sh,Bijnens:1999hw},
the large number of unknown low-energy constants (LECs) appearing
at this order puts a limit to the achievable accuracy.
The determination of these $\chi$PT couplings is compulsory
to further progress in our understanding of strong interactions
at low energies. Various techniques have allowed  the determination  of some
$\cO(p^6)$ LECs: direct comparison of next-to-next-leading order~(NNLO) $\chi$PT computations and
experiment~\cite{Bijnens:2006zp,Strandberg:2003zf},
sum rules and dispersion relations~\cite{GonzalezAlonso:2008rf,GonzalezAlonso:2010xf,Kampf:2006bn},
Pad\'e approximants~\cite{Masjuan:2008fr,Masjuan:2008fv},
resonance chiral Lagrangians~\cite{Ecker:1988te,Ecker:1989yg},
Dyson-Schwinger equation~(DSE)~\cite{Jiang:2009uf,Jiang:2010wa}.

%

In this paper we study the LECs in a class of holographic theories, which was first proposed
in ref.~\cite{Son:2003et}, based on early ideas of dimensional deconstruction
\cite{ArkaniHamed:2001ca,Hill:2000mu} and hidden local symmetry \cite{Bando:1987br}. Later, an explicit model of this
kind was constructed from string theory~, in which chiral symmetry breaking was implemented geometrically through the embedding of the
flavor branes~\cite{Sakai:2004cn}. At the same time, it was realized that chiral symmetry breaking can actually be induced by different boundary conditions~(b.c.) in the infrared~\cite{Hirn:2005nr}. In this kind of models, the chiral Lagrangian up to order $p^4$ is automatically accommodated, both of the even intrinsic parity sector~\cite{Hirn:2005nr} and of the odd sector~\cite{Sakai:2004cn,Sakai:2005yt}. Moreover, the predictions for $\cO(p^4)$ LECs have slight model dependence~\cite{Hirn:2005nr,Becciolini:2009fu}, and agree quite
well with the experimental data. This success at $\cO(p^4)$ has motivated the present
study of the $\cO(p^6)$ LECs within the same class of models. There are also other
holographic calculations of the LECs~\cite{DaRold:2005zs} in another framework involving
the quark condensate~\cite{Erlich:2005qh,DaRold:2005zs}.


The other motivation for our study is that this kind of models have recently led to an interesting relation
between the left--right quark current correlator $\Pi_{LR}(Q^2)$
and the transverse part $w_T(Q^2)$ of the anomalous $AVV$ Green's function, the so-called Son-Yamamoto relation~\cite{Son:2010vc}.
The relation does not depend on the details of the different models among this class, showing some kind of universality. When
extrapolated to the large momentum region and combined with the results of the operator product expansion in QCD,
the relation gives the same prediction of the magnetic susceptibility as the early prediction by Vainshtein \cite{Vainshtein:2002nv}.
However, while the power correction on the $\Pi_{LR}(Q^2)$ side can be included~\cite{Son:2010vc,Colangelo:2011xk,Iatrakis:2011ht}
through the dual scalar field of the quark condensate~\cite{Erlich:2005qh,DaRold:2005zs}, the holographic description of the power corrections
in $w_T(Q^2)$ is still unclear, making the naive extrapolation ambiguous. For recent studies along this line, see \cite{Cappiello:2010tu,Domokos:2011dn,Alvares:2011wb,Gorsky:2012ui}.


The corresponding analysis at low energies yields
a relation between the $\cO(p^4)$ even--parity $\chi$PT coupling $L_{10}$
and the $\cO(p^6)$ odd-intrinsic-parity coupling  $C_{22}^W$  which, respectively, rule
the left--right and $AVV$ Green's functions at large
distances~\cite{Knecht:2011wh,Kampf:2011th}. Numerically this relation is reasonably well satisfied
at low energies~\cite{Colangelo:2011xk}. Through a detailed analysis of the $\cO(p^6)$ predictions for the LECs of the odd-intrinsic parity sector in this kind of models, we find further relations
between the remaining LECs $C_j^W$ and the $\cO(p^4)$ chiral couplings from the
even-parity sector. A few relations among the $\cO(p^6)$ LECs in the even sector have also been found.

At this stage, one may wonder whether these relations between even
and odd sector LECs hint a more general interplay between even-parity
and anomalous QCD amplitudes. Here we have focused on
the odd couplings $C_{22}^W$ and $C_{23}^W$, which can be directly related to the transition of a pion into
two photons and two axial-vector currents, respectively. We shall show that the former amplitude can be related to the vector form factor of the pion
and the latter to the axial-vector form factor into three pions~\cite{GomezDumm:2003ku,Dumm:2009va}. The relation involving
 the vector form factor has already been found in two specific models \cite{Grigoryan:2008cc,Stoffers:2011xe}.

In Sec.~\ref{sec.Hologr-model} we provide the details of the class of holographic models employed in our analysis,
and review the previous results for the $\cO(p^2)$ and $\cO(p^4)$
LECs.  In Sec.~\ref{sec.op6-LECs} the NNLO chiral couplings are computed, with the help of a series of
novel sum rules involving the resonance couplings.
In addition, we shall check how well these sum rules are
saturated by the lightest resonances.
The two relations between the the form factors of the pion in the even and odd sector are derived
 in Sec.~\ref{sec.amplitude-rel}.
Our conclusions are gathered in Sec.~\ref{sec.conclusions}.
Some technical details about the different holographic models are collected
in Appendix~\ref{app.holographic-models}.
In Appendix~\ref{app.1loop-SY-rel} we study the proposal in
ref.~\cite{Gorsky:2012ui} that the Son-Yamamoto relation could hold at the loop level:
we show that the amplitudes in general do not match beyond tree-level.

\section{The holographic model}
\label{sec.Hologr-model}

\subsection{The 5D  action}

In the kind of models studied in this paper, with the chiral symmetry broken through b.c.'s,
the action is composed by the Yang--Mills (YM) and Chern--Simons (CS) terms, describing
the even and anomalous QCD sectors,
respectively~\cite{Son:2003et,Hirn:2005nr,Sakai:2004cn,Sakai:2005yt}:
\begin{eqnarray}
  S &=& S_{\rm YM}+S_{\rm CS}
  \label{eq.5Daction}
  \\
  \label{eq:YM}
  S_{\rm YM} &=& -\int\! d^5x \tr \left[-f^2(z){\cal F}_{z\mu}^2
  + \frac{1}{2g^2(z)}{\cal F}_{\mu\nu}^2 \right], \\
  \label{eq:CS}
  S_{\rm CS} &=& -\kappa \int\! \tr
  \left[{\cal AF}^2+\frac{i}{2}{\cal A}^3{\cal F}-\frac{1}{10}{\cal A}^5
\right].
\end{eqnarray}
The fifth coordinate $z$ runs from $-z_0$ to $z_0$ with $0<z_0\le+\infty$. ${\cal A}(x,z)={\cal A}_M dx^M$ is the 5D $\U(N_f)$ gauge field and
${\cal F}=d{\cal A}-i{\cal A} \wedge {\cal A}$ is the field strength.
They are decomposed as ${\cal A}= {\cal A}^a t^a$ and
${\cal F}= {\cal F}^a t^a$, with the normalization of the generators Tr$\{t^at^b\}=\delta^{ab}/2$.
The coefficient $\kappa=N_C/(24\pi^2)$, with $N_C$ the number of colors, is fixed by the chiral anomaly of
QCD~\cite{Wess:1971yu,Witten:1983tw,Witten:1983tx}.
The functions $f^2(z)$ and $g^2(z)$ are  invariant under the reflection $z\to -z$
so that parity can be properly defined in the model.
In Appendix~\ref{app.holographic-models} we provide their explicit definitions
for the models we analyze here: flat metric~\cite{Son:2003et}, {``}Cosh'' model~\cite{Son:2003et},
hard wall~\cite{Hirn:2005nr} and the Sakai-Sugimoto~(SS) model~\cite{Sakai:2004cn,Sakai:2005yt}.


As first shown in ref. \cite{Son:2003et}, chiral symmetry can be realized
as a 5D gauge symmetry localized on the two boundaries at $z=\pm z_0$.
The gauging of the chiral symmetry allows one to naturally introduce
the corresponding right and left current sources, respectively $r_\mu(x)$ and $\ell_\mu(x)$.
The Goldstone bosons are contained in the gauge component ${\cal A}_z$
and can be parameterized through the chiral field $U$ as
\begin{equation}
U(x^\mu)=\mbox{P} \exp\left\{i\int^{+z_0}_{-z_0} {\cal A}_z(x^\mu,z') dz'\right\},
\end{equation}
which transforms as
\begin{equation}
U(x)\to g_R(x) U(x) g_L^\dagger(x)
\end{equation}
with $g_L(x)$ and $g_R(x)$ the gauge transformations located at $z=-z_0$
and $z=z_0$, respectively.  The vector/axial-vector resonances are contained in the
gauge field components ${\cal A}_\mu$,
\begin{equation}
{\cal A}_\mu(x,z)=\ell_\mu(x) \psi_-(z)+r_\mu(x) \psi_+(z)
+\sum_{n=1}^\infty B_\mu^{(n)}(x) \psi_n(z)\, ,
\label{eq.Amu-decomposition1}
\end{equation}
with the UV boundary conditions
\be
\mA_\mu(x, -z_0)=\ell_\mu(x)\, ,\qquad\qquad
\mA_\mu(x,z_0)=r_\mu(x)\, .
\label{eq.Amu-bcs}
\ee
Note that the above boundary conditions are different from those in
refs.~\cite{Sakai:2005yt,Son:2010vc}, and correspondingly the sign
of the CS action is different.

The resonance wave-functions $\psi_n(z)$ are provided by
the normalizable eigenfunctions of the equation of motion for the transverse part of the gauge field,
\begin{equation}
-g^2(z)\, \partial_z[f^2(z)\partial_z {\cal A}_\mu(q ,z)]
\, =\, q^2 {\cal A}_\mu(q ,z) \, ,
\label{eq.5D-EoM}
\end{equation}
with the resonance masses given by the eigenvalues  $q^2=m_n^2$,
with b.c.'s $\psi_n(\pm z_0)=0$. The 4D metric signature $(+,-,-,-)$ is assumed all along the article.
In order to have canonically normalized kinetic terms for the resonance fields
in the later derivation, the orthogonal wave-functions  are chosen to be normalized in the form
\begin{equation}
\int_{-z_0}^{+z_0} \frac{1}{g^2(z)}\psi_n(z)\psi_m(z)~ \mathd z=\delta_{nm}\, ,
\label{eq.norma-psin}
\end{equation}
leading to the completeness condition,
\begin{equation}
\sum_{n=1}^\infty \frac{1}{g^2(z)}\psi_n(z)\psi_n(z')=\delta(z-z')\, .
\label{eq:completeness}
\end{equation}

In addition, one has $\psi_\pm(z)=\frac{1}{2}(1\pm\psi_0(z))$,
where $\psi_0(z)$ is the solution of the EoM at $q^2=0$
with   b.c.'s   $\psi_0(\pm z_0)= \pm 1$.
It is  non-normalizable and will provide the chiral Goldstone wave-function.

The solutions  $\psi_n(z)$ are even (odd) functions of $z$ when $n$ is odd (even).
Thus,  the modes with odd $n$ describe vector
excitations $v_\mu^n=B_\mu^{(2n-1)}$ with $m_{v^n}^2=m_{2n-1}^2$,
and those with even $n$ correspond to axial-vector resonances in
an analogous way.


We consider the convenient 5D gauge ${\cal A}_z=0$,
which can be achieved through the transformation
${\cal A}_M \to g {\cal A}_M g^{-1} +i g\partial _M g^{-1}$ with
\begin{equation}
g^{-1}(x,z)= P \exp\left\{i
\int_{0}^z dz'\, {\cal A}_z(x,z')\right\} \ .
\end{equation}
The value of this transformation  on the UV branes is given by
$\xi_{R (L)}(x) \equiv g^{-1}(x,\pm z_0)$, which now encode
the original information from $\mA_z$.
The chiral Goldstones can be then  non-linearly realized  through the coset
representative $\xi_R(x) =\xi_L(x)^\dagger \equiv u(\pi)=
\exp\{ i \pi^a t^a  /f_\pi\}$~\cite{Bijnens:1999sh,Bijnens:1999hw,Ecker:1988te,Ecker:1989yg}.
Conversely, after the gauge transformation the space-time  components of the  5D field
in eq.~(\ref{eq.Amu-decomposition1})   takes  the form
\begin{equation}
{\cal A}_\mu(x,z)=i\Gamma_\mu(x)+\frac{u_\mu(x)}{2}\psi_0(z)
+\sum_{n=1}^\infty v_\mu^n(x)\psi_{2n-1}(z)
+\sum_{n=1}^\infty a_\mu^n(x)\psi_{2n}(z)\, ,
\label{eq.Amu-decomposition2}
\end{equation}
where the commonly used tensors from $\chi$PT~\cite{Bijnens:1999sh,Bijnens:1999hw,Ecker:1988te,Ecker:1989yg},
 $u_\mu(x)$ and $\Gamma_\mu(x)$,   show up naturally:
\begin{eqnarray}
  u_{\mu} \left( x \right)  & \equiv & \mathi \left\{ \xi_R^{\dag}\left( x \right)  \left( \partial_{\mu} - \mathi r_{\mu}
  \right) \xi_R\left( x \right) - \xi_L^{\dag}\left( x \right)  \left( \partial_{\mu} - \mathi \ell_{\mu}
  \right) \xi_L\left( x \right) \right\} \\
    \Gamma_{\mu} \left( x \right) & \equiv & \frac{1}{2}  \left\{ \xi_R^{\dag}\left( x \right)
  \left( \partial_{\mu} - \mathi r_{\mu} \right) \xi_R\left( x \right) + \xi^{\dag}_L\left( x \right)  \left(
  \partial_{\mu} - \mathi \ell_{\mu} \right) \xi_L\left( x \right) \right\}.
\end{eqnarray}

%

Sometimes it is more convenient to work with the bulk-to-boundary propagators
for the transverse part of the gauge field,
rather than in the meson decomposition. These propagators are
the solutions of eq.~(\ref{eq.5D-EoM}) for a general
Euclidean squared momentum $Q^2\equiv -q^2$, with b.c.'s $V(Q,\pm z_0)=1$ and
$A(Q,\pm z_0)=\pm 1$~($Q=\sqrt{Q^2}$).
They can be nonetheless expressed by means of  the on-shell state decomposition
in terms of the infinite summation
\begin{eqnarray}
V(Q,z)&=&\sum_{n=1}^\infty \frac{g_{v^n} \psi_{2n-1}(z)}{Q^2+m_{v^n}^2}\, ,
\qquad\qquad A(Q,z)=\sum_{n=1}^\infty \frac{g_{a^n} \psi_{2n}(z)}{Q^2+m_{a^n}^2}
\label{eq:decomposition}
\end{eqnarray}
with the decay constants given as
\begin{eqnarray}
g_{v^n}&=&-f^2(z)\partial_z \psi_{2n-1}(z)\mid^{+z_0}_{-z_0}\, ,
\qquad \qquad g_{a^n} = - f^2(z)\psi_0(z)\partial_z \psi_{2n}(z)\mid^{+z_0}_{-z_0} \, .
\label{eq.gv-def}
\end{eqnarray}
When $Q^2$ goes to $0$, we have $V(Q,z)\to 1$ and $A(Q,z)\to \psi_0(z)$, which have been used before.
Alternatively, one can use the Green function $G(Q^2;z,z')$,
which satisfies the equation
\begin{equation}
g^2(z) \partial_z[f^2(z)\partial_z G(Q^2;z,z')]-Q^2 G(Q^2;z,z')=-g^2(z) \delta(z-z')\label{eq:GFeq},
\end{equation}
together with the same UV boundary conditions as the $\psi_n$.
With the completeness condition of the resonances, the Green function can be solved as
\begin{equation}
G(Q^2;z,z')= \sum_{n=1}^\infty \frac{\psi_n(z)\psi_n(z')}{Q^2+m_n^2}.
\label{eq.G-GreenFun}
\end{equation}
Therefore, the bulk-to-boundary propagators defined above are related to the Green function as
\begin{eqnarray}
V(Q,z)&=&-f^2(z')\partial_{z'} G(Q^2;z,z')|_{z'=-z_0}^{z'=+z_0}\nonumber\\
A(Q,z)&=&-f^2(z')\psi_0(z')\partial_{z'} G(Q^2;z,z')|_{z'=-z_0}^{z'=+z_0}.
\end{eqnarray}
Contrarily, once the propagators are known the Green function can be immediately obtained:
\begin{eqnarray}
G(Q^2,z,z')&=&\frac{1}{2W(Q^2)}\left[(V(Q,z)-A(Q,z))(V(Q,z')+A(Q,z'))\theta(z-z')\right.\nonumber\\
&&+ \left.(V(Q,z)+A(Q,z))(V(Q,z')-A(Q,z'))\theta(z'-z)\right]\, .\label{eq:GFsolution}
\end{eqnarray}
Here $W(Q^2)$ is the Wronskian of eq.~(\ref{eq.5D-EoM})
\begin{equation}
W(Q^2)=f^2(z)[V(Q,z)\partial_z A(Q,z)-A(Q,z)\partial_z V(Q,z)],
\end{equation}
which is independent of $z$. At zero momentum $W(0)=f_\pi^2/2$ and the Green function simplifies
\begin{equation}
G(0,z,z')=\frac{1}{f_\pi^2}\left[(1-\psi_0(z))(1+\psi_0(z'))\theta(z-z')+ (1-\psi_0(z'))(1+\psi_0(z))\theta(z'-z)\right]\, .\label{eq:GFsolution0}
\end{equation}

\subsection{The chiral couplings at $\cO(p^2)$ and $\cO(p^4)$}

Once we have rewritten the 5D fields in terms of the chiral Goldstones and
vector and axial-vector resonances, the derivation of the meson Lagrangian
is rather straightforward.     We   consider the 5D gauge $\mA_z=0$
and   substitute the   $\mA_\mu$  decomposition
provided in eq.~(\ref{eq.Amu-decomposition2})  in the 5D action~(\ref{eq.5Daction}).
The Lagrangian is now expressed in terms of meson fields
($\xi_{R,L}(x)$,$v^n_\mu(x)$, $a^n_\mu(x)$)
and the    left and right current sources  ($r_\mu(x)$, $\ell_\mu(x)$).
Each of these terms of the action shows then a factorization into an
integration over the  fifth dimension, which provides the meson coupling,
and an integration over the space-time components, which provides
the chiral structure of the operator in the effective 4D  action.

Hence, after substituting the $\mA_\mu$ decomposition~(\ref{eq.Amu-decomposition2})
in the YM action, one gets the even-parity action without resonance
fields~\cite{Hirn:2005nr,Sakai:2004cn}:
\begin{eqnarray}
S_2[\pi]+S_4[\pi]&=& \Int d^4x\, \bigg[
\frac{f_\pi^2}{4}<u_\mu u^\mu>\nonumber\\
        &&\qquad\qquad +L_1 <u_\mu u^\mu>^2+L_2 <u_\mu u_\nu><u^\mu u^\nu> +L_3 <u_\mu u^\mu u_\nu u_\nu>\nonumber\\
        &&\qquad\qquad  -iL_9<f_{+\mu\nu}u^\mu u^\nu>
+\frac{L_{10}}{4}<f_{+\mu\nu}f_+^{\mu\nu}-f_{-\mu\nu}f_-^{\mu\nu}>
\nn\\
        &&\qquad\qquad +\frac{H_1}{2}<f_{+\mu\nu}f_+^{\mu\nu}+f_{-\mu\nu}f_-^{\mu\nu}>\,
        \bigg]\, ,
\end{eqnarray}
with the covariant tensors $f_{\pm}^{\mu\nu}\equiv
\xi_L ^{-1}\ell^{\mu\nu}\xi_L\pm \xi_R^{-1}r^{\mu\nu}\xi_R$
containing the field-strengths $\ell^{\mu\nu}$ and $r^{\mu\nu}$ of the left and right
sources respectively.
At low energies these terms provide the Goldstone interaction at the dominant orders
and produce the  $\cO(p^2)$ and $\cO(p^4)$  $\chi$PT Lagrangian
with the corresponding LECs given by the 5D integrals:
\begin{eqnarray}
f_\pi^2&=&4\left(\int_{-z_0}^{z_0}\frac{\mathd z}{f^2(z)}\right)^{-1} \, ,
\\
L_1&=&\frac{1}{2} L_2=-\frac{1}{6}L_3=\frac{1}{32}\int_{-z_0}^{z_0} \frac{(1-\psi_0^2)^2}{g^2(z)}  \, \mathd z \, ,
\nn\\
L_9&=&-L_{10}=\frac{1}{4}\int_{-z_0}^{z_0} \frac{1-\psi_0^2}{g^2(z)}  \, \mathd z \, ,
\nn\\
H_1&=&-\frac{1}{8}\int_{-z_0}^{z_0} \frac{1+\psi_0^2}{g^2(z)}   \, \mathd z \, .
\nn
\end{eqnarray}
The functions $f^2(z)$ and $g^2(z)$ have to satisfy some
properties in order for these constants to be finite, with the exception of $H_1$.
The numerical results for the O$(p^4)$ low energy constants in four different models
are collected in Table~\ref{tab:LEC4}. In the last two lines of Table~\ref{tab:LEC4} we also provide the 5D integrals
\begin{eqnarray}
Y &\equiv &  \int_{-z_0}^{+z_0} \frac{ (1-\psi_0^4)^2}{48g^2(z)}  \, \mathd z \,
\nn\\
Z &\equiv& \int_{-z_0}^{+z_0} \frac{\psi_0^2(1-\psi_0^2)^2}{4g^2(z)}  \, \mathd z \,
\label{eq.Z-def}
\end{eqnarray}
which will appear in the next sections.
%


\begin{table}[t!]
\centering
\begin{tabular}{|c|c|c|c|c|c|}
\hline
&   ~~~~~~flat~~~~~~       &   ~~~~Cosh~~~~  &  ~~~~hard-wall~~~~
&  Sakai-Sugimoto   &  ~~~$\chi$PT~~~
\\
&  \cite{Son:2003et} & \cite{Son:2003et}   & \cite{Hirn:2005nr}
& \cite{Sakai:2004cn,Sakai:2005yt}    &     ~~~~~~~\cite{Gasser:1983yg,Gasser:1984gg,Gasser:1984ux}~~~~~~~
 \\ \hline
 $~10^{3}~L_1~$       &    $0.5$ &   $0.5$  &   $0.5$  &  $0.5$  &   $0.9\pm 0.3$
 \\
$~10^{3}~L_2~$        &    $1.0$ &  $1.0$ &  $1.0$ &   $1.0$  &    $1.7\pm 0.7$
 \\
 $~10^{3}~L_3~$       &    $-3.1$ &   $-3.2$  &   $-3.1$  &  $-3.1$  &   $-4.4\pm 2.5$
 \\
$~10^{3}~L_9~$        &    $5.2$ &  $6.3$ &  $6.8$ &   $7.7$  &  $7.4\pm 0.7$
 \\
$~10^{3}~L_{10}~$        &    $-5.2$ &  $-6.3$ &  $-6.8$ &   $-7.7$  &   $-6.0\pm 0.7$
 \\ \hline
$~10^{3}~Y~$            & $0.5$  &   $0.5$   &  $0.5$  & $0.6$  &     ---
 \\
$~10^{3}~Z~$            & $0.6$  &   $0.8$   &  $1.0$  & $1.0$  &     ---
 \\
\hline
\end{tabular}
\caption{{\small
Predictions for the $\cO(p^4)$ low energy constants in various holographic models.
The 5D integrals $Y$ and  $Z$ are defined in eq.~(\ref{eq.Z-def}) and
will be  related to appropriate sum rules and the odd-sector $\cO(p^6)$ LECs.
The  experimental  determinations in $\chi$PT at $\cO(p^4)$
are provided in the last column for  comparison~\cite{Gasser:1983yg,Gasser:1984gg,Gasser:1984ux}.
}} 
 \label{tab:LEC4}
 \end{table}

Conversely, the substitution of the $\mA_\mu$ decomposition~(\ref{eq.Amu-decomposition2})
in the CS action  produces some operators without resonances, with only Goldstone bosons,
which give rise to the Wess-Zumino-Witten (WZW) action~\cite{Sakai:2004cn,Sakai:2005yt}:
\begin{equation}
S_{\rm{WZW}}=-\frac{i N_C}{48\pi^2}\int_{M^4} \mathcal{Z}
-\frac{i N_C}{240\pi^2} \int_{M^4\times\mathbb{R}} \tr(g\mathd g^{-1})^5
\end{equation}
with the source-dependent terms
\begin{eqnarray}
\mathcal{Z}&=&\langle  (\ell d \ell +d \ell  \ell  -i \ell ^3)(i U^{-1}r U -U^{-1}d U)
-d \ell  d U^{-1}r U-i\ell (d U^{-1}U)^3-\frac{1}{2}(\ell d U^{-1}U)^2
\nonumber\\
& &
\qquad -U\ell U^{-1}rd U d U^{-1}-i \ell d U^{-1}U\ell U^{-1}r U+\frac{1}{4}(\ell U^{-1}r U)^2\rangle
-{\rm{p.c.}},
\end{eqnarray}
where``p.c.'' stands for the interchanges
$U\leftrightarrow U^{-1}$ and $\ell \leftrightarrow r$.

\section{Chiral couplings at $\cO(p^6)$}
\label{sec.op6-LECs}

\subsection{Resonance lagrangian}
\begin{figure}[!t]
\begin{center}
\includegraphics[angle=0,clip,width=4cm]{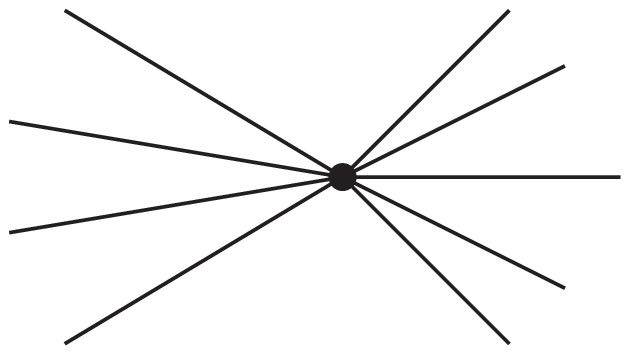}
\includegraphics[angle=0,clip,width=5cm]{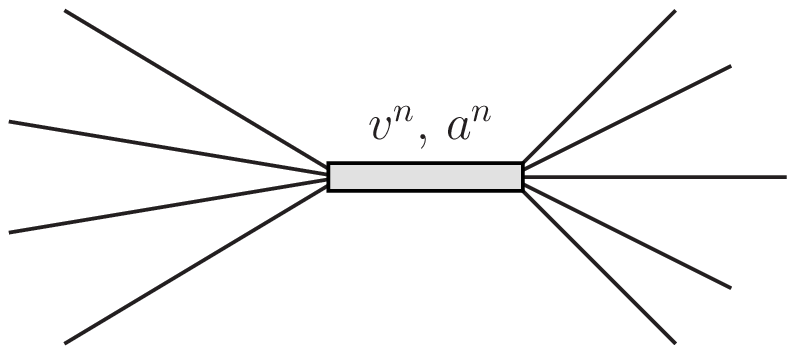}
\includegraphics[angle=0,clip,width=4cm]{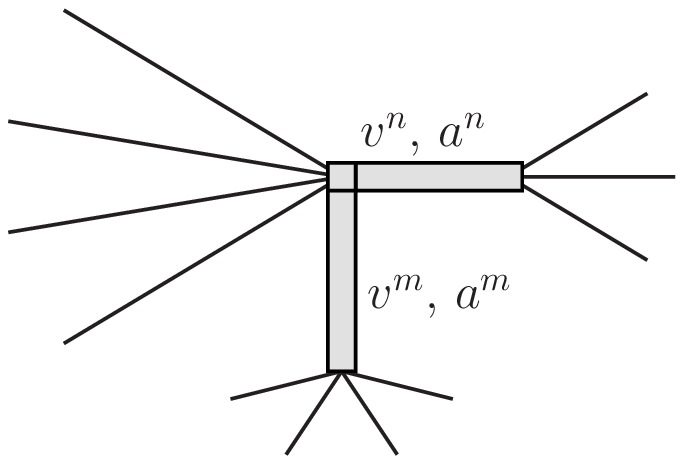}
\caption{{\small
Examples for the different possible diagram contributing to the Goldstone interactions
at low energies:
a) $\cO(p^2)$ and $\cO(p^4)$  local interaction,
b) one-resonance exchange (contribution starting at $\cO(p^6)$),
c) two or more resonance exchanges (contribution  starting at
$\cO(p^8)$).  The single lines represent Goldstones and the double ones
resonances.
}}
\label{fig.diagrams}
\end{center}
\end{figure}

We now proceed to compute the $\cO(p^6)$ couplings at low energies.
After substituting $\mA_\mu$
in terms of the meson decomposition~(\ref{eq.Amu-decomposition2}),
the 5D action~(\ref{eq.5Daction}) given by the YM and CS terms  contains
operators with only Goldstones of at most $\cO(p^4)$. In general we denote an operator as $\cO(p^k)$ when it contains a number $k$
of derivatives or external vector and axial-vector sources ($v_\mu,a_\mu \sim \partial_\mu$).
The $\cO(p^6)$ LECs are generated by the intermediate heavy resonance exchanges.
More precisely, we need just the one-resonance exchanges as diagrams with
a higher number of intermediate resonances will contribute to the
low-energy $\chi$PT action at $\cO(p^8)$ or beyond
(see Fig.~\ref{fig.diagrams}).
Hence, the only  operators in the Lagrangian that interest us
are those containing one resonance field:
\begin{eqnarray}
S_{\rm{YM}}\bigg|_{{\rm Kin.}}&=&\sum_n \int \mathd ^4x
\bra    -\frac{1}{2} (\nabla_\mu v_\nu^n-\nabla_\nu v_\mu^n)^2
+m_{v^n}^2 {v_\mu^n}^2
-\frac{1}{2}(\nabla_\mu a_\nu^n-\nabla_\nu a_\mu^n)^2
+m_{a^n}^2 {a_\mu^n}^2\ket\, ,
\label{eq.S-kin} \\
\nn\\
S_{\rm{YM}}\bigg|_{{\rm 1-res.}}&=&\sum_n \int \mathd ^4x
\, \bigg\{\,
-\,  \bra \frac{f^{\mu\nu}_+}{2}
\bigg[    (\nabla_\mu v_\nu^n-\nabla_\nu v_\mu^n)a_{Vv^n}-\frac{i}{2}([u_\mu,a_\nu^n]
\, -\, [u_\nu,a_\mu^n])a_{Aa^n}  \bigg] \ket
\label{eq.S-1res-even} \\
&&
\qquad  -\, \bra
\frac{i}{4}[u^\mu,u^\nu]
\bigg[    (\nabla_\mu v_\nu^n-\nabla_\nu v_\mu^n)b_{v^n\pi\pi}
\, -\, \frac{i}{2}([u_\mu,a_\nu^n]-[u_\nu,a_\mu^n])b_{a^n\pi^3}]\ket
\nn\\
&&  \qquad  +\,   \bra  \frac{f^{\mu\nu}_-}{2}
\bigg[  (\nabla_\mu a_\nu^n-\nabla_\nu a_\mu^n)a_{Aa^n}
\, -\, \frac{i}{2}([u_\mu,v_\nu^n]-[u_\nu,v_\mu^n])(a_{Vv^n}-b_{v^n\pi\pi})\bigg]\ket
\,\bigg\}\, ,
\nn\\
\nn\\
S_{\rm{CS}}\bigg|_{{\rm 1-res.}}&=&
\sum_n \int \mathd ^4x \,
\bigg\{
-\frac{N_C}{32\pi^2}\, c_{v^n}\epsilon^{\mu\nu\alpha\beta}
\bra u_\mu\{v_\nu^n,f_{+\alpha\beta}\}\ket
+ \frac{N_C}{64\pi^2}\, c_{a^n}\epsilon^{\mu\nu\alpha\beta}
\bra u_\mu\{a_\nu^n,f_{-\alpha\beta}\}\ket
\nn \\
&&\qquad \qquad \qquad +\frac{i N_C}{16\pi^2}\, (c_{v^n}-d_{v^n})\epsilon^{\mu\nu\alpha\beta}
\bra v_\mu^n u_\nu u_\alpha u_\beta\ket \,\bigg\} \, ,
\label{eq.S-1res-odd}
\end{eqnarray}
with $v_\mu^n = t^a v_\mu^{n,\, a}$, $a_\mu^n = t^a a_\mu^{n,\, a}$, and the summation over any possible $n$
implicitly assumed.
The covariant derivative is defined with the connection
$\Gamma_\mu$ as $\nabla_\mu ~.=\partial_\mu+[\Gamma_\mu,.]$.
The couplings are defined as
\begin{eqnarray}
a_{Vv^n}&=&
\int_{-z_0}^{z_0} \frac{\psi_{2n-1}}{g^2(z)} \mathd z\, , \qquad
b_{v^n\pi\pi}=\int_{-z_0}^{z_0} \frac{\psi_{2n-1}(1-\psi_0^2)}{g^2(z)}\mathd z
\nn\\
a_{Aa^n}&=&
\int_{-z_0}^{z_0} \frac{\psi_0\psi_{2n}}{g^2(z)} \mathd z\, ,  \qquad
b_{a^n\pi^3}=\int_{-z_0}^{z_0} \frac{\psi_0\psi_{2n}(1-\psi_0^2)}{g^2(z)}\mathd z\, ,
\nn\\
c_{v^n}&=&-\frac{1}{2}\int _{-z_0}^{z_0}\psi_0\psi_{2n-1}' \mathd z\, ,\qquad
d_{v^n}=\frac{1}{2}\int_{-z_0}^{z_0}\psi_0^2\psi_0'\psi_{2n-1}\mathd z
\nn\\
c_{a^n}&=&-\frac{1}{2}\int _{-z_0}^{z_0}\psi_0^2\psi_{2n}' \mathd z\, .
\label{eq.res-coupling-def}
\end{eqnarray}
Notice that the $a_{Vv^n}$ are related to the decay constants $g_{v^n}$,  defined
previously from the bulk-to-boundary propagator in eq.~(\ref{eq.gv-def}),
in the form $g_{v^n}=m_{v^n}^2a_{Vv^n}$. The same happens for $a_{Aa^n}$.
Likewise, by means of the $\psi_n$  EoM's and  their b.c.'s one can  extract the relations
\begin{eqnarray}
c_{v^n}&=&\frac{m_{v^n}^2}{2f_\pi^2 }b_{v^n\pi\pi}\, , \qquad\qquad
c_{a^n}=\frac{m_{a^n}^2}{3f_\pi^2 }b_{a^n\pi^3}\, ,
\label{eq:cvca}
\label{eq.b-c-rel}
\\
d_{v^n}&=&\frac{m_{v^n}^2}{12f_\pi^2 }\int_{-z_0}^{+z_0} \frac{\psi_{2n-1}(1-\psi_0^4)}{g^2(z)} \mathd z \label{eq:dv}.
\end{eqnarray}
At the level of the generating functional, in order to compute the diagrams with intermediate
resonances one must perform the functional integration over the resonance configurations.
At the tree level this means that one has to find the classical solution
$v_{\mu}^{n}[\pi,\ell_\mu,r_\mu]$ and $a_{\mu}^{n}[\pi,\ell_\mu,r_\mu]$
for the resonance fields
in terms of the external sources and the Goldstone fields, and then substitute it in the
resonance action. In the LEC analysis in this article only the terms~(\ref{eq.S-1res-even})
and~(\ref{eq.S-1res-odd}) are relevant.
Moreover, since we are interested in the Goldstone interaction at long distances,
we need the resonance field solutions in the low-energy limit. Thus,
the resonance action yields for vector and axial-vector mesons the EoM's:
\begin{eqnarray}
v_\mu^n&=&-\frac{1}{2m_{v^n}^2}\left(a_{Vv^n}\nabla^\rho f_{+\rho\mu}
+\frac{i}{2}b_{v^n\pi\pi}\nabla^\rho [u_\rho,u_\mu]
-\frac{i}{2}(a_{Vv^n}-b_{v^n\pi\pi})[f_{-\rho\mu},u^\rho]\right.
\nonumber\\
&&+\left.\frac{N_C}{32\pi^2}c_{v^n} \epsilon_{\mu\nu\alpha\beta}\{u^\nu,f_+^{\alpha\beta}\}
+\frac{i N_C}{16\pi^2}(c_{v^n}-d_{v^n})\epsilon_{\mu\nu\alpha\beta}u^\nu u^\alpha u^\beta\right)
\, ,
\nonumber\\
\nn\\
a_\mu^n&=& \frac{1}{2m_{a^n}^2}\left(a_{Aa^n}\nabla^\rho f_{-\rho\mu}+\frac{1}{4}b_{a^n\pi^3}[[u_\rho,u_\mu],u^\rho]-\frac{i}{2}a_{Aa^n}[f_{+\rho\mu},u^\rho]\right.\nonumber\\
&&+\left.\frac{N_C}{64\pi^2}c_{a^n}\epsilon_{\mu\nu\alpha\beta}
\{u^\nu, f_-^{\alpha\beta}\}\right)  \, ,
\label{eq.res-EoMs}
\end{eqnarray}
where terms $\cO(p^5/m_{R^n}^4)$ (with the derivatives and external vector
and axial-vector sources counting as $\cO(p)$)  and two or more resonance fields
are neglected on the right-hand side of the equations.

By substituting the classical field solutions~(\ref{eq.res-EoMs})
back into the resonance action from
eqs.~(\ref{eq.S-kin})--(\ref{eq.S-1res-odd}),
one integrates out the one-resonance intermediate exchanges  at the generating functional
level and retrieves  the corresponding  $\cO(p^6)$ Goldstone operators  from the
$\chi$PT Lagrangian.
However, the obtained $\cO(p^6)$ Goldstone terms do not show up directly
in the chiral basis of operators commonly used in $\chi$PT~\cite{Bijnens:1999sh,Bijnens:1999hw,Bijnens:2001bb}.
In order to express our outcome in this form  some care needs to be paid
as the  scalar/pseudoscalar source is absent
in our holographic model.
First we expand our
operators using the whole operator set in refs.~\cite{Bijnens:1999sh,Bijnens:1999hw,Bijnens:2001bb}
and then we eliminate the terms of those operators involving the scalar/pseudoscalar
source at the final step.
In the even-parity sector we will see that in the absence of scalar/pseudoscalar sources
it is possible to further reduce  the size of $\cO(p^6)$ operator basis.
Conversely, from the study in~\cite{Ebertshauser:2001nj} we
find that in the odd-parity sector the $\cO(p^6)$
basis chosen in refs.~\cite{Bijnens:2001bb,Ebertshauser:2001nj} in the case when there are no
scalar/pseudoscalar sources, is already minimal.

\subsection{Even and odd sector sum rules}



Now we discuss a series of  sum rules which constrain the resonance couplings,
some of which will be needed in the calculation of the LECs. They can be obtained through the 5D equation of motion (\ref{eq.5D-EoM}) and the completeness condition~(\ref{eq:completeness}). Indeed, by means of the expressions~(\ref{eq.res-coupling-def})
for the resonance couplings, it is possible to introduce a double 5D integral
in  the infinite summation of resonance couplings.
Then, thanks to the completeness condition~(\ref{eq:completeness}),
it is possible to eliminate the infinite resonance summation
and one of the 5D integrals.  We are left with a 5D integral
of appropriate products of wave-function $\psi_0$, which can be in general related to the
integrals that provide the $\cO(p^2)$ and $\cO(p^4)$
chiral couplings~\cite{Hirn:2005nr,Sakai:2005yt}.  Only in a few cases we were left with
two new 5D integrals, $Y$ and $Z$ defined in~(\ref{eq.Z-def}).
For the resonance couplings in the YM part we have:
\begin{eqnarray}
&&\sum_{n=1}^\infty a_{Vv^n}^2 m_{v^n}^2
- \sum_{n=1}^\infty a_{Aa^n}^2 m_{a^n}^2=f_\pi^2,\,\,\qquad
\sum_{n=1}^\infty a_{Vv^n}^2-\sum_{n=1}^\infty a_{Aa^n}^2= -4 L_{10} \, ,
\label{eq.sum-rule1}
\label{eq.sum-rule-PiLR}
\\
&&\sum_{n=1}^\infty a_{Vv^n} b_{v^n\pi\pi} m_{v^n}^2=2f_\pi^2,\,\,~\qquad
\sum_{n=1}^\infty a_{Vv^n} b_{v^n\pi\pi}=4L_9\, ,
\label{eq.sum-rule-VFF}
\\
&&\sum_{n=1}^\infty b_{v^n\pi\pi}^2  m_{v^n}^2=\Frac{4 f_\pi^2}{3} ,\,\,~~~~~~\qquad
\sum_{n=1}^\infty b_{v^n\pi\pi}^2 =32 L_1\, ,
\label{eq.sum-rule-scat}
\\
&&\sum_{n=1}^\infty a_{Aa^n} b_{a^n\pi^3} m_{a^n}^2=2f_\pi^2,\,\,~\qquad
\sum_{n=1}^\infty a_{Aa^n}b_{a^n\pi^3}=4L_9-32L_1\, ,
\label{eq.sum-rule-AFF}
\\
&&\sum_{n=1}^\infty   b_{a^n\pi^3}^2  m_{a^n}^2= \Frac{ 4 f_\pi^2}{5}\, ,\,\,~~~~~\qquad
\sum_{n=1}^\infty b_{a^n\pi^3}^2=4 Z\, ,
\label{eq.sum-rule-Z}
\end{eqnarray}
with $Z$ defined in eq.~(\ref{eq.Z-def}).
The sum rules~(\ref{eq.sum-rule-PiLR}) are related to the VV-AA
correlator~\cite{Son:2003et}.
The relations~(\ref{eq.sum-rule-VFF}),
(\ref{eq.sum-rule-scat}), (\ref{eq.sum-rule-AFF}) and (\ref{eq.sum-rule-Z}) stem from the
$\pi\pi$ vector form factor~(VFF)~\cite{Hirn:2005nr}, the $\pi\pi$--scattering~\cite{Hirn:2005nr},
 the $\pi\pi\pi$ axial-vector form factor~(AFF) and the $\pi\pi\pi\to\pi\pi\pi$ scattering respectively.
The $VV-AA$ correlator, VFF and $\pi\pi$--scattering sum rules have been already
studied within the holographic framework in previous works~\cite{Hirn:2005nr,Sakai:2005yt}.
The others involving the axial resonances are new. Especially we want to emphasize the two sum rules in eq.~(\ref{eq.sum-rule-AFF}). They are related to the AFF into $\pi\pi\pi$, which will be studied
in Sec.~\ref{sec.amplitude-rel}.
They can serve to  improve the current phenomenological analyses on the
$\pi\pi\pi$--AFF, which  employ similar high-energy sum rule
constraints~\cite{GomezDumm:2003ku,Dumm:2009va}.
The fifth couple of sum rules~(\ref{eq.sum-rule-Z}) would  allow
us to provide a more physical meaning  to the constant $Z$.
%
Indeed, in the case  when the summation is well defined
one can use eq.~(\ref{eq.sum-rule-Z}) as an alternative definition of $Z$, in parallel with $L_1$ in eq.~(\ref{eq.sum-rule-scat}).
Notice that all the sum rules involving $b_{v^n \pi\pi}$ and $b_{a^n \pi^3}$ can
be reexpressed in terms of $c_{v^n}$ and $c_{a^n}$ through~(\ref{eq.b-c-rel}), and then are relevant for some anomalous processes.

For the other sum rules involving the anomalous couplings, only those containing $c_{v^n}-d_{v^n}$ are
 independent due to the resonance coupling relations~(\ref{eq.b-c-rel}). They are found to be
\begin{eqnarray}
&&\sum_{n=1}^\infty \Frac{c_{v^n} (c_{v^n}-d_{v^n})}{m_{v^n}^2} \,
=\, \Frac{4}{15 f_\pi^2} \, ,\,\,\,  \qquad
\sum_{n=1}^\infty \Frac{c_{v^n} (c_{v^n}-d_{v^n})}{m_{v^n}^4} \,
=\, \Frac{1}{6 f_\pi^4}\, (40 L_1 -Z)\, ,
\label{eq.dv-sum-rule-V4pi}
\\
&&  \sum_{n=1}^\infty \Frac{  (c_{v^n}-d_{v^n})^2}{m_{v^n}^2} \,
=\, \Frac{68}{315 f_\pi^2} \, ,\,\,\,  \qquad
\sum_{n=1}^\infty \Frac{ (c_{v^n}-d_{v^n})^2}{m_{v^n}^4} \,
=\, \Frac{1}{3 f_\pi^4}\, (16 L_1 -Z+Y)\, ,
\label{eq.dv-sum-rule-6pi}
\\
&&\sum_{n=1}^\infty  a_{Vv^n}(c_{v^n}-d_{v^n})  \,=\, \frac{2}{3}
\, ,\qquad
\sum_{n=1}^\infty  \frac{a_{Vv^n}  (c_{v^n}-d_{v^n})   }{m_{v^n}^2} \,=\,
\frac{4}{3f_\pi^2}(2L_1+L_9) \, .
 \label{eq.odd-sum-rules2}
\end{eqnarray}
The first two sum rules are related to a set of diagrams in
the vector form factor into four pions, or $\pi\pi\to\pi\pi\pi$ scattering replacing $c_{v^n}$ with $b_{v^n \pi\pi}$.
The second two appear, for instance, in the
scattering $\pi\pi\pi\to\pi\pi\pi$, while the last two can be related to the vector form factor into three pions.


These sum rules perform a summation over the infinite tower of resonances,
and in principle one may wonder how well the first terms of the series reproduce
the full result.
The reason is that, in spite of the fact of having an infinite number of hadrons
in the large--$N_C$ limit of QCD~\cite{'tHooft:1973jz,'tHooft:1974hx,Witten:1979kh}, in the majority of the hadronic
analyses only the lightest states are taken into account, introducing a {``}truncation" error. Many authors have investigated
the importance of the lightest resonances
in the summation~\cite{Knecht:1997ts,Peris:1998nj,Golterman:2001nk,Golterman:2006gv,Masjuan:2008fr}.
Since the development of the Weinberg sum rules~\cite{Weinberg:1967kj,Shifman:1978bx},
the relations~(\ref{eq.sum-rule1})--(\ref{eq.odd-sum-rules2}) have been truncated
at their lowest orders and have been used to make predictions on hadronic
phenomenology~\cite{Ecker:1989yg}.  Some authors have nonetheless argued
about the relevance of the tail of the series and the numerical (and also theoretical)
impact of the higher terms of the series on the QCD amplitudes at low and intermediate
energies~\cite{Golterman:2006gv,Masjuan:2008fr}.
For this reason, we consider that the study of the saturation of the presented sum rules
may be useful for most of the phenomenological studies, which only include the lightest
hadrons.
Indeed, the first  constraints~(\ref{eq.sum-rule-PiLR})--(\ref{eq.sum-rule-scat})
have been previously obtained in the case of  resonance Lagrangians from the
analysis of the high-energy behavior  of the $VV-AA$  correlator
($F_V^2-F_A^2=f_\pi^2$)~\cite{Weinberg:1967kj,Shifman:1978bx,Ecker:1989yg},  the $\pi\pi$ VFF
($F_V G_V=f_\pi^2$)~\cite{Ecker:1989yg}
and the $\pi\pi$--scattering~($3~G_V^2=f_\pi^2$)~\cite{Guo:2007ff,Guo:2007hm},
where the corresponding couplings are related to those used in this paper through $F_V=a_{V\rho}m_\rho$,
$F_A=a_{Aa_1}m_{a_1}$ and $G_V=b_{\rho\pi\pi} m_\rho/2$.

\begin{table}[t!]
\centering
\begin{tabular}{|c|c|c|c|c|}
\hline
&   ~~~~~~Flat~~~~~~       &   ~~Cosh~~  &  ~~Hard-wall~~
&  Sakai-Sugimoto
 \\ \hline
$1-\Frac{a_{V\rho}^2  m_\rho^2 - a_{A a_1}^2 m_{a_1}^2}{
\sum_{n} (a_{Vv^n}^2  m_{v^n}^2 - a_{A a^n}^2 m_{a^n}^2)
}$
&  99 \%  &   300 \%   &   350 \%   &   840 \%
\\
$1-  \Frac{a_{V\rho}^2   - a_{A a_1}^2  }{\sum_n (a_{Vv^n}^2   - a_{A a^n}^2 )}$
& 8 \%  &   34 \%   &    41 \%    &   110 \%
\\
\hline
$1-\Frac{a_{V\rho} b_{\rho\pi\pi} m_\rho^2 }{
\sum_n  a_{Vv^n} b_{v^n\pi\pi} m_{v^n}^2
}$
&  18 \%   &   0   &   -11 \%    &   -30 \%
\\
$ 1- \Frac{a_{V\rho} b_{\rho \pi\pi} }{
\sum_n a_{Vv^n} b_{v^n\pi\pi}
} $          
&    0.8 \%   
&   0
&    -2 \%    
&     -6 \%  
\\
\hline
$1-\Frac{3  b_{\rho\pi\pi}^2  m_\rho^2}{
\sum_n 3  b_{v^n\pi\pi}^2  m_{v^n}^2
}$
&  0.7 \%  &  0   &   0.7 \%   &  2 \%
\\
$1-\Frac{  b_{\rho\pi\pi}^2  }{
\sum_n b_{v^n \pi\pi}^2
}$          
&   -0.6 \%    
&   0
&   0.2 \%  
&    1 \%   
\\
\hline
$1-\Frac{a_{Aa_1}  b_{a_1\pi^3}  m_{a_1}^2}{
\sum_n  a_{Aa^n}  b_{a^n\pi^3}  m_{a^n}^2
}$
&  39 \%   &   0    &   -11 \%  &  -54 \%
\\
$1-\Frac{a_{Aa_1}  b_{a_1\pi^3}   }{
\sum_n a_{Aa^n}  b_{a^n\pi^3}
}$
&  8 \%     &  0   &   -3.4 \%    &   -15 \%
\\
\hline
$1-\Frac{5  b_{a_1\pi^3}^2  m_{a_1}^2}{
\sum_n 5  b_{a^n\pi^3}^2  m_{a^n}^2
}$
&   8 \%    &  0    &    0.8 \%    &   6 \%
\\
$1-\Frac{   b_{a_1\pi^3}^2 }{
\sum_n   b_{a^n\pi^3}^2
}$
&  2 \%   &   0   &   0.2 \%    &   2 \%
\\
\hline
$1-\Frac{c_\rho (c_\rho-d_\rho)/m_\rho^2}{
\sum_n c_{v^n} (c_{v^n}-d_{v^n})/m_{v^n}^2
}$
&   2  \%    &    0   &    1.5  \%    &   4 \%
\\
$1-\Frac{c_\rho (c_\rho-d_\rho)/m_\rho^4}{
\sum_n  c_{v^n} (c_{v^n}-d_{v^n})/m_{v^n}^4
}$
&    0.3   \%   &   0   &    0.3  \%    &   1.4 \%
\\
\hline
$1-\Frac{  (c_\rho-d_\rho)^2/m_\rho^2}{
\sum_n  (c_{v^n}-d_{v^n})^2/m_{v^n}^2
}$
&    3 \%    &  0.1  \%    &    3  \%    &   6 \%
\\
$1-\Frac{  (c_\rho-d_\rho)^2/m_\rho^4}{
 (c_{v^n}-d_{v^n})^2/m_{v^n}^4
}$
&   0.6   \%   &    -0.6  \%    &    0.6 \%    &     2 \%
\\
\hline
$1-\Frac{a_{V\rho} (c_\rho-d_\rho) }{
\sum_n  a_{Vv^n} (c_{v^n}-d_{v^n})
}$
&   - 30   \%    &   - 20 \%    &    -30 \%    &   -50  \%
\\
$1-\Frac{a_{V\rho} (c_\rho-d_\rho)/m_\rho^2}{
\sum_n a_{Vv^n} (c_{v^n}-d_{v^n})/m_{v^n}^2
}$
&     - 5 \%   &   - 3 \%   &    -5 \%    &   -10 \%
\\
\hline
\end{tabular}
\caption{{\small
Analysis of the saturation  of the sum rules~(\ref{eq.sum-rule1})--(\ref{eq.odd-sum-rules2})
by the first resonance multiplets $\rho(770)$ and $a_1(1260)$.
The  denominators  $\sum_n (...)$  represent the resummed expression
for the summations from $n=1$ up to $\infty$
provided in eqs.~(\ref{eq.sum-rule1})--(\ref{eq.odd-sum-rules2}).
}} 
 \label{tab.SR-saturation}
 \end{table}
In Table~\ref{tab.SR-saturation} we check how well the
sum rules~(\ref{eq.sum-rule1})--(\ref{eq.odd-sum-rules2})
are saturated by the lightest resonance multiplets of vectors and axial-vectors.
The second sum rule in every line
in eqs.~(\ref{eq.sum-rule1})--(\ref{eq.odd-sum-rules2}) is always much better saturated
by the lightest meson   than the  first one, as it contains an extra $1/m_{R^n}^2$ suppression.
One can see that, in general, the first term of the series
already provides a reasonable approximation.
The only exception is the $VV-AA$ Weinberg sum rules  in eq.~(\ref{eq.sum-rule1}),
which is found to be badly convergent.   Indeed,
in the asymptotically anti-de Sitter~(AdS) models we are studying (``Cosh''~\cite{Son:2003et} and hard-wall~\cite{Hirn:2005nr})
the resonance masses and couplings scale, respectively,   as  $m_{v^n}^2, m_{a^n}^2 \sim n^2$
and $a_{Vv^n}^2, a_{Aa^n}^2 \sim n^{-1}$: the first WSR
--first constraint in eq.~(\ref{eq.sum-rule1})--   is not convergent
and, even though the alternate series in the second sum rule in eq.~(\ref{eq.sum-rule1}) is convergent,
it is not absolutely convergent~\footnote{
The convergence behavior is slightly better
in the case of  flat metric~\cite{Son:2003et} and worse in
the Sakai-Sugimoto model~\cite{Sakai:2004cn,Sakai:2005yt}.  In the flat model~\cite{Son:2003et}
one has the high-energy behavior $\Pi_{VV}, \Pi_{AA} \sim 1/Q$
and  $a_{Vv^n}^2, a_{Aa^n}^2 \sim n^{-2}$ for large $n$.
On the other hand, in the Sakai-Sugimoto case~\cite{Sakai:2004cn,Sakai:2005yt}
one has  $\Pi_{VV}, \Pi_{AA} \sim  Q$ and $a_{Vv^n}^2, a_{Aa^n}^2 \sim n^{0}$.
In all the  holographic theories considered here~\cite{Son:2003et,Hirn:2005nr,Sakai:2004cn,Sakai:2005yt}
the resonance masses behave like $m_{v^n}^2, m_{a^n}^2 \sim n^2$ for high $n$.
The dependence of the masses and couplings on the excitation number $n$ in the different models can be found in Appendix.~\ref{app.holographic-models}. }.
This kind of pathologies are softened in the case of soft-wall
models with a quadratic dilaton~\cite{Karch:2006pv,Colangelo:2008us,Zuo:2009dz},
as they recover the linear Regge trajectories $m_{v^n}^2, m_{a^n}^2 \sim n$ and
$a^2_{Vv^n},a^2_{Aa^n}\sim n^{-1}$.

\subsection{Odd-intrinsic parity LECs}

Now we are ready to show the results for the $\cO(p^6)$ LECs in the odd sector. After
expressing the outcome in the $\cO(p^6)$ basis of odd-parity operators
provided in ref.~\cite{Bijnens:2001bb}, we obtain the predictions:
\begin{eqnarray}
C_{12}^W&=&-\frac{N_C}{128\pi^2} S_{\pi^5}\nonumber\\
C_{13}^W&=&\frac{N_C}{64\pi^2}(\tilde S_{V\pi^3}-S_{V\pi^3}-S_{\pi VV})\nonumber\\
C_{14}^W&=&\frac{N_C}{128\pi^2}(\tilde S_{V\pi^3}-S_{\pi VV})\nonumber\\
C_{15}^W&=&\frac{N_C}{128\pi^2}(2\tilde S_{V\pi^3}+S_{V\pi^3}-S_{\pi VV})\nonumber\\
C_{16}^W&=&\frac{N_C}{128\pi^2}(3S_{\pi^5}-2 S_{V\pi^3}-S_{A\pi^4})\nonumber\\
C_{17}^W&=&\frac{N_C}{512\pi^2}(6 \tilde S_{V\pi^3}-4 S_{\pi VV}-S_{A\pi^4})\nonumber\\
C_{19}^W&=&-\frac{N_C}{128\pi^2} (2S_{V\pi^3}+S_{\pi AA})\nonumber\\
C_{20}^W&=&\frac{N_C}{256\pi^2} (6\tilde S_{V\pi^3}-4S_{\pi VV}-S_{\pi AA})\nonumber\\
C_{21}^W&=&\frac{N_C}{256\pi^2} (2S_{V\pi^3}-S_{\pi AA})\nonumber\\
C_{22}^W&=&\frac{N_C}{64\pi^2}S_{\pi VV}\nonumber\\
C_{23}^W&=&\frac{N_C}{128\pi^2}S_{\pi AA},\label{eq:Codd}
\end{eqnarray}
with the summations over resonances,
\begin{eqnarray}
S_{\pi VV}&=&\sum_{n=1}^\infty \frac{a_{Vv^n}c_{v^n}}{m_{v^n}^2}\, ,\qquad\qquad
S_{\pi AA}=\sum_{n=1}^\infty \frac{a_{Aa^n}c_{a^n}}{m_{a^n}^2}
\nonumber\\
S_{V\pi^3}&=&\sum_{n=1}^\infty \frac{a_{Vv^n}}{m_{v^n}^2}(c_{v^n}-d_{v^n})\,,\qquad\qquad
\tilde S_{V\pi^3}=\sum_{n=1}^\infty \frac{c_{v^n}b_{v^n \pi\pi}}{m_{v^n}^2}
\nonumber\\
S_{\pi^5}&=&\sum_{n=1}^\infty \frac{b_{v^n \pi\pi}}{m_{v^n}^2}(c_{v^n}-d_{v^n})\, , \qquad
S_{A\pi^4}=\sum_{n=1}^\infty \frac{c_{a^n}b_{a^n \pi^3}}{m_{a^n}^2}\, ,\label{eq:Sodd}
\end{eqnarray}
which one may find in the sum rules~(\ref{eq.sum-rule1})--(\ref{eq.odd-sum-rules2})
with the help of the resonance coupling relations~(\ref{eq.b-c-rel}).
Hence, all the odd intrinsic-parity LECs  can be expressed through $L_1$, $L_9$
and the constant  $Z$ (defined in~(\ref{eq.Z-def})):
\begin{eqnarray}
C_{12}^W&=&-\frac{N_C}{384\pi^2f_\pi^2}(40L_1-Z)\label{eq:C12}
 \\
C_{13}^W&=&\frac{5N_C}{96\pi^2f_\pi^2}(4L_1-L_9)\label{eq:C13}
 \\
C_{14}^W&=&\frac{N_C}{64\pi^2f_\pi^2}(8L_1-L_9)\label{eq:C14}
 \\
C_{15}^W&=&\frac{N_C}{192\pi^2f_\pi^2}(52L_1-L_9)\label{eq:C15}
 \\
C_{16}^W&=&\frac{N_C}{384\pi^2f_\pi^2}(104L_1-8L_9-7Z)\label{eq:C16}
 \\
C_{17}^W&=&\frac{N_C}{384\pi^2f_\pi^2}(72L_1-6L_9-Z)\label{eq:C17}
 \\
C_{19}^W&=&\frac{N_C}{96\pi^2f_\pi^2}(4L_1-3L_9)\label{eq:C19}
 \\
C_{20}^W&=&\frac{N_C}{192\pi^2f_\pi^2}(80L_1-7L_9)\label{eq:C20}
 \\
C_{21}^W&=&\frac{N_C}{192\pi^2f_\pi^2}(12L_1+L_9)\label{eq:C21}
 \\
C_{22}^W&=&\frac{N_C}{32\pi^2f_\pi^2}L_9\label{eq:C22}
 \\
C_{23}^W&=&\frac{N_C}{96\pi^2f_\pi^2}(L_9-8L_1).
\label{eq:C23}
\label{eq.CWk-Lj-rel}
\end{eqnarray}
The $1/N_C$ subleading coupling $C_{18}^W$
and those related to operators with scalar/pseudoscalar sources
do not appear. This set of equations provides a relation between the even and odd-parity sectors of QCD.
Actually, taking into account that $L_9=-L_{10}$~\cite{Hirn:2005nr,Sakai:2005yt},
we recover the result $C_{22}^W= - \frac{N_C}{32\pi^2 f_\pi^2} L_{10}$ previously
derived through the low-energy expansion
of the Son-Yamamoto relation ~\cite{Knecht:2011wh}.

We provide in Table~\ref{tab.AdS-model-op6LECs} the numerical results
for the four holographic models mentioned before.
Among them,  the {``}Cosh'' model accommodates better
the low-energy inputs and the required gauge coupling value $g_5$ needed to recover
the right coefficient for the pQCD log of the two-points correlators~\cite{Son:2003et}.
Thus we take this as our preferred set of estimates.
Nonetheless, it is worthy to remark that the LEC predictions from the various
models considered here are relatively stable,
with variations between one and another
of the order of $\sim 1\cdot 10^{-3}$~GeV$^{-2}$, similar to the
renormalization $\mu$ scale dependence of the $\cO(p^6)$ LECs in $\chi$PT~\cite{Bijnens:2001bb}.
In cases like  $C_{12}^W$, $C_{15}^W$ or $C_{21}^W$ the shift is pretty small,
while for others like $C_{13}^W$, $C_{19}^W$ or $C_{20}^W$ one sees larger oscillations depending
on the models at hand.

\begin{table}[t!]
\centering
\begin{tabular}{|c|c|c|c|c|}
\hline
&   ~~~~~~Flat~~~~~~       &   ~~{``}Cosh"~~  &  ~~Hard-wall~~
&  Sakai-Sugimoto
\\ \hline
~~~$  ~C_{12}^W~~~$   & $-2.1$ & $-2.1$ & $-2.1$ & $-2.1$ \\
$  ~C_{13}^W$ &  $-6.5$  & $-8.8$ & $-9.9$ & $-11.9$  \\
$  ~C_{14}^W$  &  $-0.6$  & $-1.3$ & $-1.7$ & $-2.3$  \\
$  ~C_{15}^W$  &  $4.5$  & $4.4$ & $4.2$ & $4.0$  \\
$  ~C_{16}^W$  &  $0.9$  & $-0.2$ & $-0.8$ & $-1.6$  \\
$  ~C_{17}^W$  &  $0.6$  & $-0.1$ & $-0.5$ & $-1.1$  \\
$  ~C_{19}^W$  &  $-5.6$  & $-7.0$ & $-7.7$ & $-8.8$  \\
$  ~C_{20}^W$  &  $1.1$  & $-0.4$ & $-1.3$ & $-2.3$  \\
$  ~C_{21}^W$  &  $2.4$  & $2.6$ & $2.7$ & $2.9$  \\
$  ~C_{22}^W$  &  $6.5$  & $7.9$ & $8.6$ & $9.7$  \\
$  ~C_{23}^W$  &  $0.4$  & $0.9$ & $1.1$ & $1.5$  \\
\hline
\end{tabular}
\caption{{\small
Numerical results for the $\cO(p^6)$ low energy constants in the odd sector
from different holographic models.  The $\cO(p^6)$ LECs
are in units of $10^{-3}$~GeV$^{-2}$.
}} 
 \label{tab.AdS-model-op6LECs}
 \end{table}

\begin{table}[t!]
\centering
\begin{tabular}{|c|c|c|c|c|c|c|c|c|}
\hline
~~~~~~~&
 ~~~{``}Cosh"~~~   &  ~~~DSE~~~~   &  ~~~$\chi$PT~~~
 & ~~~CQM~~~ &        Res. Lagr.            
 & ~~~VMD~~~
 \\
 & &  \cite{Jiang:2010wa}  &
 &  \cite{Strandberg:2003zf}  &
 &
 \\
 \hline
~~~$  ~C_{12}^W~~~$   &   $-2.1$   & $-5.13^{+0.15}_{-0.25}$
& & & $-4.3 \pm 0.3$~\cite{Benayoun:2009im}
&
\\
$  ~C_{13}^W$ &                  $-8.8$     &     $-6.37^{+0.18}_{-0.31}$
& $-70\pm 60$~\cite{Strandberg:2003zf}  &  $14\pm 15$  & &   $-20.0$~\cite{Strandberg:2003zf}
\\
& & &   $-10 \pm 70$~\cite{Strandberg:2003zf}  &   $-7\pm 20$  &  &
\\
$  ~C_{14}^W$  &                 $-1.3$     &     $-2.00^{+0.06}_{-0.10}$
& $ 30\pm 11$~\cite{Strandberg:2003zf}   &   $10\pm 8$  &   &  $-6.0$~\cite{Strandberg:2003zf}
\\
& & & $1\pm 15$~\cite{Strandberg:2003zf}    &   $-1\pm 10$  &  &
\\
$  ~C_{15}^W$  &                 $4.4$       &     $4.17^{+0.12}_{-0.20}$
& $-25\pm 24$~\cite{Strandberg:2003zf}  & $20\pm 7$  &   &   $2.0$~\cite{Strandberg:2003zf}
\\
& & & $-3\pm 29$~\cite{Strandberg:2003zf}   &  $9\pm 10$    &  &
\\
$  ~C_{16}^W$  &                 $-0.2$    &      $3.58^{+0.10}_{-0.17}$  &  &  &  &
\\
$  ~C_{17}^W$  &                 $-0.1$     &   $1.98^{+0.06}_{-0.10}$     &  &  &  &
\\
$  ~C_{19}^W$  &                $-7.0$     &    $0.29^{+0.01}_{-0.01}$  &  &  &  &
\\
$  ~C_{20}^W$  &               $-0.4$      &  $1.83^{+0.05}_{-0.09}$  &  &  &  &
\\
$  ~C_{21}^W$  &               $2.6$      &    $2.48^{+0.07}_{-0.12}$  &  &  &  &
\\
$  ~C_{22}^W$  &              $7.9$      &   $5.01^{+0.14}_{-0.24}$
& $ 6.5\pm 0.8$~\cite{Strandberg:2003zf}     &   $3.9\pm 0.4$  &
$ 8.0$~\cite{Kampf:2011ty}
&  $8.0$~\cite{Bijnens:1989jb,Pallante:1992qe,Moussallam:1994xp}
\\
& & & $5.1\pm 0.7$~\cite{Strandberg:2003zf}  &  &
$6.5$~\cite{RuizFemenia:2003hm}  &
\\
& & & $5.4\pm 0.8$~\cite{Unterdorfer:2008zz}  &  & $8.1\pm 0.8$~\cite{Masjuan:2012wy}
&
\\
& & &  $7.0^{+1.0}_{-1.5}$~\cite{Mateu:2007tr}
&  &   &
\\
$  ~C_{23}^W$  &              $0.9$    &   $2.74^{+0.08}_{-0.13}$   &  &  &  &
\\
\hline
\end{tabular}
\caption{{\small
Numerical results for the odd-intrinsic parity $\cO(p^6)$ low energy constants
within the ``Cosh'' model compared to alternative determinations in other frameworks:
Dyson-Schwinger Equation (DSE)~\cite{Jiang:2010wa},
$\chi$PT~\cite{Strandberg:2003zf,Unterdorfer:2008zz},
Constituent Chiral Quark Model (CQM)~\cite{Strandberg:2003zf},
Hidden Local Symmetry~\cite{Benayoun:2009im},
Resonance Chiral Theory~\cite{Kampf:2011ty,RuizFemenia:2003hm,Mateu:2007tr}, rational approximations~\cite{Masjuan:2012wy}
and Vector Meson Dominance (VMD)~\cite{Strandberg:2003zf,Bijnens:1989jb,Pallante:1992qe,Moussallam:1994xp}.
The $\cO(p^6)$ LECs are given in units of $10^{-3}~$GeV$^{-2}$.
}} 
 \label{tab:odd-LECs-vs-exp}
 \end{table}
Along the years, many analyses have been devoted to the study of QCD amplitudes
under the minimal hadronical approximation~\cite{Knecht:1997ts,Peris:1998nj,Golterman:2001nk}, where one  considers
just the lightest multiplets of resonances entering in the problem at hand.
A recent  study of the low-energy contributions from a general odd-intrinsic
resonance  Lagrangian with only the lightest meson multiplets
has  led to the relation~\cite{Kampf:2011ty}
\be
\Frac{F_V^2}{2 G_V} C_{12}^W \, -\, F_V (C_{14}^W-C_{15}^W) \, =\, G_V C_{22}^W
\,\,\,\,\,\,\, .
\label{eq.Kampf-rel}
\ee
One can verify that this is a direct consequence of our results (\ref{eq:Codd}) when keeping only the lightest resonance multiplets in
the sums (\ref{eq:Sodd}) involved. Since all the sums are well saturated by the lowest multiplets, as shown in the previous subsection,
this relation should be reasonably satisfied with the exact values of the couplings.
We have checked that this relation is pretty well fulfilled 
for various holographic models, with deviations that vary from $1\%$ in the
flat background to $11\%$ in the Sakai-Sugimoto model.

One can also compare our predictions with previous phenomenological  determinations
in various frameworks. The full set of $\cO(p^6)$ LECs was computed
in ref.~\cite{Jiang:2010wa}  by means of the DSE.
Other approaches are based on the analysis of experimental decays
and anomalous processes directly through $\chi$PT~\cite{Strandberg:2003zf,Unterdorfer:2008zz},
although in general the determinations are not very precise except for $C_{22}^W$.
Alternatively, several authors have   estimated LECs by
considering just the lightest multiplet of resonances
in the effective Lagrangian~\cite{Benayoun:2009im,Kampf:2011ty,RuizFemenia:2003hm}, rational approximations~\cite{Masjuan:2012wy}
and the assumption of vector meson  dominance (VMD)~\cite{Strandberg:2003zf,Bijnens:1989jb,Pallante:1992qe,Moussallam:1994xp}.
All these results are compared  in Table~\ref{tab:odd-LECs-vs-exp}.
There are some couplings
($C_{16}^W\, ...\, C_{21}^W$, $C_{23}^W$)  that,
except for the DSE computation~\cite{Jiang:2010wa},  are completely unknown,
so that the present work may help further explorations of anomalous QCD amplitudes.
One has to take into account that variations of the renormalization scale $\mu$
induce shifts in the physical $\cO(p^6)$ chiral couplings in the form~\cite{Bijnens:2001bb}
\begin{equation}
\Frac{dC_k^W}{d\ln\mu^2} \,=\, - \, \Frac{\eta_k}{32\pi^2}\, ,
\end{equation}
with $\bigg|\frac{\eta_k}{32\pi^2}\bigg|\sim 1\cdot 10^{-3}$~GeV$^{-2}$~\cite{Bijnens:2001bb}.
Hence, no comparison to a large--$N_C$ estimate can claim an absolute precision beyond that.
In most cases, we  find a reasonable agreement with the few previous determinations. Some LECs like $C_{15}^W$ and $C_{21}^W$ agree extremely well with the DSE determinations. Conversely, the predictions for other couplings like $C_{19}^W$ seem
to be far more spread.
In any case, with the exception of  $C_{22}^W$ which is relatively well know from
the $\pi^0\to\gamma\gamma^*$ decay, the remaining estimates carry large uncertainties,
if known at all, so our results are expected to be of help for the development of the study of the anomalous sector of QCD.

\subsection{Even-intrinsic parity LECs}

For the even-parity sector we proceed in a similar way.
%
Our results for the constants in the three-flavor notation  are listed
in Tab. \ref{tab:LECe}.

\begin{table}[t!]
\centering
\begin{tabular}{|c|c|c|c|}
\hline
$C_1$ &  $-\frac{1}{4}S_{A\pi^3}+\frac{1}{32}S_{\pi^4}$ &  $C_{59}$  & $\frac{1}{4}S_{AA}-\frac{1}{16}S_{V\pi\pi}-\frac{1}{4}S_{VV}-\frac{N_C^2}{3072\pi^4f_\pi^2}$\\\hline
 $C_3$  & $\frac{1}{16}S_{A\pi^3}$ &  $C_{66}$ &  $-\frac{1}{16}S_{A\pi^3}+\frac{3}{32}S_{\pi^4}-\frac{1}{16}S_{V\pi\pi}$  \\\hline
 $C_4$ &  $-\frac{3}{16}S_{A\pi^3}+\frac{1}{32}S_{\pi^4}$ &  $C_{69}$  & $\frac{1}{16}S_{A\pi^3}-\frac{3}{32}S_{\pi^4}+\frac{1}{16}S_{V\pi\pi}$\\\hline
$C_{40}$  & $\frac{1}{4}S_{A\pi^3}-\frac{1}{32}S_{\pi^4}-\frac{1}{32}S_{\pi^6}$ &  $C_{70}$ &  $-\frac{1}{8}S_{AA}+\frac{3}{8}S_{A\pi^3}-\frac{1}{32}S_{\pi^4}-\frac{1}{16}S_{V\pi\pi}+\frac{1}{8}S_{VV}+\frac{N_C^2}{46080\pi^4f_\pi^2}$ \\\hline
$C_{42}$ &  $\frac{1}{8}S_{A\pi^3}-\frac{1}{32}S_{\pi^4}-\frac{1}{32}S_{\pi^6}+\frac{17N_C^2}{80640\pi^4f_\pi^2}$ & $C_{72}$  & $\frac{1}{8}S_{AA}-\frac{3}{16}S_{A\pi^3}+\frac{1}{16}S_{V\pi\pi}-\frac{1}{8}S_{VV}+\frac{N_C^2}{46080\pi^4f_\pi^2}$\\\hline
$C_{44}$  & $-\frac{1}{2}S_{A\pi^3}+\frac{1}{16}S_{\pi^4}+\frac{1}{8}S_{\pi^6}-\frac{17N_C^2}{40320\pi^4f_\pi^2}$
&  $C_{73}$  & $\frac{1}{8}S_{AA}-\frac{1}{2}S_{A\pi^3}+\frac{1}{8}S_{V\pi\pi}-\frac{1}{8}S_{VV}-\frac{N_C^2}{23040\pi^4f_\pi^2}$  \\\hline
$C_{46}$  & $-\frac{1}{8}S_{A\pi^3}-\frac{17N_C^2}{80640\pi^4f_\pi^2}$ &  $C_{74}$  & $\frac{1}{8}S_{AA}+\frac{3}{8}S_{A\pi^3}-\frac{5}{16}S_{\pi^4}-\frac{1}{8}S_{V\pi\pi}-\frac{1}{8}S_{VV}$ \\\hline
 $C_{47}$ &  $\frac{1}{4}S_{A\pi^3}-\frac{1}{16}S_{\pi^6}+\frac{17N_C^2}{40320\pi^4f_\pi^2}$ &  $C_{76}$  & $-\frac{1}{8}S_{AA}+\frac{1}{4}S_{A\pi^3}+\frac{1}{8}S_{\pi^4}+\frac{1}{8}S_{VV}-\frac{N_C^2}{46080\pi^4f_\pi^2}$ \\\hline
  $C_{48}$  & $\frac{1}{16}S_{A\pi^3}-\frac{1}{32}S_{\pi^4}+\frac{N_C^2}{1920\pi^4f_\pi^2}$ &  $C_{78}$ &  $-\frac{1}{4}S_{AA}+\frac{1}{16}S_{V\pi\pi}+\frac{1}{4}S_{VV}$  \\\hline
   $C_{50}$ &  $\frac{1}{4}S_{A\pi^3}+\frac{1}{8}S_{V\pi\pi}+\frac{N_C^2}{960\pi^4f_\pi^2}$ &  $C_{79}$  & $-\frac{1}{8}S_{AA}-\frac{1}{16}S_{V\pi\pi}+\frac{1}{8}S_{VV}$  \\\hline
   $C_{51}$  & $\frac{3}{8}S_{A\pi^3}-\frac{1}{16}S_{\pi^4}+\frac{1}{8}S_{V\pi\pi}$ &  $C_{87}$ &  $-\frac{1}{8}S_{AA}+\frac{1}{8}S_{VV}$ \\\hline
    $C_{52}$ &  $-\frac{1}{8}S_{V\pi\pi}-\frac{N_C^2}{1920\pi^4f_\pi^2}$ &  $C_{88}$  & $-\frac{1}{8}S_{V\pi\pi}$   \\\hline
     $C_{53}$  & $\frac{3}{16}S_{AA}-\frac{1}{16}S_{V\pi\pi}-\frac{3}{16}S_{VV}+\frac{N_C^2}{3072\pi^4f_\pi^2}$ &  $C_{89}$ &  $-\frac{1}{4}S_{AA}+\frac{3}{8}S_{V\pi\pi}+\frac{1}{4}S_{VV}$ \\\hline
   $C_{55}$ &  $-\frac{3}{16}S_{AA}+\frac{1}{16}S_{V\pi\pi}+\frac{3}{16}S_{VV}+\frac{N_C^2}{3072\pi^4f_\pi^2}$ & $C_{92}$ &  $S_{VV}$  \\\hline
   $C_{56}$  & $-\frac{3}{8}S_{AA}-\frac{1}{8}S_{V\pi\pi}+\frac{3}{8}S_{VV}-\frac{N_C^2}{1536\pi^4f_\pi^2}$ &  $C_{93}$  & $-\frac{1}{4}S_{VV}$  \\\hline
   $C_{57}$ &  $-\frac{1}{8}S_{AA}+\frac{1}{4}S_{V\pi\pi}+\frac{1}{8}S_{VV}$ &    &  \\\hline
\end{tabular}
\caption{Holographic predictions for the $\cO(p^6)$ LECs in the even sector.}
\label{tab:LECe}
\end{table}
Here the chiral couplings are given in terms of the summations over
resonance exchanges,
\begin{eqnarray}
 S_{VV}&=&\sum_{n=1}^\infty\frac{a_{Vv^n}^2}{m_{v^n}^2}\, ,\qquad
 S_{V\pi\pi}=\sum_{n=1}^\infty\frac{a_{Vv^n}b_{v^n\pi\pi}}{m_{v^n}^2}\, ,\qquad
 S_{\pi^4}=\sum_{n=1}^\infty \frac{b_{v^n \pi\pi}^2}{m_{v^n}^2}\, ,
 \nonumber\\
 S_{AA}&=&\sum_{n=1}^\infty\frac{a_{Aa^n}^2}{m_{a^n}^2}\, ,\qquad
 S_{A\pi^3}=\sum_{n=1}^\infty\frac{a_{Aa^n}b_{a^n\pi^3}}{m_{a^n}^2}\, ,\qquad
 S_{\pi^6}=\sum_{n=1}^\infty \frac{b_{a^n \pi^3}^2}{m_{a^n}^2}\, .
 \label{eq:sum}
 \end{eqnarray}
With the help of the decomposition (\ref{eq.G-GreenFun}) of the Green function,
we can write these sums  in the form
\begin{eqnarray}
 S_{VV}&=&\int_{-z_0}^{+z_0}\mathd z \int_{-z_0}^{+z_0}\mathd z' \frac{G(0,z,z')}{g^2(z)g^2(z')}\nonumber\\
 S_{V\pi\pi}&=&\int_{-z_0}^{+z_0}\mathd z \int_{-z_0}^{+z_0}\mathd z' \frac{G(0,z,z')(1-\psi_0(z)^2)}{g^2(z)g^2(z')},\nonumber\\
 S_{\pi^4}&=&\int_{-z_0}^{+z_0}\mathd z \int_{-z_0}^{+z_0}\mathd z' \frac{G(0,z,z')(1-\psi_0(z)^2)(1-\psi_0(z')^2)}{g^2(z)g^2(z')}\nonumber\\
 S_{AA}&=&\int_{-z_0}^{+z_0}\mathd z \int_{-z_0}^{+z_0}\mathd z' \frac{G(0,z,z')\psi_0(z)\psi_0(z')}{g^2(z)g^2(z')}\nonumber\\ S_{A\pi^3}&=&\int_{-z_0}^{+z_0}\mathd z \int_{-z_0}^{+z_0}\mathd z' \frac{G(0,z,z')\psi_0(z')\psi_0(z)(1-\psi_0(z)^2)}{g^2(z)g^2(z')}\nonumber\\ S_{\pi^6}&=&\int_{-z_0}^{+z_0}\mathd z \int_{-z_0}^{+z_0}\mathd z' \frac{G(0,z,z')\psi_0(z)(1-\psi_0(z)^2)\psi_0(z')(1-\psi_0(z')^2)}{g^2(z)g^2(z')}.
\end{eqnarray}
They can be calculated through the expression (\ref{eq:GFsolution0}) once $\psi_0$ is known.
In deriving the low energy constants in Tab. \ref{tab:LECe} we have also used
the sum rules~(\ref{eq.dv-sum-rule-V4pi}) and~(\ref{eq.dv-sum-rule-6pi}),
which allow us to express all the contributions from the odd intrinsic parity resonance sector in terms of just $f_\pi$.
These odd-sector terms were not considered in previous
R$\chi$T estimates of   the $\cO(p^6)$ chiral couplings~\cite{Cirigliano:2006hb,Kampf:2006yf}.

In Tab. \ref{tab:LECe} we find contributions for all the even-sector $\chi$PT operators at $\cO(p^6)$
which do not contain scalar/pseudoscalar sources except for a few of them:
\bear
0\,\,\,=\,\,\, C_2\,=\, C_{41}\,=\, C_{43}\,=\, C_{45}\, =\, C_{49}\, =\, C_{54}\, = \,C_{58} \,
\nn\\
=\, C_{60}=\, C_{67}\,=\, C_{68}\, =\, C_{71}\, =\, C_{75}\, =\, C_{77}  \,  .
\label{eq.Ck=0}
\eear
They correspond to multi-trace operators and are suppressed by $1/N_C$.
In addition, we obtain $C_{90}=0$, 
which we want to discuss here in more details. If scalar-pseudoscalar sources $\chi$  are not included in $\chi$PT ($\chi=0$),
the basis of $\cO(p^6)$ even operators can be further simplified beyond
the trivial simplifications $   \mO_{5,\, ... 39}= \mO_{ 61,\, ... 65}
=  \mO_{80,\, ...  86} =  \mO_{ 91,\, 94}=0 $.
In the usual $\chi$PT computation with scalar-pseudoscalar sources  we have
the $SU(3)$ operator relation~\cite{Bijnens:1999sh,Bijnens:1999hw}
\bear
-\, \bra f_{+\,\mu\nu} [\chi_-^\mu,u^\nu]  \ket  &=&
\bigg[\, \mO_{50} + \mO_{51} -\mO_{52}
-\frac{1}{2} \mO_{53} +\frac{1}{2}\mO_{55} -\mO_{56} + 2\mO_{57}   -\frac{1}{2}\mO_{59}
\nn\\
&&
\qquad -\frac{1}{2}\mO_{70} +\frac{1}{2}\mO_{72} +\mO_{73} -\frac{1}{2}\mO_{76}
+\frac{1}{2}\mO_{78} -\frac{1}{2}\mO_{79} -\mO_{88} -\mO_{90} \,\bigg]
\nn\\
&& +\, \bigg[  \,
\frac{1}{2}\mO_{63} +\mO_{65}  +\frac{1}{4} \mO_{104} \,
\bigg]\, ,
\label{eq.op6-rel-with-chi}
\eear
with $\chi_-^\mu = \nabla^\mu \chi_-   - \frac{i}{2}\{\chi_+, u^\mu\}$~\cite{Bijnens:1999sh,Bijnens:1999hw}.
The operators in the second bracket on the right-hand side of~(\ref{eq.op6-rel-with-chi})
contain the  tensor $\chi$ and, hence,    in the absence
of scalar-pseudoscalar  sources  the basis of $\cO(p^6)$ operators can be
simplified through the relation\bear
0    &=&
\bigg[\, \mO_{50} + \mO_{51} -\mO_{52}
-\frac{1}{2} \mO_{53} +\frac{1}{2}\mO_{55} -\mO_{56} + 2\mO_{57}   -\frac{1}{2}\mO_{59}
\nn\\
&&
\qquad -\frac{1}{2}\mO_{70} +\frac{1}{2}\mO_{72} +\mO_{73} -\frac{1}{2}\mO_{76}
+\frac{1}{2}\mO_{78} -\frac{1}{2}\mO_{79} -\mO_{88} -\mO_{90} \,\bigg]\, .
\label{eq.op6-rel-without-chi}
\eear
Therefore,  when one describes QCD matrix elements in the chiral limit without
scalar-pseudoscalar sources ---as we do in the present article---
the number of independent operators is actually smaller than that one would naively
assume.  Thus, for instance, if one removes the operator $\mO_{90}$ from the minimal
basis,   we find that the only  physical combinations of $\cO(p^6)$ couplings
one may determine in the $\chi=0$ case are
\bear
&  C_{50}+C_{90}\, , \qquad   C_{51}+C_{90} \, ,\qquad  C_{52}- C_{90}\, ,\qquad
C_{53} -\Frac{1}{2} C_{90}\, , \qquad  C_{55} +\Frac{1}{2} C_{90}\, ,   &
\nn\\
&    C_{56}-C_{90}\, ,\qquad C_{57} + 2 C_{90} \, ,\qquad
C_{59} -\Frac{1}{2}C_{90}\, , \qquad C_{70}-\Frac{1}{2}C_{90} \, ,\qquad
C_{72} +\Frac{1}{2} C_{90}\, ,  &
\nn\\
&  C_{73} +C_{90}\, ,\qquad  C_{76}-\Frac{1}{2} C_{90}\, ,\qquad
C_{78}+\Frac{1}{2}C_{90} \, ,\qquad  C_{79}-\Frac{1}{2}C_{90} \, ,\qquad
C_{88}-C_{90}\, ,&
\eear
and the remaining couplings not listed in (\ref{eq.op6-rel-with-chi}).
We want to remark that our analysis of the $\cO(p^6)$ operators  in
the absence of scalar-pseudoscalar sources
is not exhaustive, and more relations might be found,
allowing a further reduction of the $\cO(p^6)$ basis.

There are combinations of couplings where the sums over the resonances
can be eliminated. In particular, in the comparison with the phenomenology
there are a few relations of interest:
\bear
&&3 C_3 + C_4  = C_1 + 4 C_3\, ,
\nn\\
&&2 C_{78} - 4 C_{87} + C_{88}  = 0\, ,
\nn\\
&&8 C_{53} + 8 C_{55} +C_{56}+C_{57}+2 C_{59} = \frac{N_C^2}{256 \pi^4 f_\pi^2} \, ,
\nn\\
&&C_{56}+C_{57}+2 C_{59}= -\frac{N_C^2}{768 \pi^4 f_\pi^2}\,,
\nn\\
&&8 C_{53} - 8 C_{55} +C_{56}+C_{57}-2 C_{59}+4 C_{78} +8 C_{87}-4 C_{88} = 0\,,
\nn\\
&&C_{56}+C_{57}-2 C_{59}-4 C_{78}= 0\,.
\eear
The first relation is connected to the $\pi\pi$ scattering, and the combination of  LECs in the second line yields the contribution to the axial-vector form factor to $\pi\gamma$, $G_A(Q^2)$, at $\cO(p^6)$  in the chiral limit~\cite{Bijnens:1999hw}. Actually, this form factor vanishes identically in this class of models, $G_A(Q^2)=0$~\cite{Hirn:2005nr}. This is one example of those universal relations in these models, which will be studied further in the next section. The relation is not in serious conflict with experiments since the branching ratio of the relevant decay $\pi^+\to l^+\nu \gamma$ is suppressed by more than two orders of magnitude
with respect  to the dominant decay channel $\pi^+\to \mu^+\nu$. At $\cO(p^4)$, it leads to the identity $L_9+L_{10}=0$ shown before.
From the view of resonance exchanges, one finds that all the decay amplitudes $a^n\to \pi \gamma$ vanish~\cite{Hirn:2005nr,Sakai:2005yt}.
These decay amplitudes receive contributions from the  $\langle f^{\mu\nu}_+[u_\mu,a_\nu^n]\rangle$ and  $\langle f^{\mu\nu}_-\nabla_\mu a_\nu^n \rangle$ terms in the resonance Lagrangian (\ref{eq.S-1res-odd}), which have the same coupling and cancel exactly with each other~\cite{Hirn:2005nr}. Actually, in the original hidden local symmetry model with the lightest resonance multiplets~\cite{Bando:1987ym}, no $a_1\to \pi \gamma$ decay occurs.
Moreover, in the holographic framework where a scalar field dual to the quark condensate is introduced, these amplitudes also vanish~\cite{DaRold:2005zs}.
 Experimentally the partial width if $a_1\to \pi \gamma$ is very
small compared to the total width, so it is supposed to be induced by some higher derivative terms or $1/N_C$ suppressed terms~\cite{Son:2003et,Hirn:2005nr,Sakai:2005yt,DaRold:2005zs}.

The other relations are connected to the photon-photon collision. For the $\gamma\gamma\to \pi^0\pi^0$ process, the following quantities are
introduced~\cite{Gasser:2005ud}:
\begin{eqnarray}
a^{00}_2&=&256\pi^4 (8 C_{53} + 8 C_{55} +C_{56}+C_{57}+2 C_{59})\,,\nn\\
b^{00}&=& -128\pi^4(  C_{56}+C_{57}+2 C_{59} )\,.
\end{eqnarray}
For $\gamma\gamma\to \pi^+\pi^-$ a similar set of quantities are defined~\cite{Gasser:2006qa}:
\begin{eqnarray}
a^{+-}_2&=&256\pi^4 (8 C_{53} - 8 C_{55} +C_{56}+C_{57}-2 C_{59}+4 C_{78} +8 C_{87}-4 C_{88})\,,\nn\\
b^{+-}&=& -128\pi^4( C_{56}+C_{57}-2 C_{59}-4 C_{78} )\,.
\end{eqnarray}
We have assumed the large--$N_C$ limit to express the $SU(2)$ definitions of the low-energy parameters $a^{00,+-}_2$ and $b^{00,+-}$~\cite{Gasser:2005ud,Gasser:2006qa} in terms of $SU(3)$ chiral couplings.
Taking into account these definitions, our relations simply predict:
\begin{eqnarray}
a^{00}_2&=&N_C^2\,,\,~~~~b^{00}=+\frac{N_C^2}{6}\,,\nn\\
a^{+-}_2&=&0\,,\,~~~~~~~~~~b^{+-}=0.
\end{eqnarray}
A simple explanation can be found for these results. From the previous discussion we know that there is no $a^n \pi\gamma$ interaction. One can
also check that the $v^n \gamma\gamma$ interaction vanishes due to the special group structure from the $\langle f^{\mu\nu}_+\nabla_\mu v_\nu^n\rangle$ term. As a result, the only contribution to $\gamma\gamma\to \pi^0\pi^0$ at $\cO(p^6)$ comes from the vector resonance exchanges in the crossed channels with two odd vertices, which turns out to be universal, i.e., independent of the 5D background. For $\gamma\gamma\to \pi^+\pi^-$ even this kind of diagram does not exist, so that $a^{+-}_2$ and $b^{+-}$ vanish. It will be interesting to calculate the corresponding amplitudes, to see if these properties remain.
More relations among the other constants seem to hold, and this requires further investigation.


In Tab. \ref{tab:LEC6e} we list the numerical results in different models,
and isolate the  odd-sector  contributions,   which are not negligible for some LECs
and are sometimes even dominant.
The different models provide similar results,
up to a variation of the order of $\pm 1\cdot 10^{-3}$~GeV$^{-2}$.
One must be aware that when we perform a tree-level estimate of the renormalized
chiral couplings  we cannot specify at what renormalization scale $\mu$  they  correspond,
and a deviation of that order of magnitude might occur as one has the running~\cite{Bijnens:1999sh,Bijnens:1999hw}
\begin{equation}
\Frac{dC_k}{d\ln\mu^2} \,=\, \Frac{1}{32\pi^2}\bigg[2 \Gamma_k^{(1)}+\Gamma_k^{(L)}(\mu)\bigg]\, ,
\end{equation}
with $\bigg|\frac{1}{32\pi^2}\bigg[2 \Gamma_k^{(1)}+\Gamma_k^{(L)}(\mu)\bigg]\bigg|\sim 1\cdot 10^{-3}$~GeV$^{-2}$~\cite{Bijnens:1999sh,Bijnens:1999hw}.

\begin{table}[!t]

\centering
\vspace*{-1.7cm}
\begin{tabular}{|c|c|c|c|c|c|}
\hline
\tiny
~~~~~&~Odd contrib.~& ~~~~~ flat~~~~~ &
 ~~~~ {``}Cosh"~~~~ & ~~~Hard-wall~~~ & Sakai-Sugimoto  \\ \hline
~~~$  ~C_1$~~~   &   ---    
& $0.5$ & $-0.3$ & $-0.9$ & $-1.9$ \\
~~~$  ~C_3$~~~   &   ---    
& $0.1$ & $0.3$ & $0.4$ & $0.7$ \\
~~~$  ~C_4$~~~   &  ---    
&  $0.6$ & $0$ & $-0.5$ & $-1.2$ \\
~~~$  ~C_{40}$~~~& ---    
& $-0.5$ & $0.2$ & $0.8$ & $1.8$ \\
~~~$  ~C_{42}$~~~ &$2.6$   & $1.9$ & $2.2$ & $2.5$ & $3.0$ \\
~~~$  ~C_{44}$~~~ &$-5.2$          & $-4.1$ & $-5.5$ & $-6.6$ & $-8.7$ \\
~~~$  ~C_{46}$~~~ &$-2.6$         & $-2.8$ & $-3.2$ & $-3.5$ & $-4.0$ \\
~~~$  ~C_{47}$~~~ &$5.2$         & $5.5$ & $6.2$ & $6.8$ & $7.8$ \\
~~~$  ~C_{48}$~~~ &$6.4$         & $5.6$ & $5.8$ & $5.9$ & $6.2$ \\
~~~$  ~C_{50}+C_{90}$~~~ &$12.7$         & $17.4$ & $19.1$ & $20.2$ & $22.2$ \\
~~~$  ~C_{51}+C_{90}$~~~ &    ---   
& $3.1$ & $5.2$ & $6.7$ & $9.2$ \\
~~~$  ~C_{52}- C_{90}$~~~ &  $-6.4$         & $-10.6$ & $-11.6$ & $-12.1$ & $-13.1$
\\
~~~$  ~C_{53}-\frac{1}{2}C_{90}$~~~ &$4.0$         & $-5.6$ & $-8.8$ & $-10.5$ & $-13.4$ \\
~~~$  ~ C_{55}+\frac{1}{2} C_{90}$~~~ & 4.0 &  13.5  & 16.7  &   18.4  &   21.3
\\
~~~$  ~C_{56}-C_{90}$~~~ &$-7.9$         & $2.6$ & $7.1$ & $9.5$ & $13.3$ \\
~~~$  ~C_{57}+2C_{90}$~~~ &   ---    
& $13.4$ & $17.2$ & $19.2$ & $22.8$ \\
~~~$  ~C_{59}-\frac{1}{2} C_{90}$~~~ &$-4.0$         & $-16.0$ & $-20.1$ & $-22.3$ & $-26.0$ \\
~~~$  ~C_{66}$~~~  &    ---    
& $0.4$ & $-0.3$ & $-0.7$ & $-1.5$ \\
~~~$  ~C_{69}$~~~  &   ---    
& $-0.4$ & $0.3$ & $0.7$ & $1.5$ \\
~~~$  ~C_{70}-\frac{1}{2}C_{90}$~~~  &$0.3$        & $2.8$ & $5.3$ & $6.9$ & $9.6$ \\
~~~$  ~C_{72}+\frac{1}{2}C_{90}$~~~  &$0.3$        & $-2.9$ & $-4.7$ & $-5.9$ & $-7.8$ \\
~~~$  ~C_{73}+C_{90}$~~~  &  -0.5 &   -2.0 & -4.4 & -6.0 & -8.7
\\
~~~$  ~C_{74}$~~~  &  ---   
& $-17.1$ & $-19.0$ & $-19.5$ & $-20.4$ \\
~~~$  ~C_{76}-\frac{1}{2}C_{90}$~~~  &$-0.3$        & $8.5$ & $11.1$ & $12.7$ & $15.3$ \\
~~~$  ~C_{78}+\frac{1}{2}C_{90}$~~~  &   ---    
& $12.0$ & $16.1$ & $18.3$ & $22.0$ \\
~~~$  ~C_{79}-\frac{1}{2}C_{90}$~~~   &   ---  
& $2.8$ & $4.1$ & $4.8$ & $6.0$ \\
~~~$  ~C_{87}$~~~   &   ---   
& $4.9$ & $6.8$ & $7.7$ & $9.3$ \\
~~~$  ~C_{88}-C_{90}$~~~   &   ---    
& $-4.2$ & $-5.2$ & $-5.8$ & $-6.7$ \\
~~~$  ~C_{89}$~~~  &---     
& $22.6$ & $29.2$ & $32.7$ & $38.8$ \\
~~~$  ~C_{92}$~~~   &---    
& $42.4$ & $68.8$ & $87.3$ & $171.0$ \\
~~~$  ~C_{93}$~~~   &     ---  
 & $-10.6$ & $-17.2$ & $-21.8$ & $-42.8$ \\
\hline
\end{tabular}
\caption{{\small
Numerical results for the $\cO(p^6)$ low-energy constants of  the even sector
from different holographic models. All the results are in units of
$10^{-3}~\mbox{GeV}^{-2}$.
In the second column we provide the contribution from the odd-parity resonance sector.
}} 
 \label{tab:LEC6e}
 \end{table}

In Tables~\ref{tab.even-comparison1},~\ref{tab.even-comparison2}
and~\ref{tab.even-comparison-DSE}   we compare our results from the {``}Cosh" model
with the outcomes from other frameworks.
In general, the different works constrain particular combinations of LECs
through the analysis of various matrix elements.
Some determinations consider the $\cO(p^6)$ determination of the amplitudes
and extract the couplings directly from the experimental data:
$\pi\pi$ and $\pi K$
scattering~\cite{Kampf:2006bn,Colangelo:2001df},
$\pi\pi$ VFF~\cite{Bijnens:2002hp}
an the $VV-AA$ correlator~\cite{GonzalezAlonso:2008rf,GonzalezAlonso:2010xf}.
Many authors were able to compute these LECs  through
rational approximants~\cite{Masjuan:2008fv,Masjuan:2008fr},
large--$N_C$ resonance estimates~\cite{Cirigliano:2006hb,Kampf:2006yf,Ecker:1989yg,Amoros:1999dp,Gasser:2006qa,Bellucci:1994eb,Gasser:2005ud,
Unterdorfer:2008zz,Geng:2003mt,Cirigliano:2004ue,Bijnens:1999hw,Bijnens:1998fm,
Kampf:2006bn,Bijnens:1997vq,Guo:2007hm,Guo:2009hi,Colangelo:2001df}
and resonance determinations at NLO in the $1/N_C$
expansion~\cite{Pich:2008jm,Pich:2010sm}.
For the study~\cite{Cirigliano:2004ue}
of the $ VAP$, the $VV-AA$, the $SS-PP$  Green's functions
and the scalar and vector form factors in R$\chi$T, we have used the inputs
$m_\rho=0.776$~GeV, $m_{a_1}=1.26$~GeV, $m_{\pi'}=1.3$~GeV.
Conversely, the analysis through the Dyson-Schwinger equation~\cite{Jiang:2009uf}
is able to determine all the LECs (Table~\ref{tab.even-comparison-DSE}).

\begin{table}[!t]
\centering
\vspace*{-2.2cm}
\begin{tabular}{|c|c|c|c|c|c|}
\hline
~~~~~~~~~&
 ~~~~ {``}Cosh"~~~~  &  DSE  & $\cO(p^6)$ $\chi$PT  &  Reson.  Lagr.
\\
&    & \cite{Jiang:2009uf}   &     &    \& Rational App.
 \\ \hline
~~~$  ~(-4 C_1+ 8 C_2   $
&   2.6
&   $6.29^{+0.28}_{-0.42}$
& $10.1\pm 2.6$~\cite{Colangelo:2001df}
& 7.2~\cite{Colangelo:2001df}
\\
~~~~~~~~~~~~~~~~$   + 5 C_3 +7 C_4)$~~~
&  & & & $5.9\pm 3.3 $~\cite{Guo:2007hm,Guo:2009hi}
\\
& & & &  7.3~\cite{Bijnens:1997vq,Bijnens:1999hw}
 \\
~~~$  ~(3 C_3 +C_4)$~~~
&   0.9
&   $2.95^{+0.11}_{-0.19}$
& $2.6 \pm 0.3$~\cite{Colangelo:2001df}
& 2.0~\cite{Colangelo:2001df}
\\
& & & $0.99\pm 0.25$~\cite{Kampf:2006bn}
& $1.00$~\cite{Kampf:2006bn}
\\
& & &  $2.10\pm 0.25$~\cite{Kampf:2006bn}
&  $1.7\pm 0.3$~\cite{Guo:2007hm,Guo:2009hi}
\\
& & &  $2.35\pm 0.23$~\cite{Kampf:2006bn}
& 2.0~\cite{Bijnens:1997vq}
\\
~~~$  ~(C_1+4 C_3) $~~~
&    0.9
&  $3.59^{+0.13}_{-0.21}$
& $2.07\pm 0.49$~\cite{Kampf:2006bn}
&   $0.72$~\cite{Kampf:2006bn}
\\
& & &  $2.81\pm 0.49$~\cite{Kampf:2006bn}
&
\\
~~~$  ~C_{2}$~~~
&   0
&  $0.00^{+0.13}_{-0.00}$
& $-0.92\pm 0.49$~\cite{Kampf:2006bn}
&   $-0.05$~\cite{Kampf:2006bn}
\\
& & &  $-0.74\pm 0.49$~\cite{Kampf:2006bn}
&
\\
~~~$  ~(C_1 +2 C_2 + 4 C_3)$~~~
&   0.9
& $3.59^{+0.13}_{-0.21}$
&   $0.23\pm 1.08$~\cite{Kampf:2006bn}
&  0.62~\cite{Kampf:2006bn}
\\
& & & $1.34\pm 1.08$~\cite{Kampf:2006bn}
&
\\
{\bf $\pi\pi$, $\pi K$ scattering }
& & & $1.88\pm 0.72$~\cite{Kampf:2006bn}
&
\\  \hline
~~~$  ~(C_{88}-C_{90})$~~~
&  -5.2
&  $-7.91^{-0.35}_{+0.57}$
&   $-7.3\pm 0.5$~\cite{Bijnens:2002hp}
&   $-6.1\pm 0.5$~\cite{Pich:2010sm}
\\
& & & &     -8.6~\cite{Bijnens:2002hp,Bijnens:1999hw}
\\
& & & &  -5.2~\cite{Cirigliano:2004ue}
\\
& & & &    -6.9~\cite{Geng:2003mt}
\\
& & & &    -6.5~\cite{Geng:2003mt}
\\
& & & &   -8.6~\cite{Geng:2003mt}
\\
{\bf $\gamma\to \pi\pi$ form factor  }
& & & &  $-6.2\pm 0.6$~\cite{Masjuan:2008fv}
\\  \hline
~~~$  ~(2 C_{78} - 4 C_{87}+C_{88})$~~~
&     0
&     $-0.73^{-0.17}_{+0.25}$
&
&     2.5~\cite{Cirigliano:2004ue}
\\
& & & &    3.6~\cite{Bijnens:1996wm,Bijnens:1999hw}
\\
& & & &    $2.3\pm 8.4$ ~\cite{Unterdorfer:2008zz}
\\
& & & &  0.8 ~\cite{Geng:2003mt}
\\
& & & &  1.8 ~\cite{Geng:2003mt}
\\
{\bf $\pi, K \to \ell\nu \gamma$ form factor }
& & & &  3.2 ~\cite{Geng:2003mt}
\\  \hline
\end{tabular}
\caption{{\small
Comparison of our results from the the {``}Cosh" model for some particular combinations of the $\cO(p^6)$ LECs in the even sector, with
those from other approaches. All the results are in units of  $10^{-3}~\mbox{GeV}^{-2}$.
The $\chi$PT and resonance
determinations~\cite{Colangelo:2001df,Guo:2007hm,Guo:2009hi,Bijnens:1997vq,Bijnens:1999hw}
of $(-4 C_1 + 8 C_2 + 5C_3 + 7 C_4)$ stem from the $SU(2)$ combination of couplings $r_5$
under the assumption that the loop correction --$1/N_C$ suppressed-- is neglected
in the relation between $SU(3)$ and $SU(2)$ couplings at large $N_C$.
A similar argument applies for the
determinations~\cite{Colangelo:2001df,Guo:2007hm,Guo:2009hi,Bijnens:1997vq}
for $(3 C_3+C_4)$, which derives from the $SU(2)$ parameter $r_6$.
}} 
 \label{tab.even-comparison1}
 \end{table}
\begin{table}[!t]
\centering
\vspace*{-1.5cm}
\begin{tabular}{|c|c|c|c|c|c|}
\hline
~~~~~~~~~&
 ~~~~ {``}Cosh"~~~~  &  DSE  & $\cO(p^6)$ $\chi$PT  &  Reson.  Lagr.
\\
&     & \cite{Jiang:2009uf}   &     &    \& Rational App.
 \\ \hline
 $  ~( 8 C_{53} + 8 C_{55}$
 &  47.7
 &  $20.7^{+0.2}_{-0.3}$
 &
 & $69\pm 19$~\cite{Bellucci:1994eb,Gasser:2005ud}
\\
~~~~~~~~~~~~$ + C_{56} + C_{57}+2C_{59} )$
& & & &
\\
$  ~(C_{56}+C_{57}+2C_{59})$
&    -15.9
&   $-17.7^{+0.6}_{-1.0}$
&
&    $-32\pm  11$~\cite{Bellucci:1994eb,Gasser:2005ud}
\\
$(8 C_{53}-8 C_{55}+C_{56}$
&    0
&    $-5.52^{-0.40}_{+0.67}$
&
&    3.7~\cite{Gasser:2006qa}
\\
~~~~~~~~~~$+C_{57} -2 C_{59} + 4 C_{78}$  &&&&
\\
~~~~~~~~~~$+ 8 C_{87} -4 C_{88} ) $     &&&&
\\
$(C_{56}+C_{57}$
&    0
&    $2.20^{+0.20}_{-0.25}$
&
&   -4.2~\cite{Gasser:2006qa}
\\
~~~~~~~~~~$-2 C_{59} - 4 C_{78}) $
&
&
&
&
\\
{\bf $\gamma\gamma\to\pi\pi$ scattering  }
& & & &
\\  \hline
~~~$  ~C_{87}$~~~
&  6.8
& $7.57^{+0.37}_{-0.60}$
& $4.9 \pm 0.2$~\cite{GonzalezAlonso:2008rf,GonzalezAlonso:2010xf}
& $4.8\pm 1.3$~\cite{Unterdorfer:2008zz}
\\
& & & &  4.0~\cite{Cirigliano:2004ue}
\\
& & & & 7.6~\cite{Amoros:1999dp}
\\
& & & &  5.3~\cite{Geng:2003mt}
\\
& & & &  6.2~\cite{Geng:2003mt}
\\
& & & &  8.6~\cite{Geng:2003mt}
\\
& & & &  $4.1\pm 1.5$~\cite{Pich:2008jm}
\\
{\bf   $VV-AA$ correlator   }
& & & & $5.7\pm 0.5$~\cite{Masjuan:2008fr}
\\  \hline
~~~$  ~\bigg(C_{78}+\frac{1}{2} C_{90}\bigg)$~~~
&   16.1
&    $18.73^{+0.83}_{-1.36}$
&
&    11.8~\cite{Cirigliano:2004ue}
\\
& & & &  14.4~\cite{Geng:2003mt}
\\
& & & &  16.6~\cite{Geng:2003mt}
\\
& & & &  23.1~\cite{Geng:2003mt}
\\  \hline
~~~$  ~C_{89}$~~~
&   29.2
&   $34.74^{+1.61}_{-2.62}$
&
&   19.6~\cite{Cirigliano:2004ue}
\\
& & & &  26.0~\cite{Geng:2003mt}
\\
& & & &  30.3~\cite{Geng:2003mt}
\\
& & & &  42.5~\cite{Geng:2003mt}
\\ \hline
~~~$  ~C_{93}$~~~
&   -17.2
&
&
&   -8.4~\cite{Ecker:1989yg,Cirigliano:2006hb,Kampf:2006yf}
\\
& & & &  -17~\cite{Amoros:1999dp}
\\   \hline
\end{tabular}
\caption{{\small
Comparison of our results from the the {``}Cosh" model for some particular combinations
of the $\cO(p^6)$ LECs in the even sector, with
those from other approaches. All the results are in units of  $10^{-3}~\mbox{GeV}^{-2}$.
The $\gamma\gamma\to\pi\pi$
determinations~\cite{Bellucci:1994eb,Gasser:2005ud,Gasser:2006qa},
based on resonance estimates, were extracted from the $SU(2)$ parameters $a_2^r$ and
$b^r$ under the assumption that the ($1/N_C$ suppressed) loop corrections are neglected
in the relation between $SU(2)$ and $SU(3)$ couplings at large $N_C$.
}} 
 \label{tab.even-comparison2}
 \end{table}
\begin{table}[!t]
\centering
\begin{tabular}{|c|c|c|c|c|c|c|}
\hline
 & ``Cosh'' & DSE~\cite{Jiang:2009uf}
 &&& ``Cosh'' & DSE~\cite{Jiang:2009uf}
\\
\hline
~~~~$C_1$~~~~
& ~~~~  -0.3 ~~~~ & ~~~~
$3.79^{+0.10}_{-0.17}$            
~~~~
& ~~  & ~~~~$C_{59}-\frac{1}{2}C_{90}$~~~~
& ~~~~ -20.1 ~~~~ &  ~~~~
$-23.71^{-1.02}_{+1.66}$               
~~~~
\\\hline
 $C_3$
 & 0.3 & $-0.05^{+0.01}_{-0.01}$
 &&  $C_{66}$ &  -0.3 &     $1.7^{+0.07}_{-0.12}$                    
 \\\hline
$C_4$
&  0  &   $3.1^{+0.09}_{-0.15}$               
&&  $C_{69}$  & 0.3 &  $-0.86^{-0.04}_{+0.06}$          
\\\hline
$C_{40}$
& 0.2 &   $-6.35^{-0.18}_{+0.32}$              
&&  $C_{70}-\frac{1}{2}C_{90}$ &  5.3 &  $0.51^{+0.11}_{-0.16}$              
\\\hline
$C_{42}$
&  2.2  &  $0.60^{+0.00}_{-0.00}$              
&& $C_{72}+\frac{1}{2}C_{90}$  & -4.7 &  $-2.08^{-0.14}_{+0.23}$                     
\\\hline
$C_{44}$
& -5.5  &  $6.32^{+0.20}_{-0.36}$               
&&  $C_{73}+C_{90}$  & -4.4  &   $2.94^{+0.05}_{-0.10}$              
\\\hline
$C_{46}$
&  -3.2  &  $-0.60^{-0.02}_{+0.04}$
&&  $C_{74}$  &  -19.0 &  $-5.07^{-0.16}_{+0.27}$                    
\\\hline
 $C_{47}$
&  6.2 &  $0.08^{+0.01}_{-0.00}$
&&  $C_{76}-\frac{1}{2}C_{90}$  &  11.1  &  $-2.66^{-0.04}_{+0.08}$                    
\\\hline
$C_{48}$
&  5.8  &  $3.41^{+0.06}_{-0.10}$
&&  $C_{78}+\frac{1}{2}C_{90}$ & 16.1  &  $18.73^{+0.83}_{-1.36}$             
\\\hline
~~~~$C_{50}+C_{90}$~~~~
&  19.1 &   $11.15^{+0.40}_{-0.66}$          
&&  $C_{79}-\frac{1}{2} C_{90}$  &  4.1 &  $-1.78^{-0.11}_{+0.17}$                        
\\\hline
$C_{51}+C_{90}$
& 5.2  &     $-9.05^{-0.20}_{+0.37}$            
&&  $C_{87}$ &  6.8  &    $7.6^{+0.4}_{-0.6}$            
\\\hline
$C_{52}-C_{90}$
&  -11.6 &  $-7.48^{-0.29}_{+0.47}$           
&&  $C_{88}-C_{90}$  &  -5.2   &          $-7.91^{-0.35}_{+0.57}$
\\\hline
$C_{53}-\frac{1}{2}C_{90}$
&  -8.8   &  $-13.21^{-0.68}_{+1.10}$                  
&&  $C_{89}$ &  29.2  &    $34.7^{+1.6}_{-2.6}$                
\\\hline
$C_{55}+\frac{1}{2}C_{90}$
& 16.7  &  $18.01^{+0.77}_{-1.26}$           
&& $C_{92}$ &  68.8  &  ---
\\\hline
$C_{56}-C_{90}$
&  7.1  &    $16.90^{+0.90}_{-1.44}$           
&&  $C_{93}$  &  -17.2 &  ---
\\\hline
$C_{57}+ 2 C_{90}$
& 17.2  &  $12.80^{+0.58}_{-0.93}$                
&&    &  &
\\\hline
\end{tabular}
\caption{{\small
Comparison of our results for $\cO(p^6)$ LECs of the even sector from the {``}Cosh'' model
and those from DSE~\cite{Jiang:2009uf}.
The chiral couplings are in units of $10^{-3}$~GeV$^{-2}$.
}}
\label{tab.even-comparison-DSE}
\end{table}

In Tables~\ref{tab.even-comparison1}
and~\ref{tab.even-comparison2} we show the results for particular combinations of the LECs that contribute
directly to some processes. Here, some of these estimates have been extracted from $SU(2)$
analyses~\cite{Colangelo:2001df,Guo:2007hm,Guo:2009hi,Bijnens:1997vq,Bijnens:1999hw,Bellucci:1994eb,Gasser:2005ud,Gasser:2006qa}
under the assumption that the $1/N_C$--suppressed loop corrections are neglected in the relation
between the $SU(3)$ and $SU(2)$ LECs at large $N_C$. One can see that some of the couplings ($C_{88}-C_{90}$,
$2C_{78}- 4 C_{87} +C_{88}$,
$C_{87}$, $C_{78}+\frac{1}{2}C_{90}$, $C_{89}$, $C_{93}$)
are in relatively fair agreement. These are related with amplitudes with a small number of external legs,
which might be dominated by the lightest vector and axial-vector mesons.
For those related to the $\pi\pi\to\pi\pi$ and $\gamma\gamma\to\pi\pi$ scattering, the agreement is a little worse, but still reasonable
taking into account possible sub-leading differences between the physical and the large--$N_C$ values of the
chiral couplings.

In Table~\ref{tab.even-comparison-DSE} we compare our results for individual LECs from the {``}Cosh" model to those from the DSE
approach~\cite{Jiang:2009uf}. For most couplings one finds a reasonable agreement, although for a few others there are larger deviations.
In some cases, the discrepancy can be attributed to the odd-odd contributions mentioned before. 
Adding these contributions to the results from the DSE approach, many couplings, e.g., $C_{42}$, $C_{46}$, $C_{47}$, $C_{52}$, $C_{53}$ and $C_{56}$, agree much better with ours. Some of the others are pushed towards our results, like $C_{44}$ and $C_{50}$.
Still, there are large differences for a few couplings, the most serious ones being $C_{51}$, $C_{74}$ and $C_{76}$.

There are other possible contributions that may be responsible of the deviations. First, one should keep in mind that the present models include
only the spin-1 resonance contributions, and that scalar mesons may play an important role in
scattering processes. 
Second, our effective holographic action in principle could also accommodate operators
of even higher dimension, e.g., of the type $(\mF_{MN})^3$,
that would modify the LECs related to processes with a larger number of external legs.
These higher dimension terms can appear from the low energy expansion of the Dirac-Born-Infeld
action~\cite{Polchinski1998}.

\section{Relations between odd and even amplitudes}
\label{sec.amplitude-rel}

\subsection{Green's function relation: the LR versus the VVA correlator}
The relations between the anomalous $\cO(p^6)$ constants and the $\cO(p^4)$ constants in the even sector indicate possible relations between correlation functions or form factors in this class of models. One example is the Son-Yamamoto relation between the transverse triangle structure function $w_T(Q^2)$ and the left-right correlator~\cite{Son:2010vc}.
Explicitly, $w_T(Q^2)$ is defined as the transverse part of the correlation function of the vector current and the axial current
in a weak electromagnetic background field $\hat F_{\mu\nu}$:
\begin{equation}
 i \int \mathd^4 x\, e^{iqx}\langle j_{\mu}^a(x)j_{\nu}^{5b}(0)
  \rangle_{\hat F} = \frac{Q^2}{8\pi^2}d^{ab}
P_{\mu}^{\alpha \perp} \left[P_{\nu}^{\beta \perp} w_T(q^2)
  + P_{\nu}^{\beta \parallel} w_L(q^2)
\right] \epsilon_{\alpha \beta \sigma \rho}
\hat F^{\sigma \rho},
\end{equation}
where $Q^2=-q^2$, $a$ and $b$ are flavor indices and $d^{ab}=(1/2)\tr({\cal Q} \{t^a,\, t^b \})$
with ${\cal Q}$ being the electric charge matrix, and $P_{\mu}^{\alpha \perp}=\eta_{\mu}^{\alpha}-q_{\mu}q^{\alpha}/q^2$ and
$P_{\mu}^{\alpha \parallel}=q_{\mu}q^{\alpha}/q^2$ are the transverse and longitudinal
projection tensors.

With the bulk-to-boundary propagators introduced before, $w_T$, $\Pi_V$ and $\Pi_A$ can be expressed in the holographic models as
\begin{eqnarray}
w_T(Q^2)&=&\frac{N_C}{Q^2}\int_{-z_0}^{z_0}\mathd z A(Q,z)\partial_z V(Q,z)\\
\Pi_V(Q^2)&=&\frac{1}{Q^2}f^2(z)V(Q,z)\partial_z V(Q,z)|_{z=-z_0}^{z=+z_0}\\
\Pi_A(Q^2)&=&\frac{1}{Q^2}f^2(z)A(Q,z)\partial_z A(Q,z)|_{z=-z_0}^{z=+z_0}.
\end{eqnarray}
Taking into account that $V(Q,z)$ and $A(Q,z)$ are two independent solutions of eq.~(\ref{eq.5D-EoM}) with different
boundary conditions, one obtains the relation~~\cite{Son:2010vc}
\begin{equation}
w_T(Q^2)=\frac{N_C}{Q^2} + \frac{N_C}{f_\pi^2}\,
[ \Pi_V(Q^2)  -  \Pi_A(Q^2) ]\, .\label{eq-SY}
\end{equation}
Taking the $Q^2\to 0$ limit on both sides, one gets the relation between
$C_{22}^W$ and $L_{10}$ consistent with our result (\ref{eq:C22}). Recent studies concerning the inclusion of the
power corrections to the relation can be found in refs.~\cite{Colangelo:2011xk,Iatrakis:2011ht,Domokos:2011dn,Alvares:2011wb,Gorsky:2012ui}.

\subsection{Form factor relation:  $\gamma^*\to\pi\pi$ versus $\pi^0\to \gamma\gamma^*$}

Now let us study more the relation (\ref{eq:C22}) between $C_{22}^W$ and $L_9$. Since $C_{22}^W$ is also related to the anomalous $\pi\gamma^*\gamma^*$ form factor~\cite{Kampf:2005tz} and $L_9$ to the vector form factor of the pion~\cite{Gasser:1984ux}, it is natural to
ask if there is some relation between these two form factors. Actually, it has been pointed out that in two specific models they are equal up to
normalization~\cite{Grigoryan:2008cc,Stoffers:2011xe}. We show that the relation is universal in the class of models considered here.
The vector form factor $F_{\pi}(Q^2)$ is defined as
\begin{equation}
\langle\pi^+(p_1)|j^{\rm EM}_\mu(0)|\pi^+(p_2)\rangle = \mF_\pi(Q^2)(p_1+p_2)_\mu \ ,
\end{equation}
with $j^{\rm EM}_\mu=\bar q\gamma_\mu {\cal Q} q$ the electromagnetic current~($\cal Q$ the charge operator), and $Q^2=-(p_2-p_1)^2$.
Holographically, it has been derived in ref. \cite{Hirn:2005nr}~(see also \cite{Grigoryan:2008cc}), and in our notation is given by
\begin{eqnarray}
\mF_\pi(Q^2)&=&1-\frac{Q^2}{2f_\pi^2}\int_ {-z_0}^{z_0} \frac{1}{g^2(z)} V(Q,z)(1-\psi_0^2) \mathd z\nonumber\\
          &=&\frac{1}{f_\pi^2}\int_{-z_0}^{z_0} f^2(z) V(Q,z) \psi'_0(z)^2  \mathd z.
\end{eqnarray}
The anomalous $\pi\gamma^*\gamma^*$ form factor is defined  by
\begin{align}
\int d^4 x \ e^{-iq_1 x }& \langle {\pi}, {p}|T\left\{J^{\mu }_{\rm EM}(x)\,J^{\nu}_{\rm EM}(0)\right\}| 0 \rangle
\\ \nonumber &= \epsilon^{\mu  \nu \alpha  \beta}q_{1 \, \alpha} q_{2\, \beta} \, \mF_{\gamma^*\gamma^*\pi^0} \left(Q_1^2,Q_2^2
\right ) \ ,
\end{align}
where $ q_{1}, q_{2} $ are the momenta of photons, and  $ q^2_{1,2} =
-Q^2_{1,2} $. For real photons
\begin{equation}
\mF_{\gamma^*\gamma^*\pi^0} \left(0,0\right)=\frac{N_C}{12\pi^2f_\pi}\label{eq-AFF}
\end{equation}
reproduces the anomaly in QCD.
From the holographic approach, the form factor was derived in ref.~\cite{Grigoryan:2008cc} in the hard wall, and the expression for a general background is
\begin{equation}
\mF_{\gamma^*\gamma^*\pi}(Q_1^2,Q_2^2)=\frac{N_C}{24\pi^2f_\pi}\int_{-z_0}^{z_0} V(Q_1,z) V(Q_2,z)\psi_0'(z)\mathd z.
\end{equation}
Taking the limit $Q_1^2=Q_2^2=0$ one recovers the value in eq.~(\ref{eq-AFF}). As shown in ref.~\cite{Grigoryan:2008cc}, this confirms the choice
of the coefficient of the CS term in the action. Employing the equation of motion for $\psi_0(z)$, one finds
\begin{equation}
\mF_{\gamma^*\gamma^*\pi}(Q^2,0)=\frac{N_C}{12\pi^2f_\pi}~ \mF_\pi(Q^2).\label{eq-VFFTFF}
\end{equation}
Taking the slope at $Q^2=0$ on both sides, one obtains the relation between
$C_{22}^W$ and $L_9$ in eq.~(\ref{eq:C22}).

\begin{figure}[htbp]
\centerline{$$\epsfxsize=1.0\textwidth\epsffile{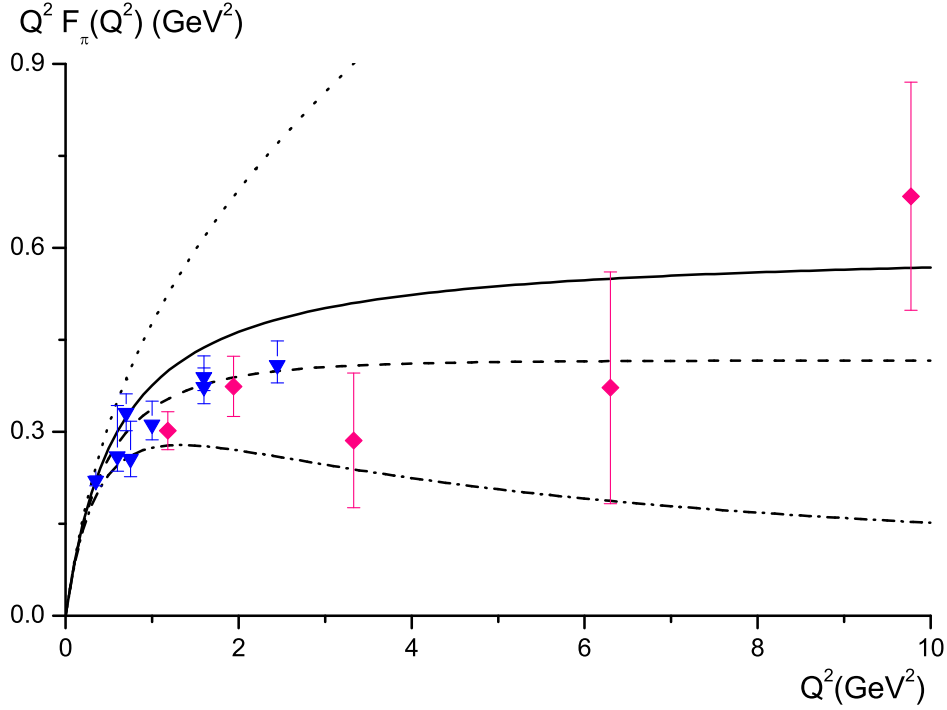}$$}
\caption{\it Vector form factor $\mF_\pi(Q^2)$ from the flat, ¡±Cosh¡±, hard wall and Sakai-Sugimoto models,
denoted by the dotted, solid, dashed and dash-dotted lines, respectively. The experimental data are
 from ref. \cite{Bebek:1977pe} (diamonds) and ref. \cite{Huber:2008id} (triangles).}\label{fig:VFF}
\end{figure}

\begin{figure}[htbp]
\centerline{$$\epsfxsize=1.0\textwidth\epsffile{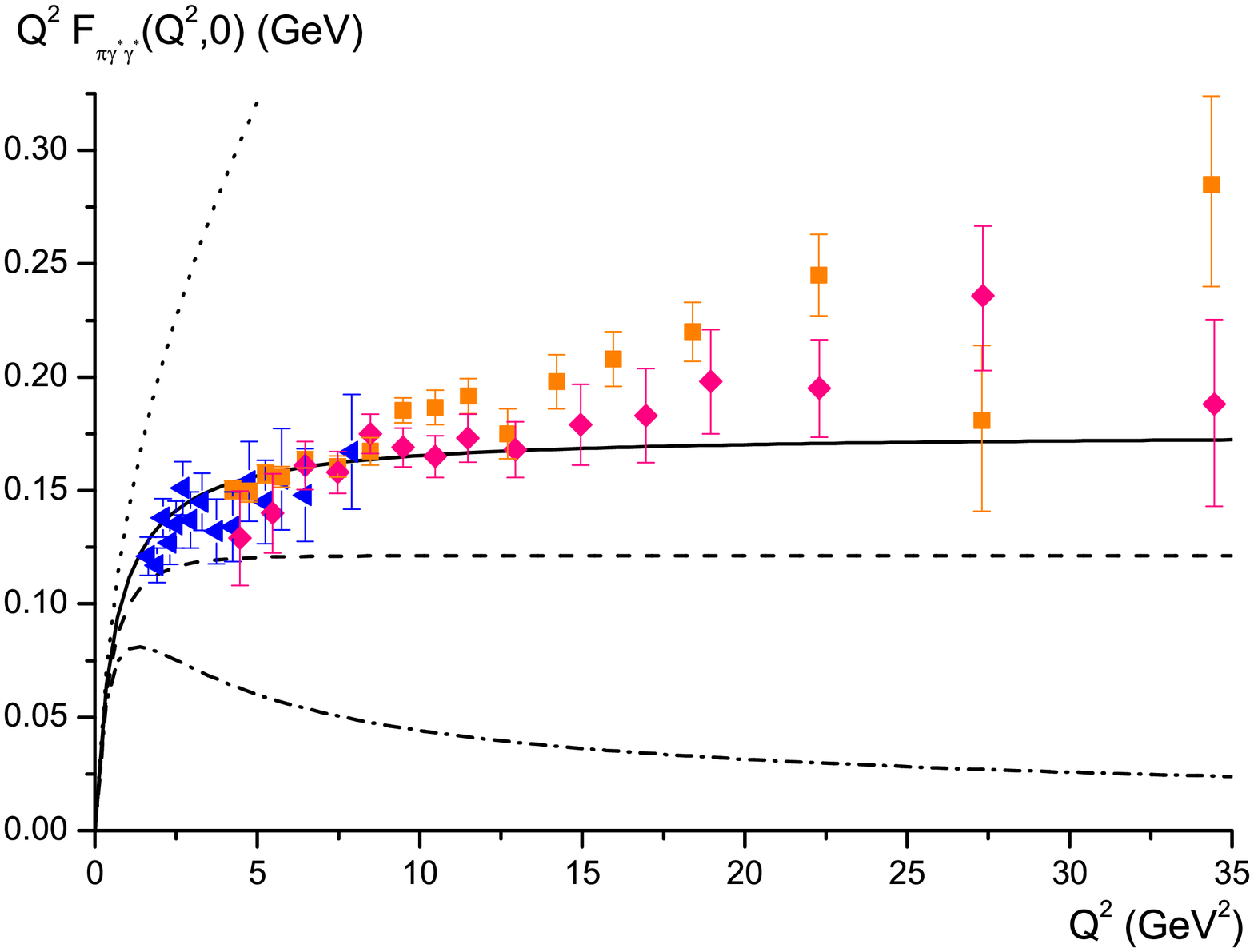}$$}
\caption{\it Anomalous $\pi\gamma^*\gamma^*$ form factor from the flat, ¡±Cosh¡±, hard wall and Sakai-Sugimoto
models, with the same notation as in fig. \ref{fig:VFF}. The experimental data are from CLEO \cite{Gronberg:1997fj} (tri-
angles), BABAR \cite{Aubert:2009mc} (squares) and BELLE collaboration\cite{Uehara:2012ag} (diamonds).}\label{fig:TFF}
\end{figure}

In the previous section we have shown that the results for the LECs do not sensibly depend
on the details of the different models. However, for the form factors, the results from different
models can differ from each other, especially in the high momentum region. Depending
on the asymptotic metric in the UV, the form factor exhibits different power behavior when $Q^2\to \infty$.
For $\mF_{\gamma^*\gamma^*\pi}(Q^2,0)$, the explicit high-energy power structure in different models is analyzed in
ref.~\cite{Zuo:2011sk}. In Fig.~\ref{fig:VFF} and Fig.~\ref{fig:TFF} we show the numerical results of both the form factors in the relation (\ref{eq-VFFTFF}) for the four different models, together with the available experimental data~\footnote{For a review and references of the $\pi\gamma^*\gamma^*$ form factor, please see ref.~\cite{TFFrefs}.}. From the figures one clearly
finds  the different power behavior of each model. To reproduce the observed $1/Q^2$
behavior for the form factors, the models need to be asymptotically AdS in the UV, and this is the case of the
¡±Cosh¡± and hard wall models. In such models, one can simplify the metric functions in the
UV region by using the Poincar\'{e} coordinates
\begin{equation}
f^2(u)=1/g^2(u)\sim 1/g^2_5 u, ~~u\to 0.
\end{equation}
Following the procedure in ref.~\cite{Grigoryan:2008up}, one finds the large-$Q^2$ behavior of the form factors~\cite{Grigoryan:2008cc}
\begin{equation}
\mF_\pi(Q^2)\to \frac{f_\pi^2 g_5^2}{Q^2},~~~~~\mF_{\gamma^*\gamma^*\pi}(Q^2,0)\to \frac{N_C g_5^2 f_\pi}{12\pi^2 Q^2}.
\end{equation}
The relation (\ref{eq-VFFTFF}) between the two form factors dictates that the leading-power
coefficients are proportional to each other. However, new experimental data for the vector
form factor in the large momentum region are needed to confirm this. It is worth mentioning
that, if the 5D coupling $g_5$ is fixed through the comparison with the perturbative
logarithmic term of the vector correlator, $g_5^2=24\pi^2/N_C$~\cite{Son:2003et}, then the above asymptotic behavior
of $\mF_{\gamma^*\gamma^*\pi}(Q^2,0)$ is the same as the perturbative result~\cite{Lepage:1979zb,Lepage:1980fj,Brodsky:1981rp}. With this, the vector form
factor has the high energy behavior $Q^2 \mF_\pi(Q^2)\to 8 \pi^2 f_\pi^2$ \cite{Grigoryan:2008cc}.

\subsection{Form factor relation: $A\to \pi\pi\pi$ versus $\pi\to AA$}

One may speculate whether there is a similar relation between the form factors
involving the axial-vector source, based on eq.~(\ref{eq:C23}).    In addition to the one-resonance
terms~(\ref{eq.S-1res-even}) and (\ref{eq.S-1res-odd}),     we need the operators
with two resonance fields:
\bear
S_{YM}\bigg|_{\rm 2-res.} &=&    i\,  c_{ a^m v^n \pi}   \,  \Int d^4x \,   \bra
(\nabla_\mu v^n_\nu -\nabla_\nu v^n_\mu) \, [u^\mu, a^{m\,\, \nu}]
+
(\nabla_\mu a^m_\nu -\nabla_\nu a^m_\mu) \, [u^\mu, v^{n\,\, \nu}]
\ket \,\,\, +\,\,\, ...
\nn\\
\nn\\
S_{CS}\bigg|_{\rm 2-res.} &=&
\Frac{N_C}{6\pi^2} \, \tilde{c}_{a^n a^m}\,
\epsilon^{\mu\nu \alpha\beta}    \,  \Int d^4x \,   \bra
u_\mu \{  a_\nu^n   ,  \nabla_\alpha a_\beta^m  \}  \ket     \,\,\, +\,\,\, ...
\eear
where the dots stand for  operators which are not relevant for the calculation
of the $\pi\to AA$ and $A\to \pi\pi\pi$ form factors.     The resonance couplings
are given by the 5D integrals
\bear
c_{a^m v^n  \pi} &=&  \Int_{-z_0}^{z_0} dz\,\Frac{1}{g^2(z)}\, \psi_0(z) \, \psi_{2m}(z)\psi_{2n-1}(z) \, ,
\nn\\
\tilde{c}_{a^n a^m} &=& \Int_{-z_0}^{z_0} dz\, \psi_0(z)\,  \psi_{2n}(z)\,  \psi_{2m}'(z) \, ,
\\
\eear
A detailed calculation of the resonance couplings from the CS term can be found in ref.~\cite{Domokos:2009cq}.

\subsubsection{Odd-sector form factor:  $\pi\to AA$ }

In parallel with the pion-photon transition form factor, we can relate $C_{23}^W$
to the form factor   involving two axial-vector sources:
\begin{eqnarray}
&&\int<\pi^c(p)|T\{j_\mu^{5a}(x)j_\nu^{5b}(0)\}|0>e^{-iq_1x} \mathd^4x\nonumber\\
&&~~=\frac{i~N_C}{24\pi^2f_\pi}~D^{abc}~\epsilon_{\mu\nu\alpha\beta}~q^{\alpha}_1q^{\beta}_2~\mF_{\pi AA}(Q_1^2,Q_2^2),
\end{eqnarray}
with $p=q_1+q_2$, $Q_1^2=-q_1^2$, $Q_2^2=-q_2^2$.
Here we are considering the axial current
$j_\mu^{5a}=\bar q \gamma_\mu \gamma_5 t^a q$ with $t^a$ the generator in
$U(N_f)$, $D^{abc}=2 \Tr (\{t^a,t^b\}t^c)$ is the corresponding fully symmetric tensor.
At low-energies, this form factor is given in    $\chi$PT by the expression
\bear
\mF_{\pi AA}(Q_1^2,Q_2^2) &=& 1 \, +\,  \Frac{ 192\pi^2 C_{23}^W}{N_C}\,
(Q_1^2 +Q_2^2)\,\,\, \,\,+\,\,\cO(E^4)\, ,
\label{eq.pi-AA-ChPT}
\eear
where higher order corrections are in the $\cO(E^4)$  term. Computing the local diagrams
from  the WZW term and the one and two axial-vector resonance
exchanges one gets:
\bear
\mF_{\pi AA}(Q_1^2,Q_2^2) &=& 1
\, -\, \sum_n \Frac{3 a_{Aa^n } c_{a^n}}{2}  \, \bigg[
\Frac{Q_1^2}{m_{a^n}^2 +Q_1^2} \, +\, \Frac{Q_2^2}{m_{a^n}^2 +Q_2^2}
 \bigg]
\nn\\
&& \qquad  +\,
\sum_{m,n}   \Frac{3  c_{a^n a^m} a_{Aa^n}    Q_1^2  }{(m_{a^n}^2 +Q_1^2)  }\,
\Frac{ a_{Aa^m} Q_2^2}{  (m_{a^m}^2 +Q_2^2) }\, .
\eear

where $c_{a^n a^m} = -\frac{1}{2} \tilde{c}_{a^n a^m}  -\frac{1}{2} \tilde{c}_{a^m a^n}$. If one sets to zero the squared momentum of one of the axial-vector sources,
the $\pi\to AA$ form factor becomes
\bear
\mF_{\pi AA}(Q^2,0) &=& 1
\, -\, \sum_{n} \Frac{3 a_{Aa^n } c_{a^n}  }{2} \,
\Frac{Q^2}{m_{a^n}^2 +Q^2} \, .
\eear

In the holographic approach, following the same procedure as the $\pi\gamma^*\gamma^*$ form factor,
we can express this form factor through the bulk-to-boundary propagator $A(Q,z)$
\begin{equation}
\mF_{\pi AA}(Q_1^2,Q_2^2)=\frac{3}{2}\int_{-z_0}^{z_0} A(Q_1,z)A(Q_2,z)\psi'_0(z) \mathd z. \label{eq:pAA}
\end{equation}

\subsubsection{Even-sector form factor: $A\to \pi\pi\pi$}

In the YM part, in parallel with the vector form factor, we can define the form factor
involving an axial-vector source and three pions
\begin{eqnarray}
&&<\pi^a(p_1)\pi^b(p_2)\pi^c(p_3)|i j_\mu^{5d}|0>\nonumber\\
&~&=f^{bce}f^{ade}~P_\mu^{\nu\perp}(q)[\mF_1(Q^2,s,t)(p_1-p_3)_\nu+\mF_2(Q^2,s,t)(p_2-p_3)_\nu]+(a\leftrightarrow c),
\end{eqnarray}
where $q=p_1+p_2+p_3$, $P_\alpha^{\mu\perp}(q)=\eta_\alpha^\mu-q_\alpha q^\mu/q^2$,
$Q^2=-q^2$, $s=(p_1+p_3)^2$, $t=(p_2+p_3)^2$, and we also use
$u=(p_1+p_2)^2$, which obeys the Mandelstam relation $s+t+u=q^2$
(the light pseudo-scalars are massless in the chiral limit considered all along this article).
Due to Bose symmetry, the two form
factors $\mF_1$ and $\mF_2$ are related through  $\mF_1(Q^2,s,t)=\mF_2(Q^2,t,s)$.
At large distances, $\chi$PT yields  the amplitude
\bear
\mF_1(Q^2,s,t) &=& \Frac{ 2}{3 f_\pi}\, \bigg[
\, 1\,\, +\,\, \Frac{2 L_9 \,q^2}{f_\pi^2}\,\,
-\,\, \Frac{16 L_1 (u + t  -s/2)}{f_\pi^2}\,\,\,\, +\,\, \cO(E^4)\,\bigg]\, .
\label{eq.3pi-AFF-ChPT1}
\eear
In the limit $s,t\to 0$~(i.e., $p^\mu_3\to 0$), hence $u\to q^2$, this expression
becomes:
\bear
\mF_1(Q^2,0,0) &=& \Frac{ 2}{3 f_\pi}\, \bigg[
\, 1\,\, +\,\, \Frac{2 q^2 \, (L_9 -8L_1)}{f_\pi^2}  \,\,\,\, +\,\, \cO(q^4)\,\bigg]  \, .
\label{eq.3pi-AFF-ChPT2}
\eear

The extraction of the form factor through the 5D action turns out to be difficult.
Instead, we choose to work in the 4D picture and include all the contributions diagram
by diagram. Summing up the diagrams with only Goldstones,   one-resonance and
two-resonance exchanges,  we obtain:
\bear
\mF_1(Q^2,s,t) &=&
 \Frac{ 2}{3 f_\pi}\, \bigg\{
\, 1\,\, +\,\, \Frac{2 L_9 \,q^2}{f_\pi^2}\,\,
-\,\, \Frac{16 L_1 (u + t  -s/2)}{f_\pi^2}   \,
+ \, \sum_n \Frac{a_{Aa^n} b_{a^n \pi^3}}{2 f_\pi^2} \,
\Frac{ q^2 (u+t-s/2)}{m_{a^n}^2-q^2}
\nn\\
&&
\qquad +\,  \sum_n \bigg[ \Frac{ (a_{Vv^n}-b_{v^n\pi\pi}) b_{v^n\pi\pi} }{4f_\pi^2}\bigg(
\Frac{ (2 u -t) s}{m_{v^n}^2 -s}\, -\, \Frac{ (u-s)t}{m_{v^n}^2 -t}
\bigg)
\nn\\
&&
\qquad \qquad\quad  +\, \Frac{ a_{Vv^n} b_{v^n\pi\pi}}{4 f_\pi^2} \bigg(
\Frac{(6s +u + 2t) s}{m_{v^n}^2-s}  \, - \,   \Frac{(s-u) t}{m_{v^n}^2 -t}
\bigg)
\, -\, \Frac{3 b_{v^n \pi\pi}^2}{8 f_\pi^2}\,\Frac{(u+t)s}{m_{v^n}^2-s}  \bigg]
\nn\\
&&
\qquad +\, \sum_{m,n}\Frac{ a_{Aa^m} c_{a^m v^n\pi} b_{v^n\pi\pi}}{4 f_\pi^2}
\, \Frac{q^2}{m_{a^m}^2 -q^2}\,
\bigg( \Frac{s (2u +t)}{m_{v^n}^2-s} +\Frac{t(s-u)}{m_{v^n}^2-t}\bigg)\,\bigg\} \, .
\eear
In the kinematic configuration $s=t=0$ 
the contribution from diagrams where two of the Goldstones are produced
through an intermediate vector resonance vanishes. Thus, the form factor
is greatly simplified into
\bear
\mF_1(Q^2,0,0) &=&
 \Frac{ 2}{3 f_\pi}\, \bigg[
\, 1\,\, +\,\, \Frac{2 q^2\, (L_9 - 8 L_1)}{f_\pi^2}
\, + \, \sum_n\Frac{a_{Aa^n} b_{a^n \pi^3}}{2 f_\pi^2} \, \Frac{ q^4}{m_{a^n}^2-q^2} \,\bigg]\, .
\eear
It is interesting to observe that,    in the case when
the resonance summations in the sum rules in eq.~(\ref{eq.sum-rule-AFF}) are convergent,
the dominant high-energy power behavior is given by
\bear
\mF_1(Q^2,0,0) &=&
 \Frac{ 2}{3 f_\pi}\, \bigg[
\, \, \, \Frac{q^2}{2 f_\pi^2}\, \bigg(   4L_9 - 32 L_1\, - \,
\sum_n a_{Aa^n} b_{a^n \pi^3}\bigg)
\nn\\
&&\qquad\qquad
 +\,\,\bigg(  1 \, -\, \sum_n \Frac{a_{Aa^n} b_{a^n \pi^3} m_{a^n}^2}{2 f_\pi^2}
 \bigg)\,\,\,\,+\,\,\, ...
 \,\bigg]\, ,
\eear
where the dots stand for contributions that vanish at high energies.
A closer look at the sum rules~(\ref{eq.sum-rule-AFF}) leads
to the prediction $\mF_1(Q^2,0,0)\stackrel{q^2\to \infty}{\longrightarrow} 0$.
This  short-distance condition can serve  to further constrain the
$A\to \pi\pi\pi$ amplitude in the case when only the lightest resonances
are taken into account~\cite{GomezDumm:2003ku,Dumm:2009va}.

The form factor  can be rewritten in terms of the bulk-to-boundary propagators and the
Green's function $G(Q^2;z,z')$ provided in~(\ref{eq.G-GreenFun}):
\begin{eqnarray}
\mF_1(Q^2,s,t)&=&\frac{2}{3f_\pi}+\frac{1}{3f_\pi}(6s+3u+t)\frac{\mF_\pi(-s)-1}{s}\nonumber\\
             &&-\frac{2}{3f_\pi^3}L_9 (3s+3u+t)+\frac{1}{3f_\pi^3}(u+t-s/2)\frac{\mF_0(Q^2)}{Q^2}\nonumber\\
             &&+\frac{1}{6f_\pi^3}(s-u)T_1(-t)-\frac{1}{18f_\pi^3}(8u-t)T_1(-s)-\frac{1}{3f_\pi^3}(s-u)T_2(Q^2,-t)
\end{eqnarray}
with
\begin{eqnarray}
\mF_0(Q^2)&\equiv &Q^2~\int_{-z_0}^{z_0}\frac{\psi_0(1-\psi_0^2)}{g^2(z)}A(Q,z) \mathd z\nonumber\\
T_1(-t)&\equiv& \int_{-z_0}^{z_0}\int_{-z_0}^{z_0} \frac{1-\psi_0(z)^2}{g^2(z)} \frac{1-\psi_0(z')^2}{g^2(z')}~tG(-t,z,z') ~\mathd z\mathd z'\nonumber\\
T_2(Q^2,-t)&\equiv& \int_{-z_0}^{z_0}\int_{-z_0}^{z_0} \frac{1-\psi_0(z)A(Q,z)}{g^2(z)} \frac{1-\psi_0(z')^2}{g^2(z')}~t G(-t,z,z') ~\mathd z\mathd z',
\end{eqnarray}
and $\mF_\pi$ the vector form factor. Since the final result for the form factor involves only those 5D quantities, it seems possible
 to derive it directly from the original YM action: this is still under investigation.
In the kinematical limit $s,t\to 0$ this expression becomes:
\be
\mF_1(Q^2,0,0) \,\,=\,\, \Frac{2}{3 f_\pi}\, \bigg[ \, 1 \, - \, \Frac{1}{2 f_\pi^2}\mF_0(Q^2)\,
\bigg]\, .
\ee

\subsubsection{Comparison of anomalous and even-sector form factors}

\begin{figure}[htbp]
\centerline{$$\epsfxsize=1.0\textwidth\epsffile{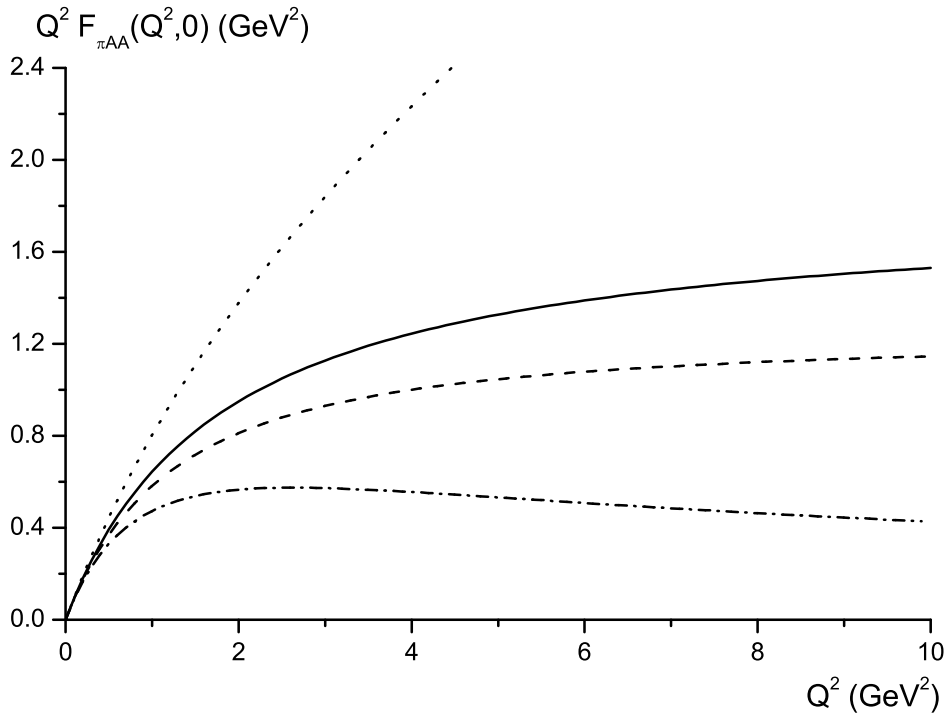}$$}
\caption{\it Results for the $\pi AA$ transition form factor in the flat~(dotted), ¡±Cosh¡±~(solid), hard wall~(dashed) and Sakai-Sugimoto~(dash-dotted) models.}\label{fig:FFpiAA}
\end{figure}

By means of the equation of motion for the bulk-to-boundary propagator $A(Q,z)$,
one can also express the anomalous form factor (\ref{eq:pAA})  in terms of the previously
defined function $\mF_0(Q^2)$.
Hence, independently  of the precise details of the models, one finds
the relation: 
\begin{equation}
\mF_1(Q^2,0,0) \,\, =  \,\,   \Frac{2}{3f_\pi}  \,  \mF_{\pi AA}(Q^2,0)\, .
\label{eq.FF-rel}
\end{equation}
This constitutes the generalization of the relation
$C_{23}^W = \frac{N_C}{96\pi^2 f_\pi^2} (L_9-8 L_1)$ in  eq.~(\ref{eq:C23})
between the odd-sector $\cO(p^6)$ LEC $C_{23}^W$  and the even-sector $\cO(p^4)$
chiral couplings $L_1$ and $L_9$. Indeed, as a final check,
if one studies~(\ref{eq.FF-rel}) at low energies with the help of their $\chi$PT
expansions~(\ref{eq.pi-AA-ChPT}) and (\ref{eq.3pi-AFF-ChPT2}),
one recovers the relation~(\ref{eq:C23}).
In Fig. \ref{fig:FFpiAA} we show the numerical results of the form factor $\mF_{\pi AA}(Q^2,0)$ in different models. One finds that
the asymptotic behavior is similar to the corresponding one for the $\pi\gamma\gamma^*$ form factor.

From eqs.~(\ref{eq:C12})--(\ref{eq:C21}) one might speculate
the possible existence of more relations of this kind between
even and odd-sector amplitudes. This will be the subject of future studies.

\section{Conclusions}
\label{sec.conclusions}
We have performed an exhaustive study of the $\cO(p^6)$ LECs
for the 5D holographic theories  which implement chiral symmetry breaking
through different boundary conditions in the infrared.
The five-dimensional action was given by the Yang-Mills and the Chern-Simons terms.
All the theoretical relations are determined for general backgrounds
$f^2(z)$ and $g^2(z)$.  Only for the numerical results we specified the precise expressions for such functions,
which were taken from four models: ``flat'' background~\cite{Son:2003et},
``Cosh'' model~\cite{Son:2003et}, hard-wall~\cite{Hirn:2005nr} and the Sakai-Sugimoto
model~\cite{Sakai:2004cn,Sakai:2005yt}.
We found that the outcome for the LECs was stable, with a weak dependence on the background.
We considered the  results from the ``Cosh'' model as our best estimate, since this model, among
the four studied ones, better matches the experimental $\rho(770)$ mass and pion decay constant
in addition to the perturbative QCD log coefficient in the $VV$ correlator.
Remarkably, this model reproduces fairly well the new experimental data
for the $\pi^0\to\gamma\gamma^*$ transition form factor.

As a previous step, we worked out several resonance sum rules
which became essential for the relations  between the odd-parity and even LEC's
derived later.  Likewise, some of these sum rules were employed in the
form factor analysis.  For the sake of completeness, we derived
(or rederived in some cases~\cite{Hirn:2005nr,Sakai:2005yt})
all the sum rules we could, regardless of whether
we used them in our later study.
As many of these sum rules are used in hadronic phenomenology to fix the
resonance parameters, we checked and found that, in general, the lowest meson exchanges
provide the dominant contributions. The only exception was the sum rules related
to the $VV-AA$ correlator, which are divergent due to the meson mass behavior
$m_{R^n}^2\sim n^2$ in this kind of holographic models.

We have computed the  $\cO(p^6)$ chiral couplings
by integrating out the heavy resonances in the generating functional.
In the odd-parity sector, we were able to express all the odd $\cO(p^6)$ LECs in terms
of even-sector $\cO(p^4)$ couplings $L_1$ and $L_9$,
and the 5D integral $Z$, which can be also defined from a sum rule.
In particular, we recovered the LEC relation
$C_{22}^W=- \frac{N_C}{32\pi^2 f_\pi^2}L_{10}$~\cite{Knecht:2011wh}
that stems from the Son-Yamamoto relation between the $LR$
and $AVV$ Green's functions~\cite{Son:2010vc}.
These relations are general for the type of holographic theories considered here,
 and do not depend on the details of the functions
$f^2(z)$ and $g^2(z)$ that specify the model.
At the numerical level, the outcomes were found to be fairly stable, suffering
little variation between the various holographic models studied here.
Not much is known about the LECs of the odd-parity sector:
we have compared our holographic determinations to those from other
approaches  (DSE, $\cO(p^6)$ $\chi$PT phenomenological analyses, resonance estimates
and rational approximations).  In general we have found a reasonable agreement,
considering that subleading $1/N_C$ corrections are not taken into account
in the holographic approach and that there is an inherent uncertainty on the renormalization scale $\mu$
at which our large--$N_C$ determinations correspond.
We also tested a theoretical relation~(\ref{eq.Kampf-rel})
derived from Resonance Chiral Theory~\cite{Kampf:2011ty},
which was pretty well satisfied by the lightest resonance multiplets.

The $\cO(p^6)$  even-parity  LECs have also been determined.  Since our framework does not include
scalar-pseudoscalar sources $\chi$, we obtained predictions only for the $\chi$PT operators
without $\chi$.  Indeed, in the $\chi=0$ limit the basis of $\cO(p^6)$ operators in
the chiral action can be further simplified, and we chose to eliminate the operator referred by
$C_{90}$.
We have also found some relations between $\cO(p^6)$ couplings in the even-parity sector.
In particular, those combinations of LECs related to the axial-vector form factor in the $P\to\ell\nu \gamma$ decay
and to  the $\gamma\gamma\to\pi\pi$ collision either vanish or
can be expressed in terms of just $f_\pi$, independently of the
holographic theory at hand.
There is a wider phenomenology on the $\cO(p^6)$ couplings of the even sector.
Nonetheless, most of them remain essentially unknown.
The agreement with former results (DSE, $\cO(p^6)$ $\chi$PT determinations,
resonance estimates and rational approximations)   is  reasonably good.
Apart from a DSE analysis, which was  able to provide the full set of LECs~\cite{Jiang:2009uf},
there are no other approaches to compare most of our chiral coupling estimates with.
The larger disagreements are found for the couplings that receive contributions
from two odd-parity resonance couplings. Taking these contributions into account,
one finds a better agreement with the corresponding DSE outcomes.
In other few cases, the disagreement cannot be explained in this way, however, the corresponding couplings
are related  to processes with a high number of external legs.
In string constructions, it has been argued about the possible presence of
cubic, $(\mF_{MN})^3$, and higher terms in the 5D action which would modify
the value of those LECs. Likewise, scalar resonances are absent in our approach,
and they may play a role in some LECs.

Some of the relations we have just found
among   couplings of different parity $\chi$PT Lagrangians
are the consequence of relations between QCD amplitudes
of the anomalous and even-parity sectors.
The original motivation for this work was the relation found by Son and Yamamoto
between the $LR$ and $AVV$ Green's functions~\cite{Son:2010vc}.
We found that, for any type of background, the
$\pi\pi$ vector form factor and the $\pi\to\gamma\gamma^*$ transition form factor
are identical for any energy up to a known overall normalization,
as it was already hinted in previous works~\cite{Grigoryan:2008up,Stoffers:2011xe}.
At large momentum, this relation dictates that the leading-power
coefficients are proportional to each other, which can be experimentally checked.
 %
 In the last section we studied the pion transition into two axial-vector currents
 and the axial-vector form factor into three pions. We found a relation between them for
 a particular kinematical configuration. In addition, we showed how this and
 the previous amplitudes could be rewritten from the 4D picture (with an infinite number
 of resonance exchanges)   into  a holographic  form
 (in terms of bulk-to-boundary propagators). However, in the case of the $\pi\pi\pi$ AFF,
 with four external legs, one also needs to include the contributions
 with bulk-to-bulk Green's function propagators connecting two points in the bulk
 $z$ and $z'$.
 All these amplitude relations, when taken to the low energy limit and compared at
 each chiral order, reproduce the relations between the odd and even-parity LECs
 we derived through the generating functional, serving as a double-check
 of our chiral coupling determinations.

 The study of other consequences of some of the relations
 obtained here both in the odd and even sectors of QCD,
 e.g.,  those in the $P\to\ell\nu\gamma$ decay or the $\gamma\gamma\to\pi\pi$ scattering, requires new dedicated analyses.
 Moreover, new experimental data on the various form factors, in particular at high energies,
 would definitely help to discern the most appropriate version among the various
 holographic models, in the search of a precise dual formulation of QCD.

\section*{Acknowledgments}

We thank F. De Fazio,  F. Giannuzzi, L. Girlanda, K. Kampf and S. Nicotri for their comments on the paper and discussions.
We are also grateful to S. Peris and A. Vainshtein for comments about the validity of the Son-Yamamoto relation in SU(3),
and Qing Wang for explaining the anomaly structure in the Dyson-Schwinger equation approach. This work is partially supported by the Italian MIUR PRIN 2009. J.J.S.C. has been partially supported by the Universidad CEU Cardenal Herrera
grants PRCEU-UCH15/10 and PRCEU-UCH35/11, and the MICINN-INFN fund AIC-D-2011-0818.
F.Z. was supported in part by the National Natural Science Foundation of China under Grant No. 11135011.

\appendix

\section{Holographic models}
\label{app.holographic-models}

Four different holographic models have been studied in the present paper, among them the ``Cosh"  and ``hard wall" background
are asymptotically AdS in the UV. In all these models there are two parameters, the 5D gauge coupling $g_5$,
 and the energy scale $\Lambda$ (or $z_0^{-1}$). We fix them in a unified way through the $\rho$ meson mass $m_\rho=0.776~\mbox{GeV}$ and the pion decay constant $f_\pi=0.087 ~\mbox{GeV}$.
Therefore, the results for $g_5$ differ from that fixed through the high-energy vector correlator in asymptotically AdS models, which gives $g_5^2=24\pi^2/N_C$~\cite{Son:2003et}. In the non-AdS models, one is able to fix it in this way because the perturbative QCD logarithm is not recovered.

\subsection{{``}Flat" background}

The model is specified by~\cite{Son:2003et}
\begin{equation}
f^2(z)=\Lambda^2 /g_5^2, ~~g^2(z)=g_5^2, ~~z_0=1.
\end{equation}
The pion decay constant, meson masses, wave functions and couplings are given by
\begin{eqnarray}
V(Q,z)&=&\frac{\cosh (Qz/\Lambda)}{\cosh (Q/\Lambda)},~~~~~A(Q,z)= \frac{\sinh (Qz/\Lambda)}{\sinh (Q/\Lambda)},\nonumber\\
f_\pi^2&=&\frac{2 \Lambda^2}{g_5^2},~~~~\psi_0(z)=z,\nonumber\\
m_n^2&=&\frac{\pi^2 \Lambda^2}{4}n^2,~~~\psi_n(z)=(-1)^{n-1} g_5\sin \bigg(\frac{n\pi}{2}(z+1)\bigg),\nn\\
a_{Vv^n}&=&a_{2n-1},~~~~a_{Aa^n}=a_{2n},~~~a_n=\frac{4}{n \pi g_5}\nonumber\\
c_{v^n}&=&\frac{2g_5}{(2n-1)\pi},~~~c_{a^n}=\frac{2g_5}{n\pi},\nonumber\\
d_{v^n}&=&\frac{2 g_5}{(2n-1)\pi}\left[1-\frac{8}{(2n-1)^2\pi^2}\right].
\end{eqnarray}

\subsection{ {``}Cosh" model}

The background functions are specified by~\cite{Son:2003et}
\begin{equation}
f^2(z)=\Lambda^2 \cosh^2(z) /g_5^2, ~~g^2(z)=g_5^2,~~ z_0=\infty.
\end{equation}
The various solutions and physical quantities are
\begin{eqnarray}
V(Q,z)&=&-\frac{\pi}{2}\csc (\nu\pi) \sqrt{1-\tanh^2z} [P_\nu^1(\tanh z)+P_\nu^1(-\tanh z)],\nonumber\\
A(Q,z)&=&\frac{\pi}{2}\csc (\nu\pi) \sqrt{1-\tanh^2 z} [P_\nu^1(\tanh z)-P_\nu^1(-\tanh z)],\nonumber\\
f_\pi^2&=&\frac{2 \Lambda^2}{g_5^2},~~~~\psi_0(z)=\tanh z,\nonumber\\
m_n^2&=&n(n+1)\Lambda^2,\nonumber\\
\psi_n(z)&=&- g_5 c_n \frac{P_n^1(\tanh z)}{\cosh z},~~~~c_n=\sqrt{\frac{2n+1}{2n(n+1)}},\\
a_{Vv^n}&=&a_{2n-1},~~~~a_{Aa^n}=a_{2n},~~~a_n=\frac{1}{g_5}\sqrt{\frac{2(2n+1)}{n(n+1)}},\nonumber\\
c_{v^n}&=&\frac{g_5}{\sqrt{3}}\delta_{n,1},~~~c_{a^n}=\frac{2g_5}{\sqrt{15}}\delta_{n,1},\nonumber\\
d_{v^n}&=&\frac{\sqrt{3}g_5}{15}\delta_{n,1}+\frac{2\sqrt{42}g_5}{105}\delta_{n,2},
\end{eqnarray}
where $\nu(\nu+1)=-Q^2/\Lambda^2$, and $P_\nu^1(z)$ is the associate Legendre function.

\subsection{Hard-wall model}

In this case the functions $f^2(z)$ and $g^2(z)$ are given by~\cite{Hirn:2005nr}
\begin{equation}
f^2(z)= \frac{1}{g_5^2(z_0-|z|)},~~ g^2(z)=g_5^2 (z_0-|z|), ~~z_0<\infty.
\end{equation}
It is more convenient to  focus on the interval $0<z<z_0$, and use the coordinate $\tilde z=z_0-z$. Then we have:
\begin{eqnarray}
V(Q,\tilde z)&=&Q\tilde z\left[K_1(Q\tilde z) +I_1(Q\tilde z) \frac{K_0(Qz_0)}{I_0(Qz_0)}\right],\nonumber\\
A(Q,\tilde z)&=&Q\tilde z\left[K_1(Q\tilde z) -I_1(Q\tilde z) \frac{K_1(Qz_0)}{I_1(Qz_0)}\right],\nonumber\\
f_\pi^2&=&\frac{4}{g_5^2 z_0^2},~~~~\psi_0(\tilde z)=1-\frac{\tilde z^2}{z_0^2},\nonumber\\
m_{v_n}&=&\frac{\gamma_{0,n}}{z_0},~~~~m_{a_n}=\frac{\gamma_{1,n}}{z_0},\nonumber\\
\psi_{v_n}(\tilde z)&=&\frac{g_5}{z_0|J_1(\gamma_{0,n})|} \tilde z~J_1\left(m_{v_n} \tilde z \right),\nonumber\\
\psi_{a_n}(\tilde z)&=&\frac{g_5}{z_0\left[-J_0(\gamma_{1,n})J_2(\gamma_{1,n})\right]^{1/2}} \tilde z~J_1\left(m_{a_n} \tilde z \right).
\end{eqnarray}
Here $K_n(x)$, $I_n(x)$ and $J_n(x)$ are Bessel functions, and $\gamma_{n,m}$ is the $m^{th}$ root of $J_n(x)$. The couplings
$a_{Vv^n},a_{Aa^n},c_{v^n},c_{a^n},d_{v^n}$ can be calculated from these wave functions. Since the integrals of the  Bessel functions in
the final results can not be further simplified, we do not list them individually.

\subsection{Sakai-Sugimoto model}

The model is determined by~\cite{Sakai:2004cn,Sakai:2005yt}
\begin{equation}
f^2(z)=\frac{\Lambda^2 (1+z^2)}{g_5^2} , ~~g^2(z)=g_5^2 (1+z^2)^{1/3}, ~~z_0=\infty.
\end{equation}
In this model one only finds analytic results for the pion decay constant and the wave function $\psi_0$:
\begin{equation}
f_\pi^2=\frac{4 \Lambda^2}{\pi g_5^2},~~~~\psi_0(z)=\frac{2}{\pi} \arctan z.\nonumber
\end{equation}
The quantities related to the resonances need to be calculated numerically, as done in ref.~\cite{Sakai:2004cn}. The results for the first
two excitations are:
\begin{eqnarray}
m_{v^1}^2&=&0.669~ \Lambda^2,~~~m_{a^1}^2=1.57 ~\Lambda^2,\nonumber\\
a_{Vv^1}&=&\frac{3.15}{g_5},~~~a_{Aa^1}=\frac{3.20}{g_5},\nonumber\\
c_{v^1}&=&0.415 ~g_5,~~~c_{a^1}=0.321 ~g_5\nonumber\\
d_{v^1}&=&0.0875 ~g_5.
\end{eqnarray}

\section{Son-Yamamoto relation at the one-loop level}
\label{app.1loop-SY-rel}

In this appendix we study the relation (\ref{eq-SY}) between $w_T$ and $\Pi_V-\Pi_A$, obtained by Son and Yamamoto~\cite{Son:2010vc},
that we rewrite here:
\be
w_T(Q^2)\, =\, \Frac{N_C}{Q^2}\,+\, \Frac{N_C}{f_\pi^2}\, (\, \Pi_V(Q^2)-\Pi_A(Q^2))\, ,
\label{eq.SY-rel-appendix}
\ee
at the one-loop level, as proposed in ref.~\cite{Gorsky:2012ui}.
This relation was  derived in the large--$N_C$ limit,  i.e., for tree-level amplitudes.

At low energies,  the comparison of the leading terms in the $\chi$PT expansion of the l.h.s. and r.h.s.
of eq.~(\ref{eq.SY-rel-appendix})
yields a relation between  the $\cO(p^6)$ odd-parity coupling $C_{22}^W$
and the $\cO(p^4)$ even-parity constant $L_{10}$~\cite{Knecht:2011wh}, corresponding to eq.~(\ref{eq:C22}), that again we write:
\be
128\pi^2 f_\pi^2 C_{22}^W \,\,=\,\, -\,  4 N_C L_{10} \, .
\label{eq.LEC-rel-appendix}
\ee

The possible validity of this relation (\ref{eq.SY-rel-appendix}) at the one-loop level
was studied in ref.~\cite{Gorsky:2012ui},
where the $AVV$  transition $A^3\to \gamma\gamma^*$ was analyzed.
Here $A^3$ refers to the $t^3$ generator for  the axial-vector current.
The Goldstone--loop contribution was computed  in the two-flavor
case including  singlet sources,
this is, in $U(2)$,    with the electric charge operator
${\cal Q}=t^3+\frac{1}{6}\mbox{\bf 1}=t^3+\frac{1}{3}t^0$.
Indeed, for $n_f=2$ the only non-zero flavor structures come from the
components $A^3\to V^0V^{3\,*}$ and  $A^3\to V^3V^{0\,*}$
in the $A^3\to\gamma\gamma^*$ transition.

The leading one-loop contribution in the $\chi$PT expansion, i.e.,
$\cO(p^6)$ in $w_T$    and    $\cO(p^4)$ in $\Pi_{V}-\Pi_{A}$,
was found to also fulfill the Son-Yamamoto relation~(\ref{eq.SY-rel-appendix}),
hence it was argued that this expression might be valid beyond large $N_C$,
at the loop level~\cite{Gorsky:2012ui}. We have investigated whether the agreement between the
one-loop corrections in~(\ref{eq.SY-rel-appendix})
is also valid at higher orders in the $\chi$PT expansion.
In particular, we have looked at the one-loop correction (single log) at $\cO(p^8)$ in $w_T$,
NLO in its $\chi$PT expansion.
We have also studied the effect of considering a higher number of light quarks  $n_f$
and different flavor decay structures $A^a\to V^b V^{c\, *}$.

\subsection{$VV-AA$ correlator and $A^3\to \gamma \gamma^*$  in $U(2)$}

Considering the $VV-AA$  correlator in the massless quark limit,
we get~\cite{Amoros:1999dp}
\bear
\Frac{N_C}{Q^2}\,+\,\Frac{N_C}{f_\pi^2}\, (\Pi_{V}-\Pi_A) &=&
\Frac{1}{f_\pi^2} \, \bigg[ \,-\,  4 N_C L_{10} \,
-\,  \Frac{n_f}{2}\,\Frac{N_C}{48\pi^2 } \ln\Frac{-q^2}{\nu^2} \,\bigg]
\label{eq-piLR}\\
&& +\, \Frac{q^2}{f_\pi^4}\,\bigg[\,8N_C C_{87}\, -\, L_9\, \times \,\Frac{n_f}{2}\, \Frac{N_C}{12\pi^2}\ln\Frac{-q^2}{\nu^2}
\,\,\,\,+\,\, \cO(N_C^0)\, \bigg]
\,\,\,\,+\,\, \cO(q^4)\, ,
\nn
\eear
where   the $U(n_f)$ singlet components    do not play here any role.
We present the results for an arbitrary number of light quarks,
even though we focus on the $n_f=2$ case.
We have used the $n_f=3$ notation
$L_{9}$,  $L_{10}$ and $C_{87}$     for the $\cO(p^4)$ and $\cO(p^6)$ LECs,
regardless of the number of light flavors.

For the transition $A^3\to \gamma\gamma^*$ with
electric charge operator ${\cal Q}=t^3+\frac{1}{3}t^0$,
the $\chi$PT calculation up to $\cO(p^8)$ yields:
\bear
w_T(Q^2) &=& \Frac{1}{f_\pi^2}\,\bigg[ \,
 128\pi^2f_\pi^2 \bigg(C_{22}^W - \Frac{\widetilde{c}_{13}}{2}\bigg)
\, - \,  \Frac{n_f}{2}\,\Frac{N_C}{48\pi^2 } \ln\Frac{-q^2}{\nu^2} \,\bigg]
\nn\\
\nn\\
&& \hspace*{-0.5cm}  +\, \Frac{q^2}{f_\pi^4}\,\bigg[\,
K  \,
-\,\bigg(  L_9  \, +\,
\Frac{16\pi^2 f_\pi^2}{N_C}\big(
- C_{13}^W +C_{14}^W+ C_{15}^W + C_{19}^W
-C_{20}^W- C_{21}^W + C_{22}^W
\big)  \bigg)\,
\nn\\
&&
\qquad \qquad\qquad\qquad\qquad
\times \,   \Frac{n_f}{2}\,\Frac{N_C}{24 \pi^2}\ln\Frac{-q^2}{\nu^2}
\,\,\,\,+\,\, \cO(N_C^0)\, \bigg]
\,\,\,\,\,+\,\, \cO(q^4)\, ,
\eear
where $K$ represents the corresponding $\cO(p^8)$ chiral low-energy constant.
For the sake of generality, we have computed the matrix element $w_T$  with the
flavor structure specified above for a general number of light flavors, from
where one can extract the expressions for the  $U(2)$ case.
Notice that in addition to the single trace operator
$C_{22}^W \epsilon^{\mu\nu\alpha\beta}
\bra  u_\mu\{ \nabla^\gamma f_{+\,\gamma \nu} , f_{+\,\alpha\beta}\} \ket$
at $\cO(p^6)$ one also needs to take into account the contribution from a
double-trace operator
$\widetilde{c}_{13} \epsilon^{\mu\nu\alpha\beta}
\bra \nabla^\gamma f_{+\, \gamma\mu}\ket\, \bra f_{+\,\nu\alpha} u_\beta\ket$
for $n_f\geq 3$~\cite{Bijnens:2001bb}, appearing only in the $U(n_f)$ theory.
This coupling  is $1/N_C$ suppressed with respect to $C_{22}^W$
and does not appear at large $N_C$. However, it is essential in order to renormalize
the various $A^a\to V^b V^{c\, *}$ flavor structures.

The last needed ingredient for the comparison with $\Pi_V-\Pi_A$  is the
value of the $\cO(p^6)$ odd-parity LECs which  multiply the logs in $w_T$
in terms of the $\cO(p^4)$ even-sector couplings.
For holographic models where the chiral symmetry is broken through
boundary conditions~\cite{Son:2003et,Hirn:2005nr,Sakai:2005yt,Sakai:2005yt},
we found the large--$N_C$ relations~(\ref{eq:C12})--(\ref{eq:C23})
for the relevant $C_k^W$ couplings, producing  the transverse amplitude:
\bear
w_T(Q^2) &=& \Frac{1}{f_\pi^2}\,\bigg[ \,
 128\pi^2f_\pi^2 \bigg(C_{22}^W - \Frac{\widetilde{c}_{13}}{2}\bigg)
\, - \,  \Frac{n_f}{2}\,  \Frac{N_C}{48\pi^2 } \ln\Frac{-q^2}{\nu^2} \,\bigg]
\nn\\
&& +\, \Frac{q^2}{f_\pi^4}\,\bigg[\,
K\,\, -\, \,\big(  L_9 - 2 L_1 \big)\,
\times \, \Frac{n_f}{2}\,  \Frac{N_C}{12\pi^2}\ln\Frac{-q^2}{\nu^2}
\,\,\,\,+\,\, \cO(N_C^0)\, \bigg]\,\, \,\,+\,\, \cO(q^4)\, .
\eear
The comparison of this result and $\Pi_{V}-\Pi_A$ in eq.~(\ref{eq-piLR}) shows that there is a disagreement at
$\cO(p^8)$ in the Son-Yamamoto relation,
and that the agreement at $\cO(p^6)$  is a coincidence.

\subsection{Comparison for fully non-singlet  transitions $A^a\to V^b V^{c\, *}$}

Indeed, the agreement found in ref.~\cite{Gorsky:2012ui}  at the lowest chiral order
only occurs for a particular choice of the
flavor structure of  vector and axial-vector currents.
In the case when all the three currents are $U(n_f)$ non-singlets,  one finds
for the $A^a\to V^b V^{c\, *}$ transition the transverse structure function:
\begin{eqnarray}
w_T(Q^2) &=& \Frac{1}{f_\pi^2}\,\bigg[ \,
 128\pi^2f_\pi^2  C_{22}^W
\, - \,  \Frac{n_f\, N_C}{72\pi^2 } \ln\Frac{-q^2}{\nu^2} \,\bigg]\,\,\,\,+\,\,\,\, \cO(q^2)\, ,
\end{eqnarray}
to be compared to the expression which derives from the $VV-AA$ correlator,
\begin{eqnarray}
\Frac{N_C}{Q^2}\, +\,\Frac{N_C}{f_\pi^2}\,(\, \Pi_V -\Pi_A\,) &=&
\Frac{1}{f_\pi^2}\,\bigg[ \,
-\, 4 N_C L_{10}
\, - \,  \Frac{n_f\, N_C}{96\pi^2 } \ln\Frac{-q^2}{\nu^2} \,\bigg]\,\,\,\,+\,\,\,\, \cO(q^2)\, ,
\end{eqnarray}
for a general number of flavors $n_f$. The result is also valid for  $n_f=2$,
although  in this case the overall group factor $d^{abc}$ in the amplitude is zero when all the  $a,b,c$
are non-singlet.
One can easily see that the leading one-loop logarithms of $w_T$ and $\Pi_V-\Pi_A$ do
not match.
This conclusion also comes from observing  that the corresponding couplings entering
at tree-level  have  different running~\cite{Bijnens:2001bb,Gasser:1983yg,Gasser:1984gg,Gasser:1984ux}
\footnote{ After the submission  of the manuscript to the arXiv we
became aware that the lack of matching
in $SU(3)\times SU(3)$   theory was also
noticed by S. Peris and M. Knecht~\cite{Peris2012}.}:
\bear
 \Frac{\mathd L_{10}}{\mathd \ln\nu^2} &=&
  -\, \Frac{\Gamma_{10}^{(n_f)}}{32\pi^2}
\,\,\,=\,\,\,    \Frac{n_f}{3}\,\Frac{1}{128\pi^2 }\, ,
\nn\\
\nn\\
-\, \Frac{32\pi^2 f_\pi^2}{N_C}\, \Frac{\mathd C_{22}^W}{\mathd\ln\nu^2} &=&
\Frac{32\pi^2 f_\pi^2}{N_C}\, \bigg( \, \Frac{\eta_{22}^{(n_f)}}{32\pi^2} \, \bigg)
\,\,\,=\,\,\,   \Frac{n_f}{3}\,\Frac{1}{96\pi^2 }\, .
\eear

\newpage


\begin{thebibliography}{10}

\bibitem{Weinberg:1978kz}
S. Weinberg.
\newblock Physica \textbf{A96} (1979): 327.

\bibitem{Gasser:1983yg}
J.~Gasser and H.~Leutwyler.
\newblock Annals Phys. \textbf{158} (1984): 142.

\bibitem{Gasser:1984gg}
J.~Gasser and H.~Leutwyler.
\newblock Nucl. Phys. \textbf{B250} (1985): 465.

\bibitem{Gasser:1984ux}
J.~Gasser and H.~Leutwyler.
\newblock Nucl. Phys. \textbf{B250} (1985): 517.

\bibitem{Bijnens:1999sh}
J. Bijnens, G. Colangelo, and G. Ecker.
\newblock JHEP \textbf{02} (1999): 020
  [\href{http://arxiv.org/abs/hep-ph/9902437}{arXiv: hep-ph/9902437}].

\bibitem{Bijnens:1999hw}
J.~Bijnens, G.~Colangelo, and G.~Ecker.
\newblock Annals Phys. \textbf{280} (2000): 100-139
  [\href{http://arxiv.org/abs/hep-ph/9907333}{arXiv: hep-ph/9907333}].

\bibitem{Bijnens:2006zp}
J. Bijnens.
\newblock Prog. Part. Nucl. Phys. \textbf{58} (2007): 521
  [\href{http://arxiv.org/abs/hep-ph/0604043}{arXiv: hep-ph/0604043}].

\bibitem{Strandberg:2003zf}

\newblock O. Strandberg (2003):
  [\href{http://arxiv.org/abs/hep-ph/0302064}{arXiv: hep-ph/0302064}].

\bibitem{GonzalezAlonso:2008rf}
M. Gonzalez-Alonso, A. Pich, and J. Prades.
\newblock Phys. Rev. \textbf{D78} (2008): 116012
  [\href{http://arxiv.org/abs/0810.0760}{arXiv: 0810.0760}].

\bibitem{GonzalezAlonso:2010xf}
M. Gonzalez-Alonso, A. Pich, and J. Prades.
\newblock Phys. Rev. \textbf{D82} (2010): 014019
  [\href{http://arxiv.org/abs/1004.4987}{arXiv: 1004.4987}].

\bibitem{Kampf:2006bn}
K. Kampf and B. Moussallam.
\newblock Eur. Phys. J. \textbf{C47} (2006): 723
  [\href{http://arxiv.org/abs/hep-ph/0604125}{arXiv: hep-ph/0604125}].

\bibitem{Masjuan:2008fr}
P. Masjuan and S. Peris.
\newblock Phys. Lett. \textbf{B663} (2008): 61
  [\href{http://arxiv.org/abs/0801.3558}{arXiv: 0801.3558}].

\bibitem{Masjuan:2008fv}
P.~Masjuan, S.~Peris, and J.J. Sanz-Cillero.
\newblock Phys. Rev. \textbf{D78} (2008): 074028
  [\href{http://arxiv.org/abs/0807.4893}{arXiv: 0807.4893}].

\bibitem{Ecker:1988te}
G.~Ecker, J.~Gasser, A.~Pich, and E.~de~Rafael.
\newblock Nucl. Phys. \textbf{B321} (1989): 311.

\bibitem{Ecker:1989yg}
G.~Ecker, J.~Gasser, H.~Leutwyler, A.~Pich, and E.~de~Rafael.
\newblock Phys. Lett. \textbf{B223} (1989): 425.

\bibitem{Jiang:2009uf}
S.-Z. Jiang, Y. Zhang, C. Li, and Q. Wang.
\newblock Phys. Rev. \textbf{D81} (2010): 014001
  [\href{http://arxiv.org/abs/0907.5229}{arXiv: 0907.5229}].

\bibitem{Jiang:2010wa}
S.-Z. Jiang and Q. Wang.
\newblock Phys. Rev. \textbf{D81} (2010): 094037
  [\href{http://arxiv.org/abs/1001.0315}{arXiv: 1001.0315}].

\bibitem{Son:2003et}
D.~T. Son and M.~A. Stephanov.
\newblock Phys. Rev. \textbf{D69} (2004): 065020
  [\href{http://arxiv.org/abs/hep-ph/0304182}{arXiv: hep-ph/0304182}].

\bibitem{ArkaniHamed:2001ca}
N. Arkani-Hamed, A.~G. Cohen, and H. Georgi.
\newblock Phys. Rev. Lett. \textbf{86} (2001): 4757
  [\href{http://arxiv.org/abs/hep-th/0104005}{arXiv: hep-th/0104005}].

\bibitem{Hill:2000mu}
C.~T. Hill, S. Pokorski, and J. Wang.
\newblock Phys. Rev. \textbf{D64} (2001): 105005
  [\href{http://arxiv.org/abs/hep-th/0104035}{arXiv: hep-th/0104035}].

\bibitem{Bando:1987br}
M. Bando, T. Kugo, and K. Yamawaki.
\newblock Phys.Rept. \textbf{164} (1988): 217.

\bibitem{Sakai:2004cn}
T. Sakai and S. Sugimoto.
\newblock Prog. Theor. Phys. \textbf{113} (2005): 843
  [\href{http://arxiv.org/abs/hep-th/0412141}{arXiv: hep-th/0412141}].

\bibitem{Hirn:2005nr}
J. Hirn and V. Sanz.
\newblock JHEP \textbf{12} (2005): 030
  [\href{http://arxiv.org/abs/hep-ph/0507049}{arXiv: hep-ph/0507049}].

\bibitem{Sakai:2005yt}
T. Sakai and S. Sugimoto.
\newblock Prog. Theor. Phys. \textbf{114} (2005): 1083
  [\href{http://arxiv.org/abs/hep-th/0507073}{arXiv: hep-th/0507073}].

\bibitem{Becciolini:2009fu}
D. Becciolini, M. Redi, and A. Wulzer.
\newblock JHEP \textbf{01} (2010): 074
  [\href{http://arxiv.org/abs/0906.4562}{arXiv: 0906.4562}].

\bibitem{DaRold:2005zs}
L. Da~Rold and A. Pomarol.
\newblock Nucl. Phys. \textbf{B721} (2005): 79
  [\href{http://arxiv.org/abs/hep-ph/0501218}{arXiv: hep-ph/0501218}].

\bibitem{Erlich:2005qh}
J. Erlich, E. Katz, D.~T. Son, and M.~A. Stephanov.
\newblock Phys. Rev. Lett. \textbf{95} (2005): 261602
  [\href{http://arxiv.org/abs/hep-ph/0501128}{arXiv: hep-ph/0501128}].

\bibitem{Son:2010vc}

\newblock D.~T. Son and N. Yamamoto (2010):
  [\href{http://arxiv.org/abs/1010.0718}{arXiv: 1010.0718}].

\bibitem{Vainshtein:2002nv}
A. Vainshtein.
\newblock Phys. Lett. \textbf{B569} (2003): 187
  [\href{http://arxiv.org/abs/hep-ph/0212231}{arXiv: hep-ph/0212231}].

\bibitem{Colangelo:2011xk}
P.~Colangelo, F.~De~Fazio, J.J. Sanz-Cillero, F.~Giannuzzi, and S.~Nicotri.
\newblock Phys. Rev. \textbf{D85} (2012): 035013
  [\href{http://arxiv.org/abs/1108.5945}{arXiv: 1108.5945}].

\bibitem{Iatrakis:2011ht}
I. Iatrakis and E. Kiritsis.
\newblock JHEP \textbf{1202} (2012): 064
  [\href{http://arxiv.org/abs/1109.1282}{arXiv: 1109.1282}].

\bibitem{Cappiello:2010tu}
L. Cappiello, O. Cata, and G. D'Ambrosio.
\newblock Phys. Rev. \textbf{D82} (2010): 095008
  [\href{http://arxiv.org/abs/1004.2497}{arXiv: 1004.2497}].

\bibitem{Domokos:2011dn}
S.K. Domokos, J.A. Harvey, and A.B. Royston.
\newblock JHEP \textbf{1105} (2011): 107
  [\href{http://arxiv.org/abs/1101.3315}{arXiv: 1101.3315}].

\bibitem{Alvares:2011wb}
R. Alvares, C. Hoyos, and A. Karch.
\newblock Phys. Rev. \textbf{D84} (2011): 095020
  [\href{http://arxiv.org/abs/1108.1191}{arXiv: 1108.1191}].

\bibitem{Gorsky:2012ui}
A.~Gorsky, P.N. Kopnin, A.~Krikun, and A.~Vainshtein.
\newblock Phys. Rev. \textbf{D85} (2012): 086006
  [\href{http://arxiv.org/abs/1201.2039}{arXiv: 1201.2039}].

\bibitem{Knecht:2011wh}
M. Knecht, S. Peris, and E. de~Rafael.
\newblock JHEP \textbf{1110} (2011): 048
  [\href{http://arxiv.org/abs/1101.0706}{arXiv: 1101.0706}].

\bibitem{Kampf:2011th}
K. Kampf.
\newblock Nucl. Phys. Proc. Suppl. \textbf{219-220} (2011): 64
  [\href{http://arxiv.org/abs/1109.4576}{arXiv: 1109.4576}].

\bibitem{GomezDumm:2003ku}
D.~Gomez~Dumm, A.~Pich, and J.~Portoles.
\newblock Phys. Rev. \textbf{D69} (2004): 073002
  [\href{http://arxiv.org/abs/hep-ph/0312183}{arXiv: hep-ph/0312183}].

\bibitem{Dumm:2009va}
D.~Gomez Dumm, P.~Roig, A.~Pich, and J.~Portoles.
\newblock Phys. Lett. \textbf{B685} (2010): 158
  [\href{http://arxiv.org/abs/0911.4436}{arXiv: 0911.4436}].

\bibitem{Grigoryan:2008cc}
H.~R. Grigoryan and A.~V. Radyushkin.
\newblock Phys. Rev. \textbf{D78} (2008): 115008
  [\href{http://arxiv.org/abs/0808.1243}{arXiv: 0808.1243}].

\bibitem{Stoffers:2011xe}
A. Stoffers and I. Zahed.
\newblock Phys. Rev. \textbf{C84} (2011): 025202
  [\href{http://arxiv.org/abs/1104.2081}{arXiv: 1104.2081}].

\bibitem{Wess:1971yu}
J.~Wess and B.~Zumino.
\newblock Phys. Lett. \textbf{B37} (1971): 95.

\bibitem{Witten:1983tw}
E. Witten.
\newblock Nucl. Phys. \textbf{B223} (1983): 422.

\bibitem{Witten:1983tx}
E. Witten.
\newblock Nucl. Phys. \textbf{B223} (1983): 433.

\bibitem{Bijnens:2001bb}
J.~Bijnens, L.~Girlanda, and P.~Talavera.
\newblock Eur. Phys. J. \textbf{C23} (2002): 539
  [\href{http://arxiv.org/abs/hep-ph/0110400}{arXiv: hep-ph/0110400}].

\bibitem{Ebertshauser:2001nj}
T.~Ebertshauser, H.W. Fearing, and S.~Scherer.
\newblock Phys. Rev. \textbf{D65} (2002): 054033
  [\href{http://arxiv.org/abs/hep-ph/0110261}{arXiv: hep-ph/0110261}].

\bibitem{'tHooft:1973jz}
G. 't~Hooft.
\newblock Nucl. Phys. \textbf{B72} (1974): 461.

\bibitem{'tHooft:1974hx}
G. 't~Hooft.
\newblock Nucl. Phys. \textbf{B75} (1974): 461.

\bibitem{Witten:1979kh}
E. Witten.
\newblock Nucl. Phys. \textbf{B160} (1979): 57.

\bibitem{Knecht:1997ts}
M. Knecht and E. de~Rafael.
\newblock Phys. Lett. \textbf{B424} (1998): 335
  [\href{http://arxiv.org/abs/hep-ph/9712457}{arXiv: hep-ph/9712457}].

\bibitem{Peris:1998nj}
S. Peris, M. Perrottet, and E. de~Rafael.
\newblock JHEP \textbf{9805} (1998): 011
  [\href{http://arxiv.org/abs/hep-ph/9805442}{arXiv: hep-ph/9805442}].

\bibitem{Golterman:2001nk}
M. Golterman and S. Peris.
\newblock JHEP \textbf{0101} (2001): 028
  [\href{http://arxiv.org/abs/hep-ph/0101098}{arXiv: hep-ph/0101098}].

\bibitem{Golterman:2006gv}
M. Golterman and S. Peris.
\newblock Phys.Rev. \textbf{D74} (2006): 096002
  [\href{http://arxiv.org/abs/hep-ph/0607152}{arXiv: hep-ph/0607152}].

\bibitem{Weinberg:1967kj}
S. Weinberg.
\newblock Phys. Rev. Lett. \textbf{18} (1967): 507.

\bibitem{Shifman:1978bx}
M.~A. Shifman, A.I. Vainshtein, and V.~I. Zakharov.
\newblock Nucl. Phys. \textbf{B147} (1979): 385.

\bibitem{Guo:2007ff}
Z.H. Guo, J.J. Sanz~Cillero, and H.Q. Zheng.
\newblock JHEP \textbf{0706} (2007): 030
  [\href{http://arxiv.org/abs/hep-ph/0701232}{arXiv: hep-ph/0701232}].

\bibitem{Guo:2007hm}
Z.H. Guo, J.J. Sanz-Cillero, and H.Q. Zheng.
\newblock Phys. Lett. \textbf{B661} (2008): 342
  [\href{http://arxiv.org/abs/0710.2163}{arXiv: 0710.2163}].

\bibitem{Karch:2006pv}
A. Karch, E. Katz, D.~T. Son, and M.~A. Stephanov.
\newblock Phys. Rev. \textbf{D74} (2006): 015005
  [\href{http://arxiv.org/abs/hep-ph/0602229}{arXiv: hep-ph/0602229}].

\bibitem{Colangelo:2008us}
P.~Colangelo, F.~De~Fazio, F. Giannuzzi, F.~Jugeau, and S.~Nicotri.
\newblock Phys. Rev. \textbf{D78} (2008): 055009
  [\href{http://arxiv.org/abs/0807.1054}{arXiv: 0807.1054}].

\bibitem{Zuo:2009dz}
F. Zuo.
\newblock Phys. Rev. \textbf{D82} (2010): 086011
  [\href{http://arxiv.org/abs/0909.4240}{arXiv: 0909.4240}].

\bibitem{Benayoun:2009im}
M.~Benayoun, P.~David, L.~DelBuono, and O.~Leitner.
\newblock Eur. Phys. J. \textbf{C65} (2010): 211
  [\href{http://arxiv.org/abs/0907.4047}{arXiv: 0907.4047}].

\bibitem{Kampf:2011ty}
K. Kampf and J. Novotny.
\newblock Phys. Rev. \textbf{D84} (2011): 014036
  [\href{http://arxiv.org/abs/1104.3137}{arXiv: 1104.3137}].

\bibitem{Bijnens:1989jb}
J. Bijnens, A. Bramon, and F. Cornet.
\newblock Z. Phys. \textbf{C46} (1990): 599.

\bibitem{Pallante:1992qe}
E.~Pallante and R.~Petronzio.
\newblock Nucl. Phys. \textbf{B396} (1993): 205.

\bibitem{Moussallam:1994xp}
B.~Moussallam.
\newblock Phys. Rev. \textbf{D51} (1995): 4939
  [\href{http://arxiv.org/abs/hep-ph/9407402}{arXiv: hep-ph/9407402}].

\bibitem{RuizFemenia:2003hm}
P.D. Ruiz-Femenia, A.~Pich, and J.~Portoles.
\newblock JHEP \textbf{0307} (2003): 003
  [\href{http://arxiv.org/abs/hep-ph/0306157}{arXiv: hep-ph/0306157}].

\bibitem{Unterdorfer:2008zz}
R. Unterdorfer and H. Pichl.
\newblock Eur. Phys. J. \textbf{C55} (2008): 273
  [\href{http://arxiv.org/abs/0801.2482}{arXiv: 0801.2482}].

\bibitem{Masjuan:2012wy}

\newblock P.~Masjuan (2012): [\href{http://arxiv.org/abs/1206.2549}{arXiv:
  1206.2549}].

\bibitem{Mateu:2007tr}
V.~Mateu and J.~Portoles.
\newblock Eur. Phys. J. \textbf{C52} (2007): 325
  [\href{http://arxiv.org/abs/0706.1039}{arXiv: 0706.1039}].

\bibitem{Cirigliano:2006hb}
V.~Cirigliano, G.~Ecker, M.~Eidemuller, R. Kaiser, A.~Pich, et~al.
\newblock Nucl. Phys. \textbf{B753} (2006): 139
  [\href{http://arxiv.org/abs/hep-ph/0603205}{arXiv: hep-ph/0603205}].

\bibitem{Kampf:2006yf}
K. Kampf, J. Novotny, and J. Trnka.
\newblock Eur. Phys. J. \textbf{C50} (2007): 385
  [\href{http://arxiv.org/abs/hep-ph/0608051}{arXiv: hep-ph/0608051}].

\bibitem{Bando:1987ym}
M. Bando, T. Fujiwara, and K. Yamawaki.
\newblock Prog.Theor.Phys. \textbf{79} (1988): 1140.

\bibitem{Gasser:2005ud}
J. Gasser, M.~A. Ivanov, and M.~E. Sainio.
\newblock Nucl. Phys. \textbf{B728} (2005): 31
  [\href{http://arxiv.org/abs/hep-ph/0506265}{arXiv: hep-ph/0506265}].

\bibitem{Gasser:2006qa}
J. Gasser, M.~A. Ivanov, and M.~E. Sainio.
\newblock Nucl. Phys. \textbf{B745} (2006): 84
  [\href{http://arxiv.org/abs/hep-ph/0602234}{arXiv: hep-ph/0602234}].

\bibitem{Colangelo:2001df}
G.~Colangelo, J.~Gasser, and H.~Leutwyler.
\newblock Nucl. Phys. \textbf{B603} (2001): 125
  [\href{http://arxiv.org/abs/hep-ph/0103088}{arXiv: hep-ph/0103088}].

\bibitem{Bijnens:2002hp}
J. Bijnens and P.~Talavera.
\newblock JHEP \textbf{0203} (2002): 046
  [\href{http://arxiv.org/abs/hep-ph/0203049}{arXiv: hep-ph/0203049}].

\bibitem{Amoros:1999dp}
G. Amoros, J. Bijnens, and P.~Talavera.
\newblock Nucl. Phys. \textbf{B568} (2000): 319
  [\href{http://arxiv.org/abs/hep-ph/9907264}{arXiv: hep-ph/9907264}].

\bibitem{Bellucci:1994eb}
S.~Bellucci, J.~Gasser, and M.E. Sainio.
\newblock Nucl. Phys. \textbf{B423} (1994): 80
  [\href{http://arxiv.org/abs/hep-ph/9401206}{arXiv: hep-ph/9401206}].

\bibitem{Geng:2003mt}
C.Q. Geng, I-Lin Ho, and T.H. Wu.
\newblock Nucl. Phys. \textbf{B684} (2004): 281
  [\href{http://arxiv.org/abs/hep-ph/0306165}{arXiv: hep-ph/0306165}].

\bibitem{Cirigliano:2004ue}
V.~Cirigliano, G.~Ecker, M.~Eidemuller, A.~Pich, and J.~Portoles.
\newblock Phys. Lett. \textbf{B596} (2004): 96
  [\href{http://arxiv.org/abs/hep-ph/0404004}{arXiv: hep-ph/0404004}].

\bibitem{Bijnens:1998fm}
J.~Bijnens, G.~Colangelo, and P.~Talavera.
\newblock JHEP \textbf{9805} (1998): 014
  [\href{http://arxiv.org/abs/hep-ph/9805389}{arXiv: hep-ph/9805389}].

\bibitem{Bijnens:1997vq}
J.~Bijnens, G.~Colangelo, G.~Ecker, J.~Gasser, and M.E. Sainio.
\newblock Nucl. Phys. \textbf{B508} (1997): 263
  [\href{http://arxiv.org/abs/hep-ph/9707291}{arXiv: hep-ph/9707291}].

\bibitem{Guo:2009hi}
Z.H. Guo and J.~J. Sanz-Cillero.
\newblock Phys. Rev. \textbf{D79} (2009): 096006
  [\href{http://arxiv.org/abs/0903.0782}{arXiv: 0903.0782}].

\bibitem{Pich:2008jm}
A.~Pich, I.~Rosell, and J.J. Sanz-Cillero.
\newblock JHEP \textbf{0807} (2008): 014
  [\href{http://arxiv.org/abs/0803.1567}{arXiv: 0803.1567}].

\bibitem{Pich:2010sm}
A. Pich, I. Rosell, and J.~J. Sanz-Cillero.
\newblock JHEP \textbf{1102} (2011): 109
  [\href{http://arxiv.org/abs/1011.5771}{arXiv: 1011.5771}].

\bibitem{Bijnens:1996wm}
J.~Bijnens and P.~Talavera.
\newblock Nucl. Phys. \textbf{B489} (1997): 387
  [\href{http://arxiv.org/abs/hep-ph/9610269}{arXiv: hep-ph/9610269}].

\bibitem{Polchinski1998}
 J. Polchinski, {\it String Theory}. {\it Vol. 1: An Introduction to the Bosonic
    String}, {\it Vol 2: Superstring Theory and Beyond}. Cambridge University Press, 1998.

\bibitem{Kampf:2005tz}
Karol Kampf, Marc Knecht, and Jiri Novotny.
\newblock Eur. Phys. J. \textbf{C46} (2006): 191
  [\href{http://arxiv.org/abs/hep-ph/0510021}{arXiv: hep-ph/0510021}].


\bibitem{Zuo:2011sk}
F. Zuo and T. Huang.
\newblock Eur. Phys. J. \textbf{C72} (2012): 1813
  [\href{http://arxiv.org/abs/1105.6008}{arXiv: 1105.6008}].

\bibitem{TFFrefs}
A.V. Radyushkin, Phys. Rev. D \textbf{80} (2009): 094009;
M.V. Polyakov, JETP Lett. \textbf{90} (2009): 228;
S.V. Mikhailov and N.G. Stefanis,
Mod. Phys. Lett. A \textbf{24} (2009): 2858;
X.~-G.~Wu and T.~Huang,
Phys. Rev. D \textbf{82}, (2010): 034024;
H.L.L. Roberts, C.D. Roberts, A. Bashir,
L.X. Guti$\acute{\rm e}$rrez-Guerrero and P.C. Tandy,
Phys. Rev. C \textbf{82} (2010): 065202;
T.N. Pham and X.Y. Pham,
Int. J. Mod. Phys. A \textbf{26} (2011): 4125;
S.S. Agaev, V.M. Braun, N. Offen, and F.A. Porkert,
Phys. Rev. D \textbf{83} (2011): 054020;
S.J. Brodsky, F.-G. Cao and G.F. de T\'{e}ramond,
Phys. Rev. D \textbf{84} (2011): 033001;
Phys. Rev. D \textbf{84} (2011): 075012.


\bibitem{Grigoryan:2008up}
H.~R. Grigoryan and A.~V. Radyushkin.
\newblock Phys. Rev. \textbf{D77} (2008): 115024
  [\href{http://arxiv.org/abs/0803.1143}{arXiv: 0803.1143}].

\bibitem{Lepage:1979zb}
G.~P. Lepage and S.~J. Brodsky.
\newblock Phys. Lett. \textbf{B87} (1979): 359-365.

\bibitem{Lepage:1980fj}
G.~P. Lepage and S.~J. Brodsky.
\newblock Phys. Rev. \textbf{D22} (1980): 2157.

\bibitem{Brodsky:1981rp}
S.~J. Brodsky and G.~P. Lepage.
\newblock Phys. Rev. \textbf{D24} (1981): 1808.

\bibitem{Bebek:1977pe}
C.J. Bebek, C.N. Brown, S.~D. Holmes, R.V. Kline, F.M. Pipkin, et~al.
\newblock Phys. Rev. \textbf{D17} (1978): 1693.

\bibitem{Huber:2008id}
G.M. Huber et~al. (Jefferson Lab Collaboration).
\newblock Phys. Rev. \textbf{C78} (2008): 045203
  [\href{http://arxiv.org/abs/0809.3052}{arXiv: 0809.3052}].


\bibitem{Gronberg:1997fj}
J.~Gronberg et~al. (CLEO Collaboration).
\newblock Phys. Rev. \textbf{D57} (1998): 33
  [\href{http://arxiv.org/abs/hep-ex/9707031}{arXiv: hep-ex/9707031}].

\bibitem{Aubert:2009mc}
B. Aubert et~al. (The BABAR Collaboration).
\newblock Phys. Rev. \textbf{D80} (2009): 052002
  [\href{http://arxiv.org/abs/0905.4778}{arXiv: 0905.4778}].

\bibitem{Uehara:2012ag}

\newblock S.~Uehara et~al. (Belle Collaboration):
  [\href{http://arxiv.org/abs/1205.3249}{arXiv: 1205.3249}].

\bibitem{Domokos:2009cq}
  S.~K.~Domokos, H.~R.~Grigoryan and J.~A.~Harvey,
  Phys. Rev. \textbf{D80} (2009): 115018
  [\href{http://arxiv.org/abs/0905.1949}{arXiv: 1905.1949}].

\bibitem{Peris2012}
 S. Peris, at ``2012 Project X Physics Study", Chicago, 14-23 June 2012;
 [\href{https://indico.fnal.gov/conferenceDisplay.py?ovw=True&confId=5276}{https://indico.fnal.gov/conferenceDisplay.py?ovw=True\&confId=5276}].



\end{thebibliography}



\end{document}